\documentclass[11pt]{article}

\usepackage{sectsty}
\usepackage{graphicx}
\usepackage{authblk}

\usepackage{amsmath}
\usepackage{amssymb}
\usepackage{natbib}
\usepackage{hyperref}

\begin{document}
\title{The 2024 outburst of the neutron star
LMXB EXO 0748--676: an investigation of bursts and eclipses with \textit{AstroSat}}

\author[1]{Aromal P\thanks{aromalp6i@gmail.com}}
\author[2]{Unnati Kashyap\thanks{ukashyap@ttu.edu}}
\author[1]{Manoneeta Chakraborty\thanks{manoneeta@iiti.ac.in}}
\author[3,4]{Sudip Bhattacharyya\thanks{sudip@tifr.res.in}}
\author[2]{Thomas J. Maccarone\thanks{Thomas.Maccarone@ttu.edu}}
\author[1]{Vijay Choudhary\thanks{vijaychoudhary9517@gmail.com}}

\affil[1]{Department of Astronomy, Astrophysics and Space Engineering\\
Indian Institute of Technology Indore, Indore 453552, India}

\affil[2]{Department of Physics and Astronomy\\
Texas Tech University, Box 41051, Lubbock, TX 79409-1051, USA}

\affil[3]{Department of Astronomy and Astrophysics\\
Tata Institute of Fundamental Research, 1 Homi Bhabha Road, Colaba, Mumbai 400005, Maharashtra, India}

\affil[4]{MIT Kavli Institute for Astrophysics and Space Research\\
Massachusetts Institute of Technology, 70 Vassar St, Cambridge, MA 02139, USA}
\date{} 
\maketitle	

\begin{abstract}
We present a detailed analysis of the Type-I (thermonuclear) X-ray bursts and eclipses observed from the neutron star low-mass X-ray binary EXO 0748–676 with AstroSat during the second known outburst of the source following a 16-year-long quiescence period. We detect three thermonuclear X-ray bursts, with two displaying simultaneous coverage in the soft X-rays. Simultaneous UV observations show evidence of reprocessed burst emissions in the far-ultraviolet band. The time-resolved spectral analysis reveals the photospheric radius expansion (PRE) nature for two bursts. We estimate the distance to the source to be $7.42\pm0.53$ kpc using the peak flux of PRE. Notably, one of the bursts exhibited a secondary peak, $\sim30$ s after the primary, particularly dominating in the softer X-rays, which reveals a correlation with the evolution of burst hotspot radius with no temperature dependence. The burst properties and corresponding flux values suggest that mixed H/He burning may have fueled the bursts. We also detect evidence of a soft excess during one burst, likely arising from the interaction of the burst photons with the surroundings. We uncover evidence for a hard X-ray deficit during the peak of all bursts and a hard lag of $\sim4$ s, which can be attributed to the Compton cooling of the corona by the burst photons. We also probe the temporal evolution and the energy dependence of the eclipses, which offer insights into the binary environment. Our study helps gain deeper insight into the physics of burst ignition, flame propagation, the burst-accretion interaction, and the evolution of LMXBs.
\end{abstract}


\pagebreak
\section{Introduction}
\label{sec:intro}
A neutron star (NS) low mass X-ray binary (LMXB) is a binary stellar system consisting of a neutron star and a companion star ( $\rm M<1\,M_\odot$) via Roche-lobe overflow \citep{bambi_low-mass_2023}.  Due to the high gravitational pull of the compact NS in this system, the matter from the companion star flows towards the NS, forming an accretion disk around the NS. The inner parts of the accretion disk are highly heated ($\sim 10^7$ K) with the emissions dominating in the X-ray regime of the electromagnetic spectrum. The matter in the accretion disk loses angular momentum and slowly spirals onto the surface of the neutron star.

The matter accumulated on the surface of the NS under certain favorable conditions undergoes an unstable nuclear reaction, immediately followed by a thermal runaway process. This manifests itself as a sudden increase in the X-ray flux from the source, several times higher than the persistent intensity level. This phenomenon is called a thermonuclear X-ray Burst (or Type-I Burst; hereafter burst) \citep{lewin_x-ray_1993, strohmayer_new_2003, galloway_multi-instrument_2020}. The burst is characterized by a fast rise of 0.5-5 s to reach the peak and an exponential or power-law decay that lasts for tens to hundreds of seconds. A single burst can radiate $10^{38}$-$10^{41}$ erg energy \citep{lewin_x-ray_1993}. An evolving Planckian function is often used to represent burst spectra with a temperature varying between 1–3 keV \citep{kuulkers_photospheric_2003}. During the peak of some bursts, an excess in low-energy photons (soft excess) is observed compared to that predicted by a simple blackbody model \citep{in_t_zand_bright_2013, buisson_discovery_2020, kashyap_broadband_2021}.

In powerful X-ray bursts where burst luminosity exceeds the Eddington limit, radiation pressure exceeds gravity, causing the photosphere of the neutron star to expand. This phenomenon, called Photospheric Radius Expansion (PRE), is observed as the temperature decreases while the inferred radius of the hotspot simultaneously increases \citep{strohmayer_new_2003}. PRE can serve as a standard candle for estimating the distance to the source, as during PRE, the source luminosity reaches the Eddington limit \citep{kuulkers_photospheric_2003}. Utilizing the properties of PRE, the equations of the state of exotic matter inside the NS can also be constrained \citep{ozel_soft_2006, bhattacharyya_measurement_2010}. After the expansion phase, the photosphere falls back onto the NS, resulting in an increase in temperature and a decrease in radius. The PRE nature can be identified by the consequent double peak in the temperature evolution near the peak of the bursts \citep{keek_nicer_2018}.

During the decay phase of some bursts, another weaker intensity peak is occasionally detected. This late secondary peak can reach an intensity comparable to that of the main peak in the lower energies. This kind of late peak is usually called a secondary peak. Secondary peaks are relatively rare, but they are observed in several sources (\citealt{guver_thermonuclear_2021,jaisawal_nicer_2019} and references therein). The origin of the secondary peak can be attributed either to the stalling of the flame front or to secondary burning. Flame front stalling can occur because of the Coriolis force \citep{bhattacharyya_thermonuclear_2007} or due to arresting by the magnetic field on the surface \citep{payne_frequency_2006}, which may last for a few seconds. Our understanding of the mechanism that triggers late secondary-peaked bursts is limited.

Another interesting emission variability feature observed in some LMXBs is the eclipse, which can offer critical insights regarding the binary system. Eclipses in binaries occur because of the companion star's occultation of the X-ray emitting regions. Full eclipses are prominently observed in systems with high inclination. Eclipses can provide the orbital period of the binary system, which is one of the most fundamental parameters for such systems. Regular monitoring of eclipses over time can be used to study the evolution associated with the binary orbit \citep{wolff_eclipse_2009,schaefer_evolutionary_2025}. Eclipses can also provide information about the binary mass function, allowing us to determine the mass and radius of the central compact object \citep{knight_eclipse_2022}.


EXO~0748$-$676 (hereafter EXO~0748) is a prolific bursting and eclipsing LMXB, discovered by the EXOSAT satellite in 1985 \citep{parmar_exo0748-676_1985}. Due to the high inclination ($i=76^{\circ}$) of the system, clear eclipses are regularly observed from EXO~0748. From eclipse analysis, the orbital period of the binary system is found to be 3.82 hr \citep{parmar_discovery_1986}. Eclipses last for nearly 500 s, with ingress and egress times ranging from less than 1 s to 30 s. EXO~0748 went into quiescence in 2008, after a relatively long outburst of 24 years \citep{degenaar_chandra_2009}. During this outburst period, 359 bursts are listed in Multi-INstrument Burst ARchive (MINBAR) \citep{galloway_multi-instrument_2020}, and 392 full eclipses were observed using the Rossi X-ray Timing Explorer (\emph{RXTE}) satellite \citep{wolff_eclipse_2009} from the source. The analysis of the PRE burst using the \emph{RXTE/PCA} data for EXO~0748 conducted by \cite{ozel_soft_2006} indicates that the NS has a mass of $2.10 \pm 0.28\rm \: M_{\odot}$ and a radius of $13.8 \pm 1.8 \: \text{km}$. Additionally, in a study conducting eclipse analysis of the system with \emph{XMM-Newton/EPIC}, \cite{knight_eclipse_2022} derived the mass of the neutron star to be $2.01^{+0.22}_{-0.21}\rm \: M_\odot$. EXO~0748 has an M-dwarf companion with a mass of $0.45 \rm\: M_\odot$ \citep{parmar_discovery_1986}. 

During its first outburst, EXO~0748 was regularly monitored and was found to exhibit several interesting observational features such as thermonuclear X-ray bursts, burst triplets \citep{boirin_discovery_2007}, reprocessing of bursts \citep{hynes_multiwavelength_2006}, eclipses, dips \citep{homan_xmm-newton_2003}, burst during eclipses \citep{knight_type_2025, rikame_thermonuclear_2025}, mHz quasi-periodic oscillations (QPOs) \citep{mancuso_discovery_2019}, kHz QPOs \citep{homan_695_2000}, burst oscillations \citep{galloway_discovery_2010, villarreal_discovery_2004}, ultraviolet burst oscillation candidate \citep{zanon_ultraviolet_2025} etc., making it a suitable candidate to study the exotic environment surrounding the central compact object. Regular eclipse monitoring was conducted for the source during the previous outburst \citep{wolff_eclipse_2009}, and an indication of magnetic activities in the companion star was inferred from it. After this long outburst lasting 24 years, EXO~0748 went into a prolonged quiescence of 16 years. Following this long quiescence period, EXO~0748 showed its second-ever outburst in 2024. This provides an ideal opportunity to study the evolution of the system following a prolonged quiescence. Investigations of the observational properties of the system during the current outburst can reveal any comparative variations in the binary system between the two active states of this source.

EXO~0748 was observed during its 2024 outburst using several observatories, such as The neutron star interior composition explorer (\emph{NICER}), Nuclear Spectroscopic Telescope Array (\emph{NuSTAR}), X-ray Multi-Mirror Mission-Newton (\emph{XMM-Newton}), and \emph{AstroSat}. Bursts and eclipses were reported from \emph{XMM-Newton} and \emph{NuSTAR} \citep{knight_simultaneous_2024, subba_eclipse_2024}. Additionally, \emph{XMM-Newton} also detected reprocessed bursts in the optical spectrum and dips in the X-ray light curve \citep{knight_simultaneous_2024, bhattacharya_xmm-newton_2024}. In this work, we present a detailed analysis of the bursts and eclipses detected from our \emph{AstroSat} observations of EXO~0748 during the 2024 outburst.

\section{Observation and Data Reduction}
\label{sec:Observation}
\begin{table*}[b]
\setlength{\tabcolsep}{4pt} 
\centering
\begin{tabular}{ c c c c c }
\hline
Observation & Orbit & Start & Stop & Exposure\\
ID & number & (dd-mm-yy -- hh:mm:ss) & (dd-mm-yy -- hh:mm:ss) & (ks)\\
\hline
9000006344 &47707-47742 &  24-07-2024 -- 08:01:03 & 26-07-2024 -- 19:37:41 &30\\
\hline
\end{tabular}
\caption{Observation details of EXO 0748--676, including observation ID, orbit number, dates of observation, start and stop time, and exposure time.
\label{tab:exo_obs}}
\end{table*}

In this study, we have utilized simultaneous data from three payloads onboard the Indian multi-wavelength mission \emph{AstroSat}: the Large Area X-ray Proportional Counter (LAXPC), the Soft X-ray Telescope (SXT), and the Ultraviolet Imaging Telescope (UVIT). The details of the observations are provided in Table~\ref{tab:exo_obs} \citep{Kashyap2024}. LAXPC data were collected in Event Analysis (EA) mode, SXT data were collected in Fast Windowed Photon Counting (FW) mode, and UVIT data were collected in Integration (IM) and Photon Counting (PC) modes.

\subsection{LAXPC}
LAXPC \footnote{\href{https://www.tifr.res.in/~astrosat_laxpc/astrosat_laxpc.html}{https://www.tifr.res.in/~astrosat\_laxpc/
astrosat\_laxpc.html}} consists of three identical and independent Xenon-based proportional counters (namely LAXPC10, LAXPC20, and LAXPC30) having an energy band pass of 3-80 keV with individual units having an effective area of 6000 cm$^2$ around 15 keV \citep{yadav_large_2017}. LAXPC has a timing resolution of 10 $\rm \mu s$, which makes it highly suitable for timing analysis. We have excluded data from LAXPC10 and LAXPC30 due to gain instability arising from gas leakage, and so in this work, we only present data from LAXPC20. 

The level 1 data from the \emph{AstroSat} Archive\footnote{\href{https://astrobrowse.issdc.gov.in/astro_archive/archive/Home.jsp}{https://astrobrowse.issdc.gov.in/astro\_archive/archive/Home.jsp}} was processed into a Level 2 event file using the {\tt LaxpcSoft} pipeline, provided by the \emph{AstroSat} Science Support Cell\footnote{\href{http://astrosat-ssc.iucaa.in/laxpcData}{http://astrosat-ssc.iucaa.in/laxpcData}} \citep{antia_calibration_2017}. We utilized this software to generate the spectrum, background spectrum, and response files. The \texttt{as1bary}\footnote{\href{http://astrosat-ssc.iucaa.in/data_and_analysis}{http://astrosat-ssc.iucaa.in/data\_and\_analysis}} tool was used to apply barycentric correction to the LAXPC Level 2 data.

\subsection{SXT}
SXT \footnote{\href{https://www.tifr.res.in/~astrosat_sxt/index.html}{https://www.tifr.res.in/~astrosat\_sxt/index.html}} is a Wolter-I type imaging telescope based on the grazing incidence principle working in the energy range 0.3–8.0 keV. SXT has an effective area of 90 cm$^2$ at 1.5 kV and has a focal length of 2 m \citep{singh_soft_2017}. Fast Windowed Photon Counting (FW) mode provides a time resolution of 0.278 s.

We obtained Level 2 data from the \emph{AstroSat} data archive. Using the \texttt{xselect} package from \texttt{HeaSoft} version \texttt{17Jul2023\_V6.SS32}, we created the spectra. A circular region with a radius of 5.8 arcminutes was defined as the source region centered around the source location at an RA of $117.35^\circ$ and a DEC of $-67.74^\circ$. For this analysis, we used the response file (sxt\_pc\_mat\_g0to12.rmf) and the blank sky background file (SkyBkg\_sxt\_LE0p35\_R16p0\_v05\_Gd0to12.pha) provided for dim sources by the SXT team. We generated the appropriate ancillary response files using the {\tt sxtARFmodule}.

\subsection{UVIT}
UVIT \footnote{\href{https://www.iiap.res.in/projects/uvit/instrument/}{https://www.iiap.res.in/projects/uvit/instrument/}} consists of two co-aligned telescopes, each comprising $f/12$ Ritchey–Chretien optics with an aperture of 375 mm, along with filters and detectors. One telescope observes in the far-ultraviolet (FUV) range, in a wavelength spanning from 1,300 to 1,800 \r{A}, while the other covers the near-ultraviolet (NUV) range of 2,000 to 3,000 \r{A} and in the visible (VIS) range of 3,200 to 5,500 \r{A} as well. A filter wheel is employed for each of the three channels to select either a filter or a grating in the FUV and NUV ranges \citep{tandon_-orbit_2017}. The UVIT (FUV) light curves were generated using the CURVFIT Python package \citep{joseph_curvit_2021}

\section{Data Analysis and Results}
\begin{figure*}[ht]
    \centering
    \includegraphics[width=\textwidth]{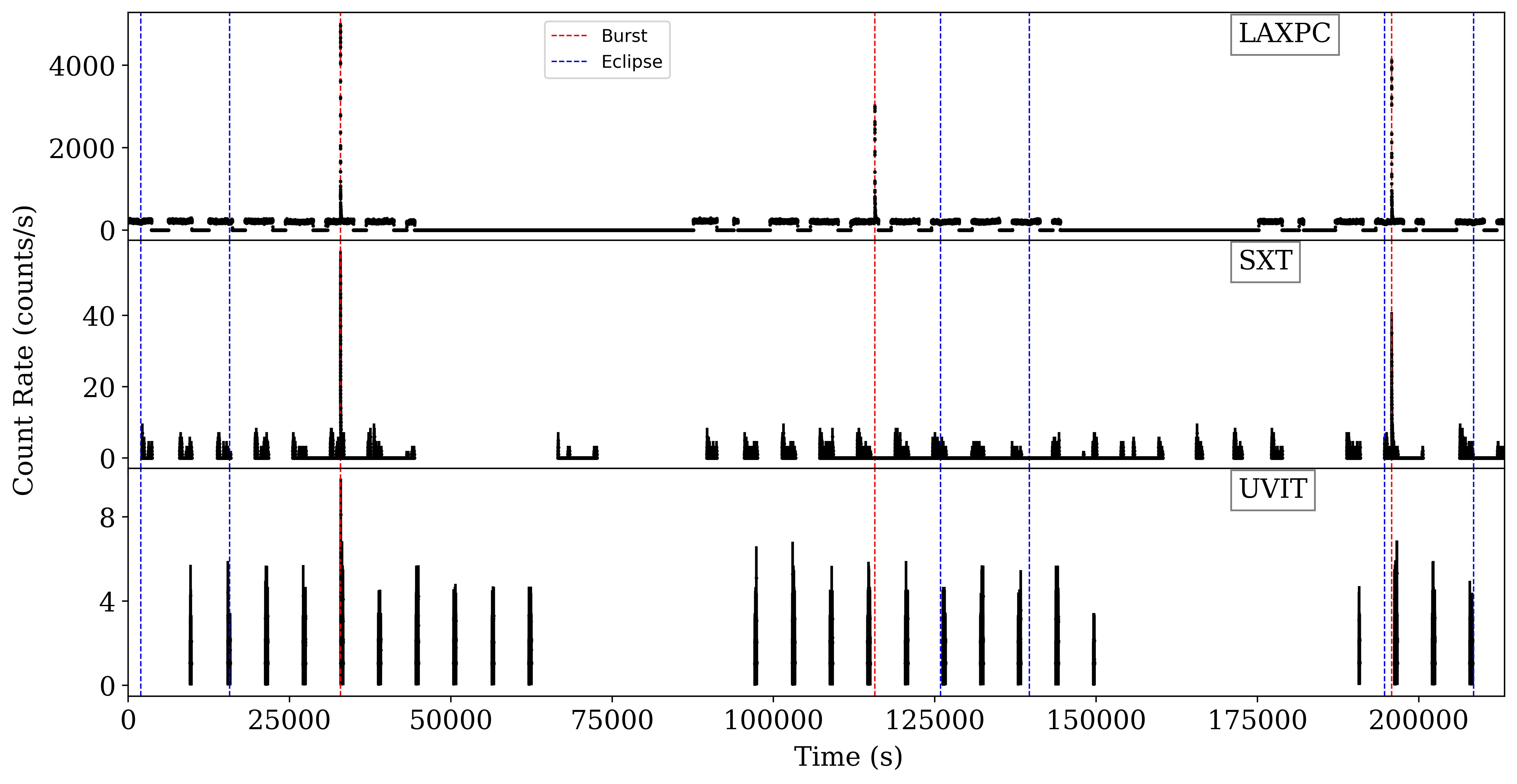}
    \caption{The light curves binned at 1~s are shown for LAXPC (top panel), SXT (middle panel), and UVIT (bottom panel) aboard \emph{AstroSat}. The red and blue vertical lines indicate the times corresponding to the bursts and full eclipses detected in the LAXPC data. The zero second in the light curves corresponds to MJD = 60515.34911006 (2024-07-24,08:01:03), which is the start time of the LAXPC observation of EXO~0748$-$676.}
    \label{fig:lc_combined}
\end{figure*}

We have generated the light curves from the cleaned Level 2 event files of LAXPC, SXT, and UVIT, which are shown in Figure~\ref{fig:lc_combined}. We inspected the LAXPC data of the entire observation for the presence of bursts and eclipses.
\subsection{Burst}

We have developed a pipeline to search for bursts, where the peaks in the total light curve are identified every 100 s using \texttt{find\_peaks} \footnote{\href{https://docs.scipy.org/doc/scipy/reference/generated/scipy.signal.find_peaks.html}{https://docs.scipy.org/doc/scipy/reference/generated\\ /scipy.signal.find\_peaks.html}} in \texttt{scipy.signal} \citep{virtanen_scipy_2020} library in Python. The pipeline assumes that the two bursts will not occur within 100 s. If the peak value count rate is 500 counts/s more than the persistent count rate in a one-second binned light curve, the peak is identified as a burst candidate. 

The burst candidate region is defined by the start time, where the count rate before the peak time first reaches 97\% of the peak count rate without subtracting the persistent count rate. The end time of the burst candidate is determined when the count rate after the peak time falls below 97\% of the peak count rate without subtracting the persistent count rate. The interval between the start time and the end time is the burst duration. This methodology extracts the burst candidate region from the total light curve. Consequently, the candidates were modelled using the fast rise and exponential decay (FRED) burst profile to confirm the burst. Bursts are also confirmed through visual inspection of the light curve in the burst candidate region.
We have identified three bursts (B1, B2, and B3) in the total light curve of LAXPC (upper panel in Figure~\ref{fig:lc_combined}). A detailed study of these bursts is mentioned in section \ref{sec:burst_lc}. 


\subsubsection{Burst Light Curve}
\label{sec:burst_lc}
We extracted all burst regions from the energy-integrated (3-80 keV) light curve of the LAXPC data. We modeled the burst with the piecewise FRED function, given in equation~\ref{eq:fred} \citep{yan_decades-long_2018}. 
\begin{equation}
\label{eq:fred}
    F_{\rm X}(t) = \left\{ 
    \begin{array}{ll}
    F_{\rm X,pers} & (t<t_s) \\ 
    \\
    \frac{t-t_\mathrm{s}}{t_\mathrm{p}-t_\mathrm{s}}\, F_{\rm X, peak} +F_{\rm X,pers} & (t_s<t\leq t_\mathrm{p}) \\
    \\
	e^{-\frac{t-t_\mathrm{p}}{\tau_{\rm decay}}}\, F_{\rm X, peak}+F_{\rm X,pers} & (t\geq t_\mathrm{p})\\
	\end{array}  \right.
\end{equation}
Where the $t_\mathrm{s}$ and $t_\mathrm{p}$ are the start and peak time of the bursts, $F_{\rm X,pers}$, $F_{\rm X, peak}$, and $\tau_{\rm decay}$ are the persistent intensity, peak intensity, and decay time-scale associated with the bursts. The characteristic properties of the bursts are listed in Table~\ref{tab:burst_char}. After removing the burst regions, the average persistent count rate of the total observation is $205.78\pm15.07$ counts/s. From the parameters estimated through the modeling, it is clear that B1 is the brightest among the three bursts (peak count rate of $5534.2\pm66.7$ counts/s) and B2 is the faintest burst among the three (peak count rate of $2874.9\pm46.5$ counts/s). B1 and B3 have almost the same $\tau_{\rm decay}$, exhibiting a secondary bump or a peak-like structure during decay. B2 decays slowly ($\tau_{\rm decay}=13.6\pm0.20 \rm \: s$) compared to the other two bursts.

\begin{table*}
    \centering
    \begin{tabular}{ccccccc}
        \hline
        Burst & Burst Start & Duration & Persistent Count Rate & Peak Count Rate & $\rm \tau_{decay}$\\
        ID & (MJD) & (s) & (counts/s) & (counts/s)  & (s)\\
        \hline
        B1 & 60515.72979 & 83 & $210.3\pm0.6 $ & $5534.2\pm66.7$ &  $11.3\pm0.05$\\
        B2 & 60516.68823 & 68 & $215.5\pm0.8$ & $2874.9\pm46.5$ &  $13.6\pm0.20$\\
        B3 & 60517.61520 & 69 & $203.8\pm0.6$ & $3983.5\pm57.1$ &  $11.3\pm0.10$\\
        \hline
    \end{tabular}
    \caption{The properties of the bursts, including burst start time, duration, Persistent Count Rate, Peak Count Rate, and $\rm \tau_{decay}$ in the LAXPC data. Persistent count rate before the onset of the burst, peak count rate and $\rm \tau_{decay}$ are found by modelling with the FRED function in Equation~\ref{eq:fred}.}
    \label{tab:burst_char}
\end{table*}

\begin{figure*}[ht]
    \centering
    \includegraphics[width=0.32\textwidth]{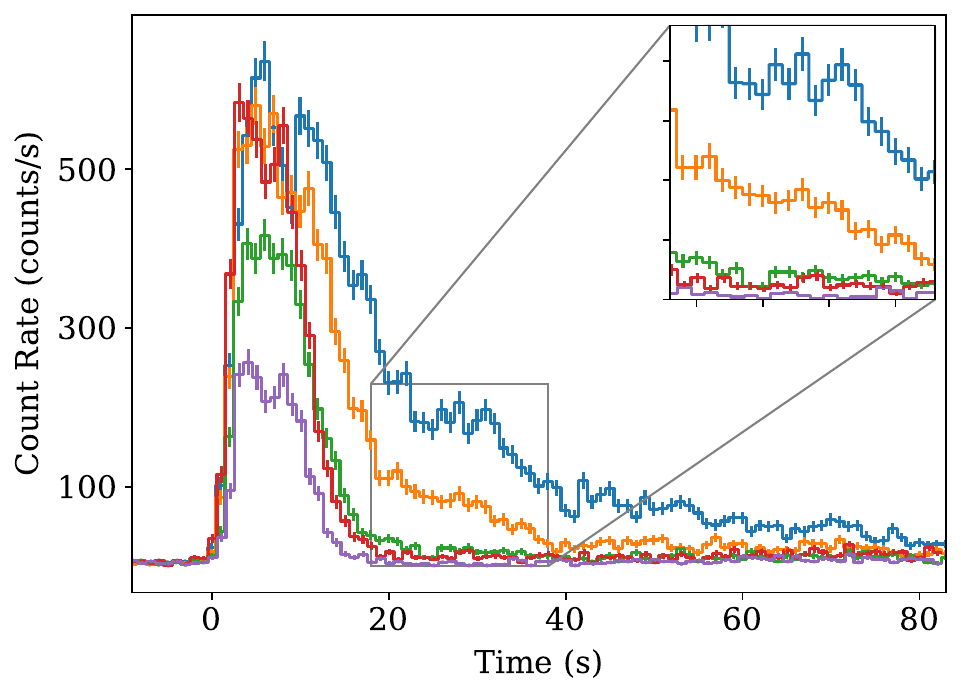}
    \includegraphics[width=0.32\textwidth]{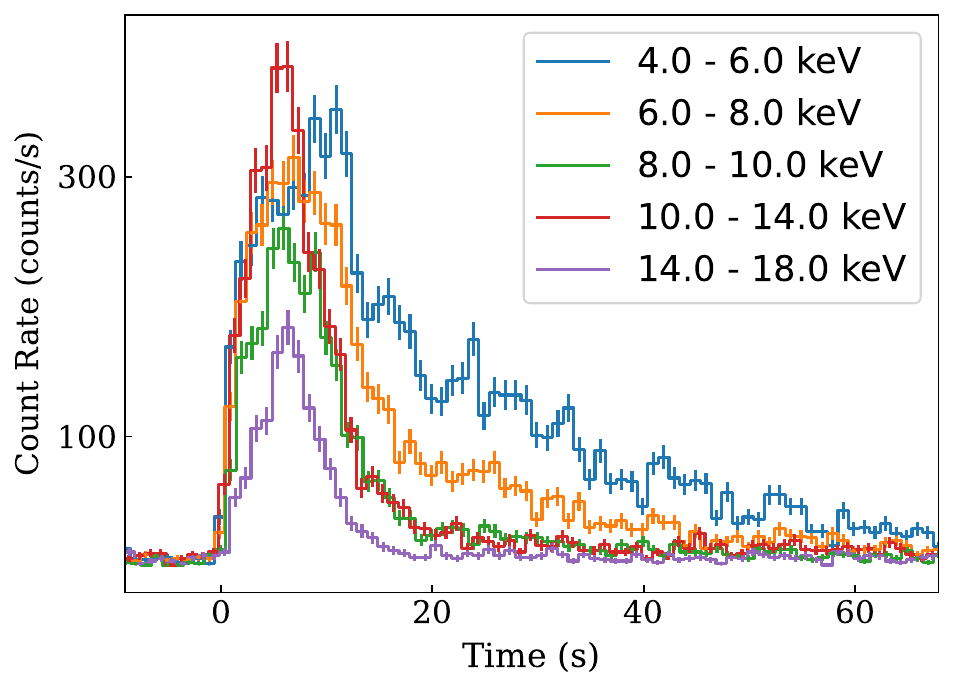}
    \includegraphics[width=0.32\textwidth]{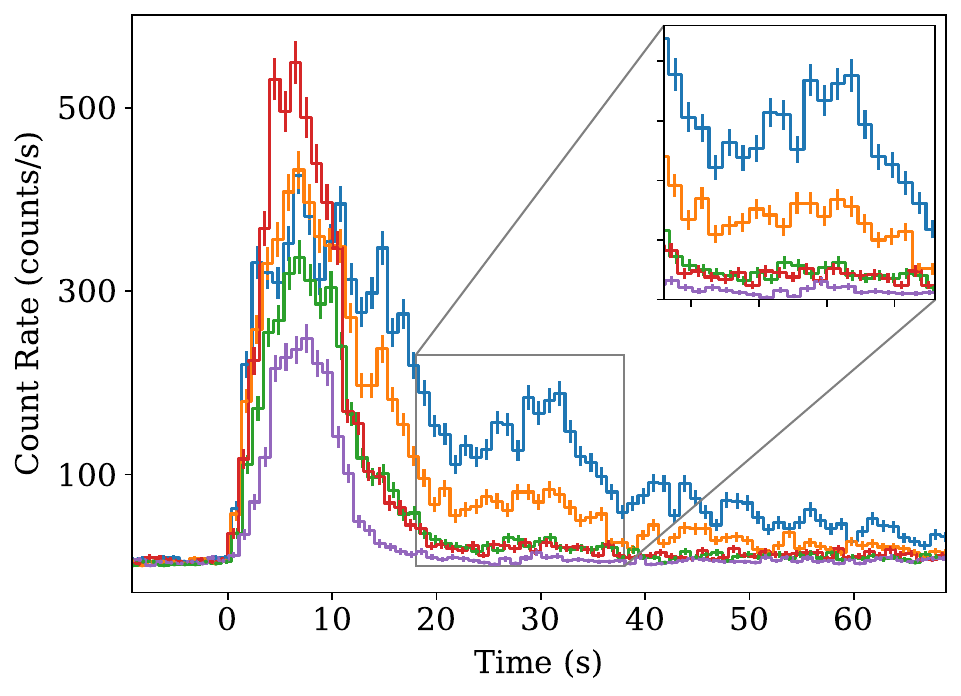}
    \caption{1~s binned X-ray light curve of B1 (left), B2 (middle), and B3 (right) respectively observed during the 2024 outburst of EXO~0748. The energy range 4.0-6.0 keV, 6.0-8.0 keV, 8.0-10.0 keV, 10.0-14.0 keV, and 14.0-18.0 keV are shown in light blue, orange, green, red, and purple, respectively. All light curves begin 5~s before the burst onset and last till the total burst duration mentioned in Table~\ref{tab:burst_char}. Zoomed-in plots in B1 and B3 show a hump-like structure and a secondary peak during the decay phase of the respective bursts. In $4.0-6.0\rm \:keV$ light curve of B3, it is evident that the secondary peak is comparable to the intensity of the primary peak.}
    \label{fig:burst_ene_res}
\end{figure*}

To further investigate the properties of the bursts, we created light curves in energy ranges: 4-6 keV, 6-8 keV, 8-10 keV, 10-14 keV, and 14-18 keV for the three bursts. It is evident from the energy-resolved light curves shown in Figure~\ref{fig:burst_ene_res} that all three bursts show energy-dependent variations. From the energy-resolved light curves, it was found that a hump-like structure (displayed in the inset of the left-most panel of Figure~\ref{fig:burst_ene_res}) is visible in B1 for the energy range of 4-6 keV. A clear secondary peak can be seen in the 4-6 keV light curve of B3, and a concurrent hump-like structure is observed in the 6-8 keV light curve of the same burst (displayed in the inset of the right-most panel of Figure~\ref{fig:burst_ene_res}). The peak count rate of the secondary peak is reaching half of the peak count rate of the primary peak in the 4-6 keV energy range. No signature of a hump-like structure was found in B2. The three bursts show a very rapid decay in the hard energies (8-18 keV) compared to the softer energies (4-8 keV). We also detected a double-peak-like structure near the peak of the burst B1 in the energy ranges 4-6 keV, 10-14 keV, and 14-18 keV. The double-peak structure was initially observed in harder energies and later appeared in the soft energy band.

\begin{figure}
\centering
   \includegraphics[width=0.45\textwidth]{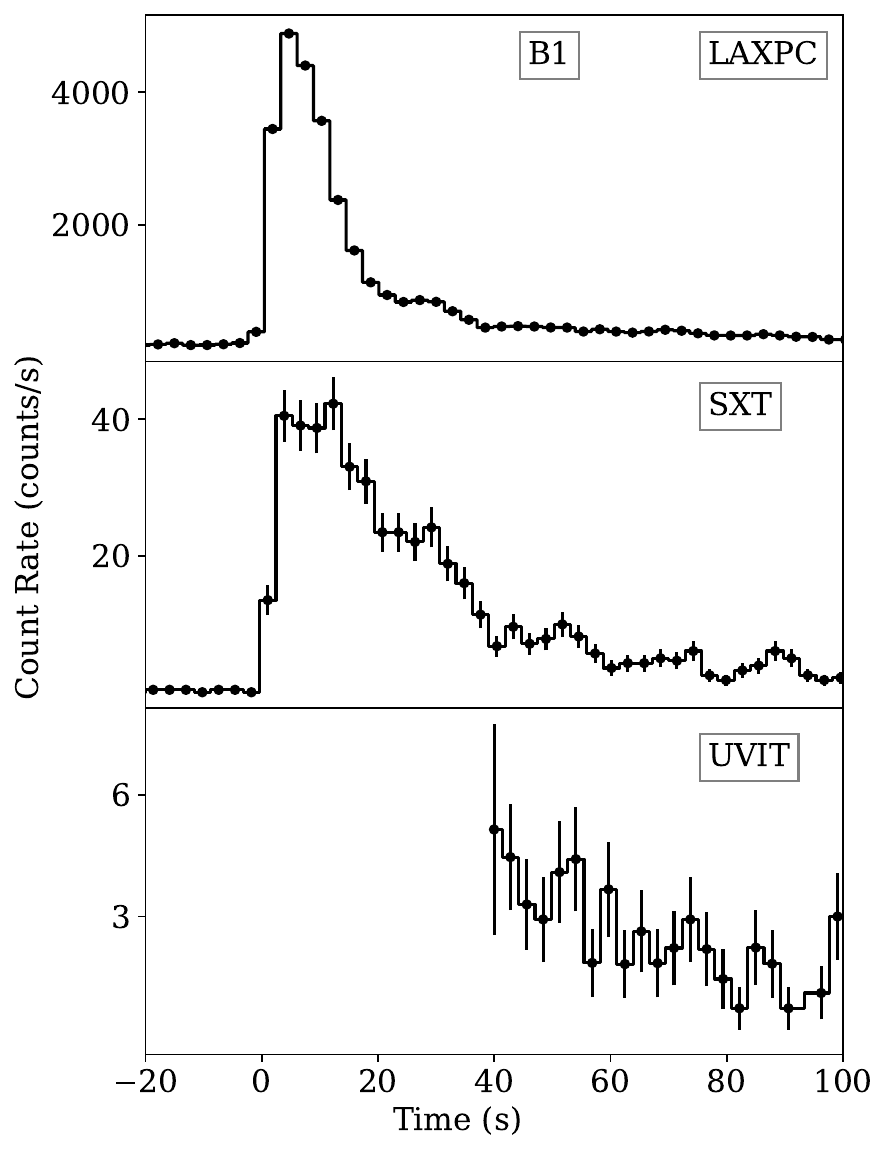}

   \includegraphics[width=0.45\textwidth]{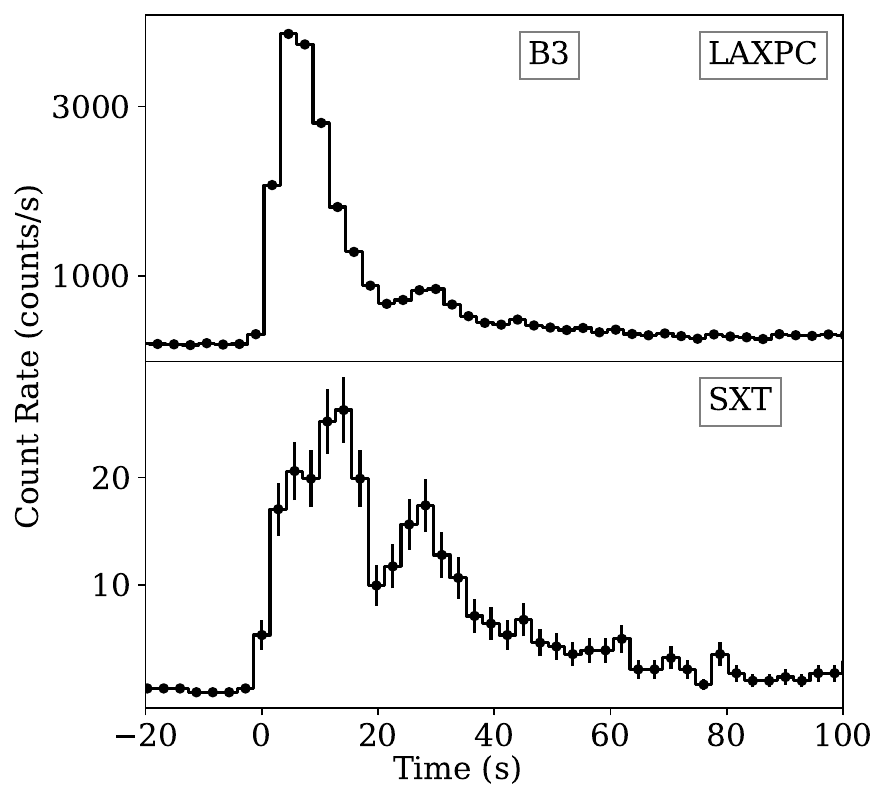}

\caption{Simultaneous bursts of EXO 0748 occurred across different detectors of \emph{AstroSat} for B1 (upper) and B3 (lower). All the light curves are 2.81 s binned. LAXPC and SXT cover an energy range of 3-80 keV and 0.7-8 keV, respectively, and UVIT works in 130-600 nm. All the light curves begin 20 s before the burst onset and continue until 100 s after the burst starts. B1 was detected in LAXPC and SXT, and the decay portion was found in UVIT. It is the first ever detection of a burst in \emph{AstroSat/UVIT}. B3 simultaneously occurred in LAXPC and SXT. The secondary peak of B3 is clearly visible in the SXT light curve, reaching an intensity of two-thirds of the primary peak.}
\label{fig:burst_sxt_laxpc}
\end{figure}

\begin{figure*}
    \centering
    \includegraphics[width=0.32\textwidth]{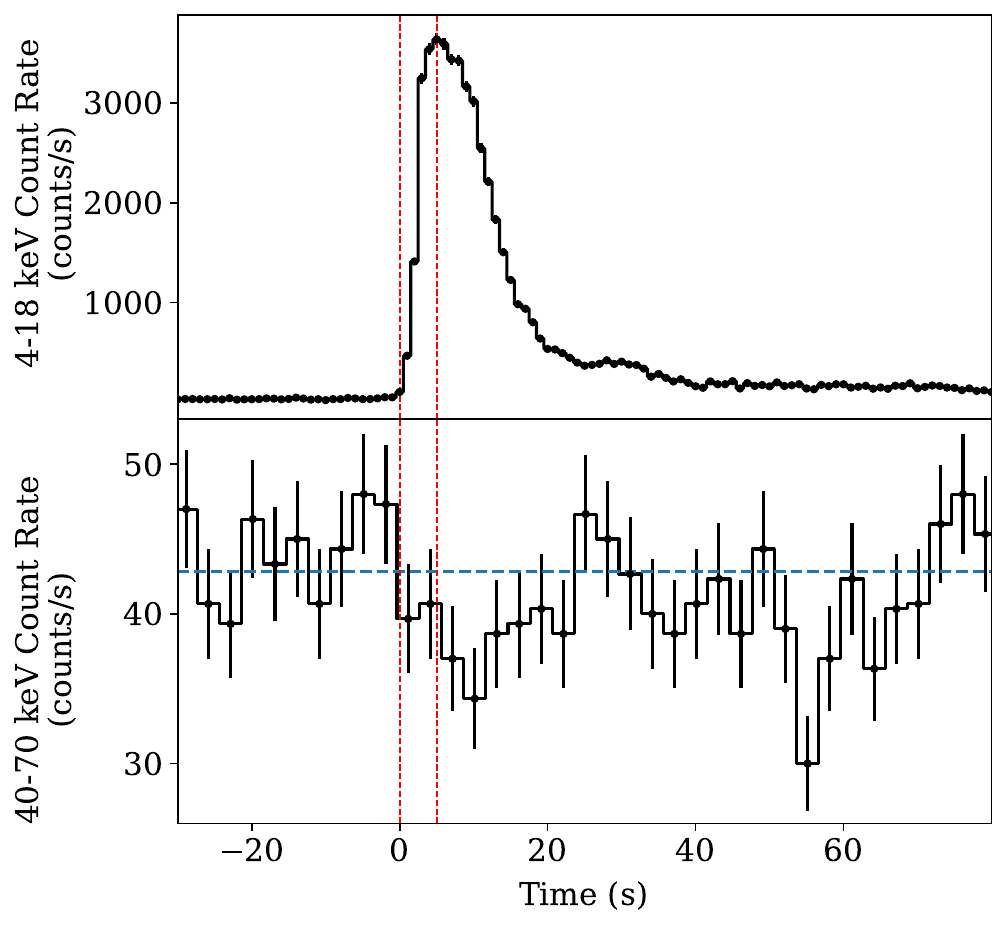}
    \includegraphics[width=0.32\textwidth]{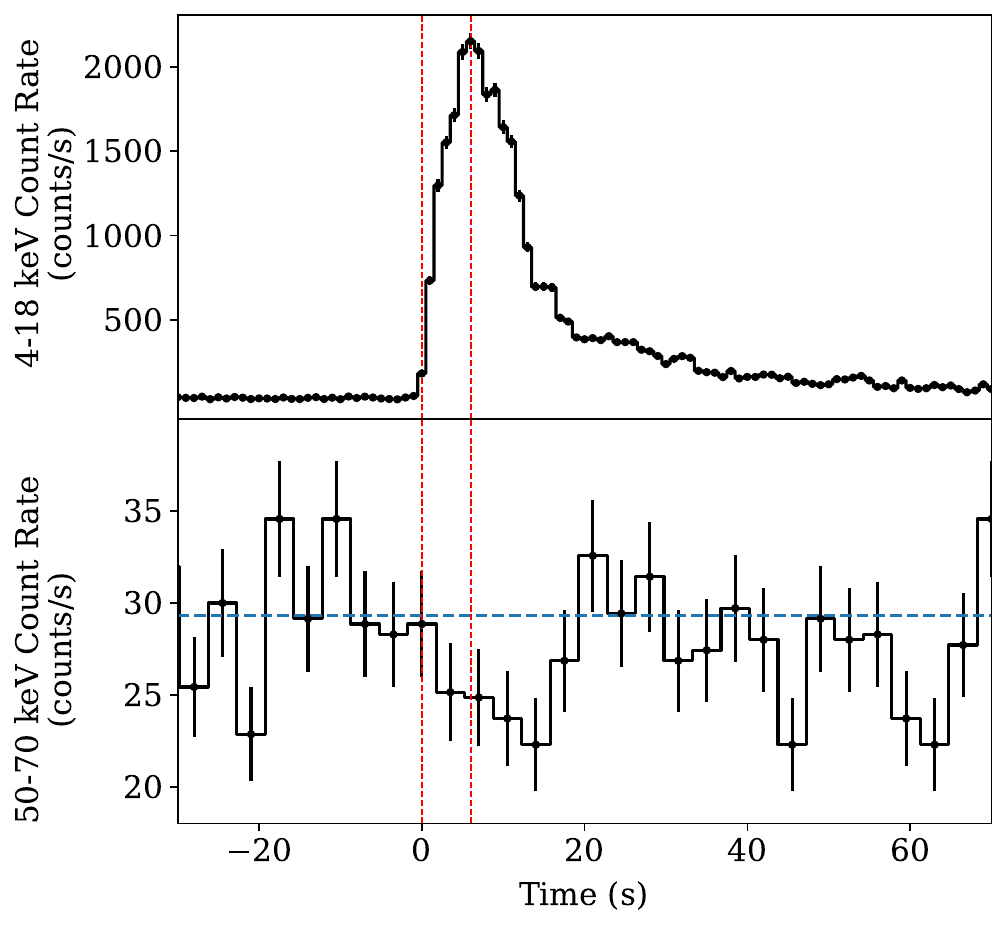}
    \includegraphics[width=0.32\textwidth]{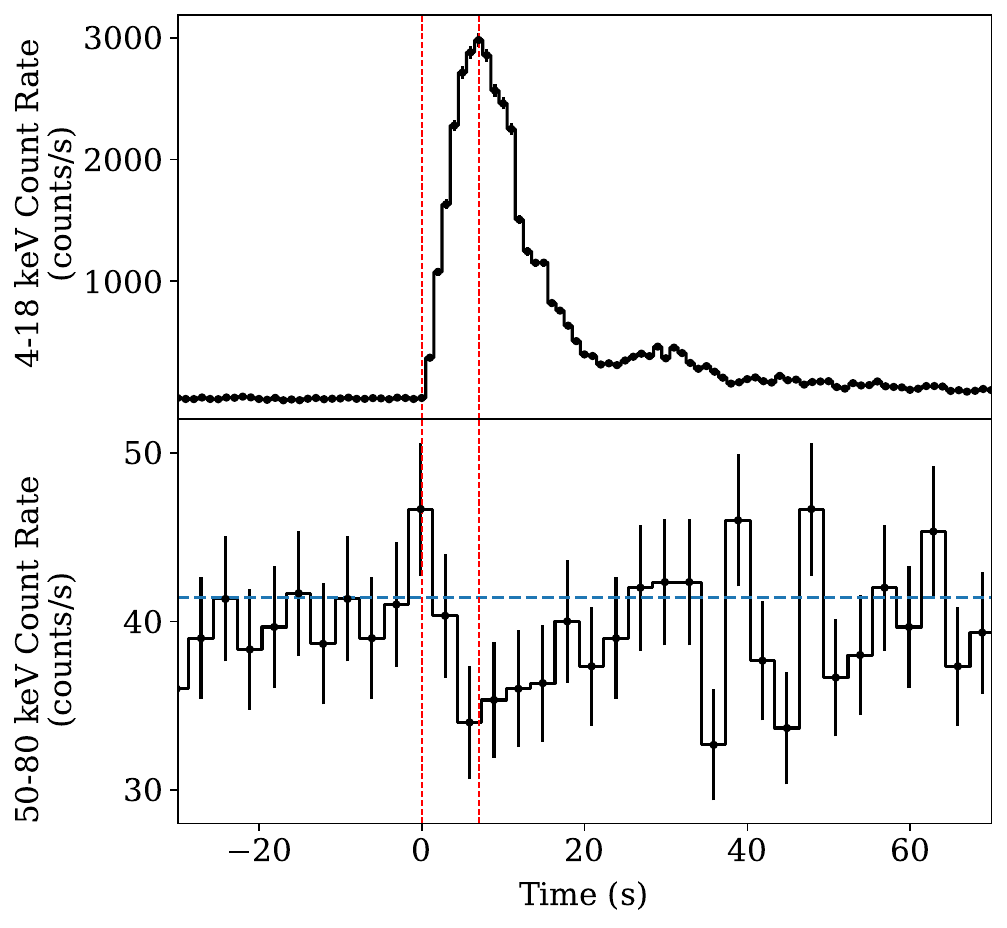}
    \caption{The light curves for B1 (left), B2 (middle), and B3 (right) are shown in two panels: the upper panel displays the low-energy light curves (4-18 keV), while the lower panel presents the simultaneous high-energy light curves. Vertical red lines indicate the burst's start and peak on the 1~s binned low-energy light curve. The high-energy light curves are coarsely binned for better representation. We have taken the coarse bin sizes as 3~s, 3.5~s, and 3~s, for B1, B2, and B3, respectively. We have chosen the high energy range and the coarser binning for the illustrative purposes. Horizontal blue lines represent the persistent count rate in the higher energy range, measured 100~s prior to the start time of the burst. All light curves begin 30~s before the onset of the burst and continue till the duration of the burst.}
    \label{fig:defi_burst}
\end{figure*}

To understand the behavior of the burst in the lower energy band, we checked for the presence of simultaneous bursts in the 0.3-8 keV SXT light curves and UVIT light curves. Both B1 and B3 are observed simultaneously in the SXT light curves, while the decay phase of B1 is also seen in the UVIT light curve, as shown in Figure~\ref{fig:burst_sxt_laxpc}. This is the first report of simultaneous observation of burst detection in the three \emph{AstroSat} payloads. Comparing the SXT and LAXPC light curves, it is evident that at lower energies, the decay of the burst is slower. In Figure~\ref{fig:burst_sxt_laxpc} for B1, we can observe that the count rate remains constant during the peak in the SXT data. Similarly, the count rate remains steady in the corresponding areas where we noted a hump during the decay of B1 in the LAXPC data. A secondary peak is clearly visible in the SXT light curve of B3. In B3, the secondary peak is reaching $2/3$ of the count rate of the primary peak. We can clearly observe that the intensity of the primary peak decays to half, followed by the secondary peak. As shown in Figure~\ref{fig:burst_sxt_laxpc}, for B3, the secondary peaks in the SXT and LAXPC light curves align together.

Following this, we studied the burst behavior in the higher energy range ($>40-80$ keV). In Figure~\ref{fig:defi_burst}, the top panel represents the 1 s binned LAXPC light curve in the energy range 4-18 keV, starting 30 s prior to the start of the burst and ending at 80 s after the burst. The lower panel shows the corresponding high-energy of the LAXPC light curve in a coarser binning to achieve a sufficient signal-to-noise (SNR) ratio. The vertical red lines represent the start and peak times of the burst, whereas the horizontal blue line in the bottom panel represents the average persistent count rate in a 100 s segment prior to the burst in the respective high-energy range. For B1, the high energy range is 40-70 keV; for B2, it is 50-70 keV; and for B3, it is 50-80 keV. Regardless of the slight variation in the energy range to enhance the SNR, all the high-energy ranges chosen represent regimes where the burst emission contribution is negligible. Clear dips can be observed in the higher energy part corresponding to the burst peak in the lower energies. The persistent count rate 100 s before the burst is $42.81\pm3.77$, $29.30\pm2.89$, $41.41\pm3.71$ counts/s in their respective high energy ranges and the minimum count rate in higher energy during the burst peak region is $34.33\pm3.38$, $22.28\pm2.52$, $34.0\pm3.36$ counts/s, respectively, for B1, B2, and B3.

\subsubsection{Pre-Burst Spectra}
\label{sec:pre_burst}

We have considered 100~s prior to 5~s before the burst onset as the pre-burst region. We carried out the spectral analysis in the 4-18 keV energy band, as the background above 18 keV significantly dominates the LAXPC burst spectrum. We have used \texttt{lxp2level2back\_shifted.spec} as the background spectra and \texttt{lx20cshm15v1.0.rmf} as the response file. The LAXPC pre-burst spectra are fitted with both single and double component models: \texttt{TBabs*(powerlaw)} and \texttt{TBabs*(powerlaw+bbodyrad)}; however, it was observed that statistically we do not require a two-component model to represent the data. We used the \texttt{wilms} abundance model \citep{wilms_absorption_2000} and considered a systematic error of 3\% for the spectral analysis. We fixed the $n_H$ value to $0.149 \times 10^{22}$ cm$^{-2}$ \citep{knight_eclipse_2022}. The results of the pre-burst spectral analysis are reported in Table~\ref{tab:pre_burst}. The pre-burst spectra for both B1 and B2 are observed to be relatively harder than B3, with a relatively higher value of the non-thermal spectral index (photon index value of $1.40\pm0.10$, $1.51\pm0.09$, and $1.90\pm0.10$, respectively, for B1, B2, and B3). The persistent flux mimics the pre-burst count rate in Table~\ref{tab:burst_char}. It is noted that B2, the weakest burst among the three, has the highest pre-burst persistent flux and count rate. We note here that we could not constrain the parameters for the joint pre-burst spectra due to the low count rate in the pre-burst SXT spectra. Thus, we proceed with the results obtained from the spectral analysis of the LAXPC data only.
\begin{table*}
    \centering
    \begin{tabular}{cccccc}
         \hline
        Burst ID & $n_H$ & $\Gamma$ & PL norm & Flux & $\chi^{2}/dof$ \\
         & ($\times 10^{22}\rm \: cm^{-2}$) &  & ($\rm photons/keV/cm^2/s$) & ($\times 10^{-11}\rm erg\: s^{-1} \: cm^{-2}$) & \\
           \hline
         B1 & 0.149 & $1.40\pm0.10$ & $0.014^{+0.003}_{-0.002}$ & $12.67\pm0.5$ & 23.03/21 \\
         B2 & 0.149 & $1.51\pm0.09$ & $0.020^{+0.004}_{-0.003}$ & $13.89\pm0.5$ & 30.24/21 \\
         B3 & 0.149 & $1.90\pm0.10$ & $0.041^{+0.009}_{-0.007}$ & $12.36\pm0.5$ & 26.54/21 \\
        \hline
    \end{tabular}
    \caption{Results of the best-fit model for the pre-burst X-ray spectra of EXO~0748--676 based on the LAXPC-only analysis.}
    \label{tab:pre_burst}
\end{table*}

\subsubsection{Time-Resolved Spectroscopy}
\subsubsection*{LAXPC Only}
\begin{figure*}[ht]
    \centering
    \includegraphics[width=0.32\textwidth]{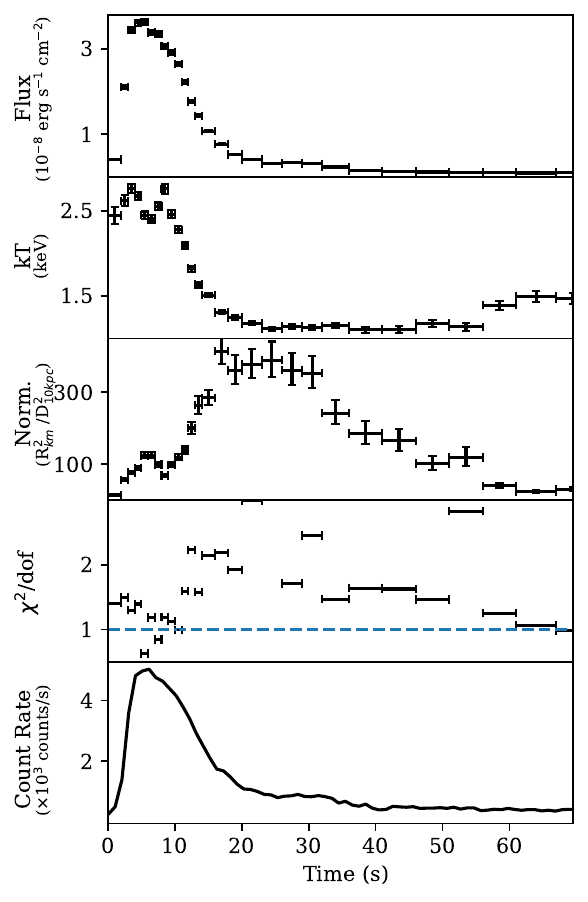}
    \includegraphics[width=0.32\textwidth]{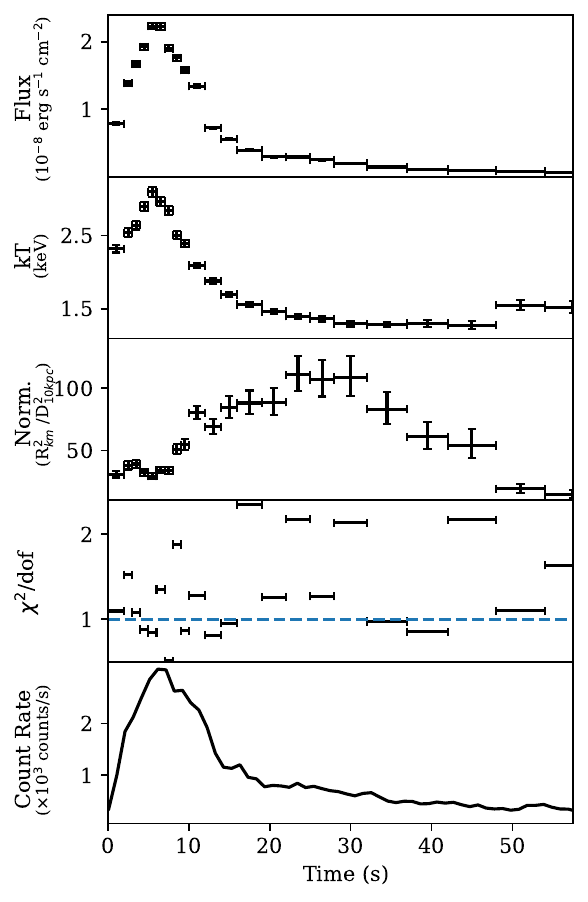}
    \includegraphics[width=0.32\textwidth]{images/trs_B3.pdf}
    \caption{The variation of the best-fit spectral parameters during the thermonuclear bursts B1 (left), B2 (middle), and B3 (right) for the observation of EXO~0748--676 is presented. The first panel illustrates the evolution of the source flux in the 4-18 keV energy range. The second panel shows the change in blackbody temperature (kT). The third panel shows the blackbody normalization, which is given by $R^{2}_{km}/D^{2}_{10\text{kpc}}$, where $R_{km}$ represents the hotspot radius in kilometers and $D_{10\text{kpc}}$ is the distance to the source, in scale of 10 kpc. The fifth panel represents the reduced chi-squared ($\chi^{2}_{\nu}$) of the fit, while the sixth panel shows the count rate. The spectra are dynamically binned, and the count rate is presented to highlight the position of the hump and the secondary burst. Bursts B1 and B3 show a double peak temperature profile, indicating the Photospheric Radius Expansion (PRE). Sudden expansion of the Normalization value can be seen during the secondary peak of B3, and the Normalization value stays constant for the hump-like feature during B1, both without changing the temperature.}
    \label{fig:trs_burst}
\end{figure*}
To understand the evolution of spectral parameters during the burst and to examine the hump/peak found during the decay phase of B1 and B3, we conducted a time-resolved spectral analysis. Using customized Python code, we first determined the time segment to create the spectra and generated dynamically binned spectra. Our criteria for dynamic binning require that each spectrum corresponds to a time segment that is an integral multiple of 1~s bins and contains more than 2,000 photon counts. The spectral analysis was conducted using the same response file, systematic error, energy range, and value of $n_H$ as mentioned in section \ref{sec:pre_burst}. We first employed the standard method where the pre-burst spectra were considered the background spectra and modeled the burst spectra using a blackbody radiation (\texttt{TBabs*BBodyrad}) model. We then applied the $f_a$ method described by \cite{worpel_evidence_2015}, where the $f_a$ value signifies the amount by which the persistent emission varies with the evolution of the mass accretion rate, possibly due to the Poynting–Robertson drag imparted by the incoming burst photons on the accretion disk matter \citep{fragile_simulating_2018, fragile_interactions_2020, guver_burstdisk_2022}. For the $f_a$ method, we used the instrumental background spectrum generated by the LAXPC software. The burst spectrum was modeled as a combination of blackbody radiation and the pre-burst emission, which was scaled by a constant factor $f_a$  (\texttt{TBabs*BBodyrad+$f_a$*(TBabs*Powerlaw)}). For both the standard and $f_a$ methods, we grouped each spectrum to a minimum of 50 counts per bin using the \texttt{grppha} command. To account for non-Gaussian effects arising from low counts, we employed Churazov weighting for the spectral fitting, following the approach by \citet{churazov_mapping_1996}. It was found that the $f_a$ method does not provide reasonable constraints to the spectral parameters. Therefore, we have considered the standard method for further discussions related to LAXPC only spectral analysis.

Figure~\ref{fig:trs_burst} represents the temporal evolution of the best-fit parameters obtained from the time-resolved spectral analysis. The first panel describes the evolution of the unabsorbed flux throughout the time of the burst, in the energy range of 4-18 keV, obtained using the \texttt{cflux} model. 
The peak flux of burst B1 ($3.63 \pm 0.06 \times 10^{-8}$ erg s$^{-1}$ cm$^{-2}$) is observed to be higher than the peak flux of B2 ($2.23\pm0.04\times 10^{-8}$ erg s$^{-1}$ cm$^{-2}$) and B3 ($3.27\pm0.06 \times 10^{-8}$ erg s$^{-1}$ cm$^{-2}$), respectively.
The second panel represents the variation in the burst blackbody temperature in keV. The double peaks observed in the temperature evolution show that B1 and B3 are Photospheric Radius Expansion (PRE) bursts. The peak temperatures of B2 and B3 ($3.09\pm0.1$ keV and $3.22\pm0.1$ keV, respectively) are comparable and are higher than that for B1 ($2.70\pm0.1$ keV). A hint of temperature rise at the end of the decay phase was also observed for all three bursts. The third panel represents the evolution of the \texttt{bbodyrad} normalization  $R_{km}^{2}/D_{10kpc}^{2}$, where $R_{km}$ represents the radius of the hotspot in km and $D_{10kpc}$ represents the distance to the source in units of 10 kpc. For B1, the hotspot radius remains relatively larger than B2 and B3 throughout the burst. In B3, we can see a sudden increase in hotspot radius near the secondary peak, and there is no change in temperature during this time. The fourth panel represents the reduced $\chi^2$ for each fit. The comparatively low statistics during the decay phase have resulted in a reduced $\chi^2$ value greater than 2 in some later time bins. The fifth panel shows the 1~s binned light curve of the corresponding burst part.

\begin{figure}[ht]
    \centering
    \includegraphics[scale=0.67]{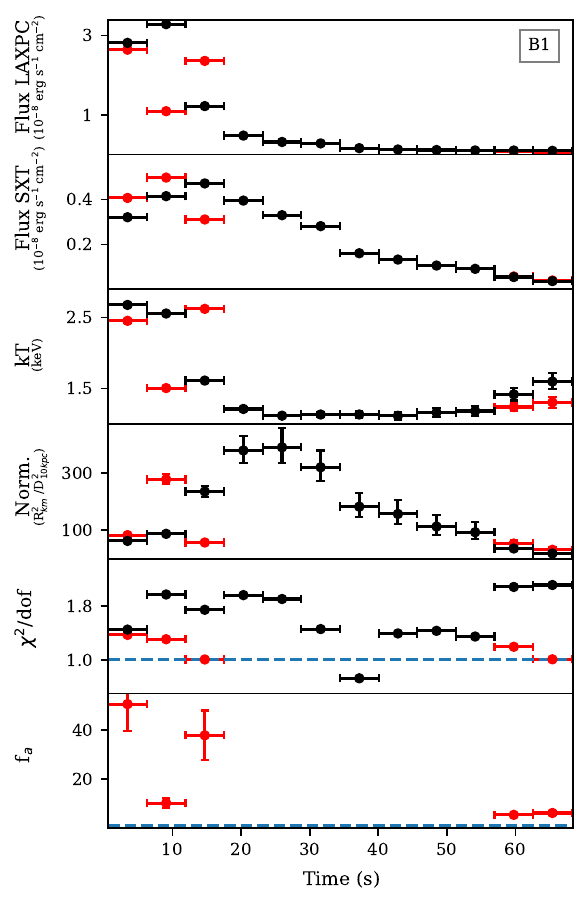}
    \includegraphics[scale=0.67]{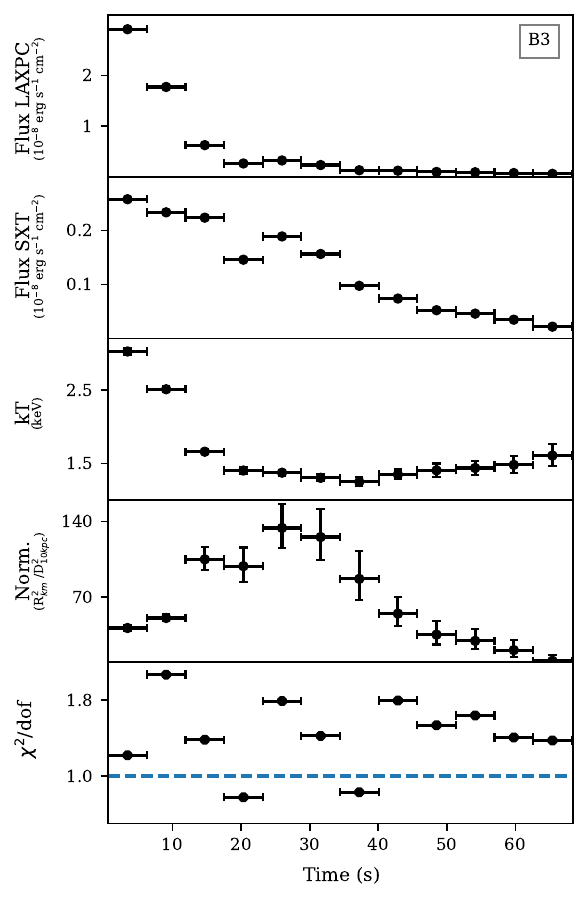}
    \caption{The variation of the best-fit spectral parameters during the thermonuclear bursts B1 (top) and B3 (bottom) for the observation of EXO~0748--676 is displayed. Each panel illustrates the evolution of the source flux in the 4-18 keV energy range measured by LAXPC, and the flux in the 0.7-4 keV energy range measured by SXT. Additionally, the panels include the blackbody temperature (kT), blackbody normalization, the reduced chi-squared ($\chi^2_\nu$) of the fit, and statistically significant $f_a$ values obtained. The red represents the $f_a$ method, while the black represents the standard method. Due to the limited time resolution of SXT and to obtain significant statistics, each spectra have a length of 5.6~s. Burst B3 shows no statistically significant changes in the $f_a$ method.}
    \label{fig:jtrs_burst}
\end{figure}

\subsubsection*{Joint Spectral Analysis}
As B1 and B3 are detected simultaneously with LAXPC and SXT, we have performed the joint time-resolved spectral analysis for both bursts to investigate the burst evolution in the broader energy range. Due to the limited time resolution of SXT and to obtain significant statistics, we created 5.6~s (20 $\times$ 0.28~s) long spectra from LAXPC and SXT data of the same burst region. SXT and LAXPC spectra are grouped with minimum counts of 5 and 50 in each energy bin, respectively. For joint spectral analysis, we have considered the energy range of $0.7 - 4.0$ keV for SXT and the energy range of $4.0 - 18.0$ keV for LAXPC. For the SXT spectra, we have used the response, ancillary file, and background available on the SXT website. Because of the low counts, we used the Churazov weighting for the spectral fitting. For the standard method, we have used the 100~s long pre-burst spectra as the background for individual burst spectra. We fitted the joint spectra using the {\tt XSPEC} model \texttt{constant*Tbabs*(bbodyrad)}. We set the LAXPC cross-calibration constant to one, allowing the SXT cross-calibration constant to vary freely. All other parameters were kept tied for LAXPC and SXT spectra during the fitting procedure. We kept $n_H$ as $0.149 \times 10^{22}$ cm$^{-2}$ along with the \texttt{wilms} abundance. A systematic error of 3\% was considered for the spectral modelling. Although the standard method provided a satisfactory reduced $\chi^2$ in most cases, it showed relatively poor statistics during the burst peaks, and hence we employed the $f_a$ method as well. We have fitted the spectra with model \texttt{constant*Tbabs*(bbodyrad)+ $f_a$*(constant*Tbabs*(powerlaw))}. We have used the parameters obtained from the pre-burst analysis of LAXPC only for the spectral fitting of the persistent spectra. The $f_a$ method improved the $\chi^2$ values and the overall statistics during the peak of B1 and B3. To find out the significance of the fitting, we performed an F-test between the $f_a$ method and the standard method. We only considered the result of the $f_a$ method valid if the probability of the improvement of $\chi^2$ occurring by random chance was below the $3\sigma$ significance level probability ($0.0027$). The peak and end portions of B1 showed valid results by the F-test for the $f_a$ method, whereas all other points in B1 and B3 showed invalid results for the $f_a$ method. Consequently, we will focus our discussion only on the points that exhibit more than $3\sigma$ significance, which are indicated with red symbols in Figure~\ref{fig:jtrs_burst}. The evolution of different parameters obtained from joint time-resolved spectroscopy is given in Figure~\ref{fig:jtrs_burst}.

We have used coarse time binning, so we cannot see short temporal variations of the spectral parameters, as they will be averaged over. However, the joint spectral fitting helps us to understand the global changes with improved accuracy and offers more reliable constraints on the spectral parameters. The first two panels in Figure~\ref{fig:jtrs_burst} represent the flux changes in each instrument in their respective energy ranges. We observed that B1 displays the highest flux value in the LAXPC and SXT energy ranges. The flux shows a slow decay in the SXT energy range compared to that in the LAXPC energy range, which is evident from the energy-resolved light curves as well (see Figures~\ref{fig:burst_ene_res} and \ref{fig:burst_sxt_laxpc}). The secondary peak of B3 is clearly visible in the SXT flux. The $f_a$ method in B1 shows a clear dip in LAXPC flux during the PRE phase, which is absent in the standard method. The third panel represents the burst temperature in keV. The LAXPC only time-resolved spectroscopy results showed signatures of PRE for both B1 and B3 (Figure~\ref{fig:trs_burst}). However, as we increased the time binning in the joint spectral analysis, the double peak signature in the temperature became undetectable in the standard method, and the double peak is visible for B1 during the $f_a$ method. The fourth panel represents the normalization of \texttt{bbodyrad} that tracks the radius of the hotspot. The hotspot radius is much larger in B1 than in B3, showing a similar evolution for both. We can see the radius increase for the $f_a$ method during the PRE phase of B1, which is absent in the standard method. Furthermore, we can clearly see that the secondary peak of B3 in SXT occurs when the corresponding hotspot radius is maximum, without any changes in the temperature. Hence, this implies that the secondary peak is more dominant in the soft energy ranges. The fifth panel represents the reduced $\chi^2$ values. The sixth panel in B1 represents the $f_a$ values for the cases where the $f_a$ method was statistically significant. We have not displayed the $f_a$ values for B3 because the $f_a$ method did not meet the significance criteria in the F-test.

\subsection{Eclipse}
We removed the bursts from the barycenter corrected LAXPC light curve and searched for eclipses using custom-made Python scripts. In a 10~s binned light curve, we calculated the mean count rate of segments without gaps separately and determined whether the count rate in any segment fell below $0.5\sigma$ of the mean for an extended duration of 150~s. If it did, we considered it a candidate for an eclipse. We then visually checked the light curves to verify whether these segments followed the expected eclipse profile and identified both full and partial eclipses. We found six complete eclipses (as shown in Figure~\ref{fig:lc_combined}) and two partial eclipses in the entire LAXPC light curve. We attempted to identify eclipses in the SXT and UVIT data. However, due to low SNR, we could not detect any evidence of eclipses in either instrument. We will focus on the complete eclipses found in LAXPC for further analysis. 

\subsubsection{Eclipse Profile Modeling}
\label{sec:eclipse_lc}
We extracted the six detected eclipses from the barycentered corrected light curve in the 4-18 keV energy range, binned over 1~s. Each eclipse was modeled using a five-parameter function in Equation~\ref{eq:eclipse}, inspired by \cite{wolff_eclipse_2009}. In this modeling, we set the ingress and egress times to equal and made the pre- and post-egress fluxes equal. The relevant time epochs and the eclipse model are given as
\begin{equation}
\label{eq:eclipse}
    F_{\rm X}(t) = \left\{ 
    \begin{array}{ll}
    F_{\rm X,norm} & (t<t_1) \\ 
    \\
    \frac{F_{\rm X,ec}-F_{\rm X,norm}}{t_2-t_1}\times (t-t_1) & (t_1<t<t_2) \\ 
    \\
    F_{\rm X,ec} & (t_2<t<t_3) \\ 
    \\
     \frac{F_{\rm X,ec}-F_{\rm X,norm}}{t_4-t_3}\times (t_4-t) & (t_3<t<t_4) \\ 
    \\
    F_{\rm X,norm} & (t>t_4) \\ 
    \\
	\end{array}  \right.
\end{equation}
\begin{align}
    t_1 = {} t_{mid}-\frac{1}{2}\Delta t_{ec}-\Delta t_{in}\\
    t_2= t_{mid}-\frac{1}{2}\Delta t_{ec}\\
    t_3= t_{mid}+\frac{1}{2}\Delta t_{ec}\\
    t_4= t_{mid}+\frac{1}{2}\Delta t_{ec}+\Delta t_{in}
\end{align}

where $t_1$ is the starting time of ingress, $t_2$ is the end time of ingress, $t_3$ is the starting time of egress, and $t_4$ is the end of egress. $F_{\rm X,norm}$ is the persistent count rate before and after eclipse (we call it the persistent count rate hereafter), and $F_{\rm X,ec}$ is the count rate during the eclipse. $t_{ec}$, $t_{in}$, and $t_{mid}$ are the duration of the eclipse, duration of the ingress (and egress), and the mid eclipse time, respectively. The best-fit modeling parameters are listed in Table~\ref{tab:eclipse_fit}. It should be noted here that Eclipse 5 showed a flaring event immediately following the eclipse, making it difficult to constrain its parameters properly.

\begin{table*}
\centering
\begin{tabular}{cccccc}
\hline
Eclipse & Persistent  & Eclipse & Ingress & Eclipse   & $T_{mid}$ \\
ID & Count Rate & Count Rate & Duration &  Duration &  \\
 & (counts/s) & (counts/s) & (s) & (s)  & (MJD) \\
 \hline
1 & $39.58\pm0.20$ & $23.71\pm0.27$ & $19.52\pm4.81$ & $477.33\pm5.13$ & $60515.3720535615$\\
2 & $41.56\pm0.26$ & $24.85\pm0.27$ & $17.99\pm4.32$ & $481.72\pm4.73$ & $60515.5313773061$\\
3 & $37.40\pm0.19$ & $22.69\pm0.25$ & $17.27\pm4.65$ & $485.52\pm5.09$ & $60516.8060839684$\\
4 & $37.13\pm0.18$ & $22.86\pm0.24$ & $12.02\pm3.79$ & $491.68\pm3.91$ & $60516.9654113277$\\
5 & $39.07\pm0.20$ & $21.97\pm0.27$ & $115.96\pm10.17$ & $415.38\pm11.29$ & $60517.602950646$\\
6 & $37.57\pm0.19$ & $22.99\pm0.25$ & $28.51\pm5.73$ & $477.69\pm6.32$ & $60517.7621153956$\\
\hline
\end{tabular}
\caption{The parameters for the eclipse include the persistent count rate, the eclipse count rate, ingress and egress durations, eclipse duration, and the mid-eclipse time ($T_{mid}$) for each observed eclipse. Eclipse modeling was performed using a 1~s binned light curve in the 4-18 keV range. The discrepancies noted in the parameters of Eclipse 5 are attributed to a flare-like event that occurred shortly after the egress.}
\label{tab:eclipse_fit}
\end{table*}

\begin{figure}[b!]
    \centering
    \includegraphics[width=.9\linewidth]{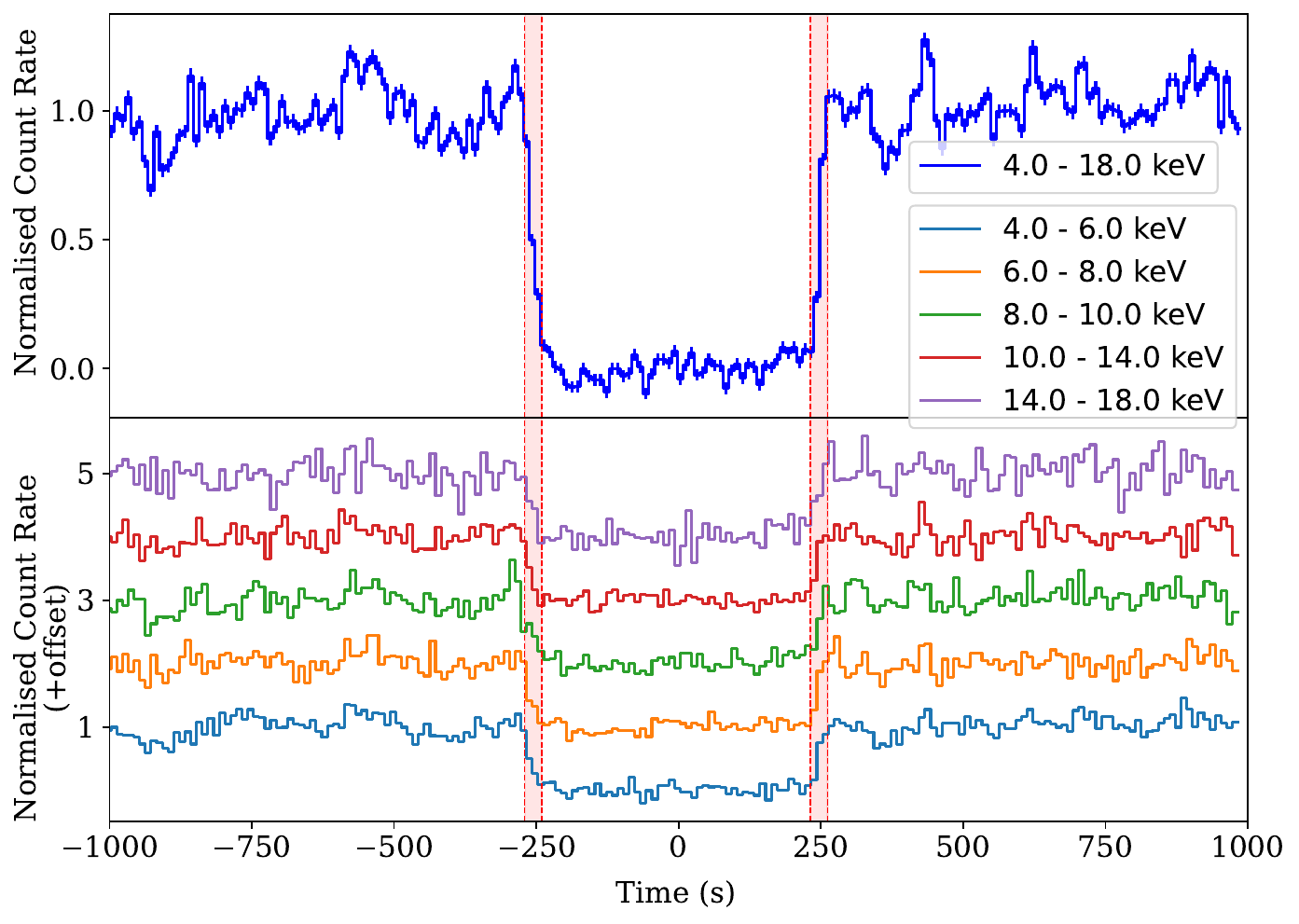}
    \caption{The $T_{mid}$ folded light curves of all the observed eclipses. The top panel y-axis represents the normalized count rate in the energy range 4.0-18.0 keV, and in the bottom panel also represents the same, but added by a constant for visualization purposes for different energy bands of $4.0-6.0\rm\:keV$, $6.0-8.0\rm\:keV$, $8.0-10.0\rm\:keV$,$10.0-14.0\rm\:keV$ and $14.0-18.0\rm\:keV$. The red shaded region shows the ingress and egress of the eclipse, and the dashed lines indicate the start and stop of the ingress and egress from the profile modeling.}
    \label{fig:eclipse_stacked}
\end{figure}

To understand how the eclipse light curves vary across different energy bands, we divided the light curves into five specific energy ranges: 4-6 keV, 6-8 keV, 8-10 keV, 10-14 keV, and 14-18 keV. We generated a 10~s binned eclipse light curve to account for the low statistics in each energy band. For each light curve, we extracted data from \( t_{mid} - 1000 \) s to \( t_{mid} + 1000 \) s, resulting in a total of approximately 2000 s of data, which includes the entire eclipse. We removed Eclipse 5 from this analysis as it showed a flaring event after the eclipse.

We modeled the eclipse profile for each energy band. However, we could not obtain sufficient statistics for individual eclipses within each energy range. To analyze the average characteristics of the eclipses, we stacked all the individual eclipses, aligning their $t_{\text{mid}}$ times, setting the mid-point to zero. In Figure~\ref{fig:eclipse_stacked}, the top panel displays the normalized stacked eclipse light curve in the 4-18 keV range. In contrast, the bottom panel shows the stacked normalized light curves for each energy band plotted with a constant shift for better visualization. The different colors in the bottom panel represent the light curves in the different energy ranges. To normalize the averaged eclipse light curve, we subtracted the modeled eclipse count rate from the modeled persistent count rate and divided by the modeled persistent count rate. We then calculated the average properties for each energy band, as summarized in Table~\ref{tab:folded_eclipse}. However, we did not observe any common energy dependency in the eclipse parameters, both for the stacked eclipse and for the individual ones.
\begin{table*}
\centering
\begin{tabular}{ccccc}
\hline
Energy & Persistent  & Eclipse  & Ingress  & Eclipse \\
& Count Rate & Count Rate & Duration & Duration \\
(keV) & (counts/s) & (counts/s) & (s) & (s) \\
\hline
$4.0-18.0$ & $48.03\pm0.14$ & $29.06\pm0.20$ & $30.37\pm3.96$ &$ 471.57\pm4.05$ \\ 
$4.0-6.0$ & $6.54\pm0.03$ & $3.53\pm0.03$ & $35.31\pm0.38$ & $467.7\pm5.40$ \\ 
$6.0-8.0$ & $5.34\pm0.03$ & $2.92\pm0.03$ & $25.63\pm0.31$ & $475.72\pm7.26$ \\ 
$8.0-10.0$ & $4.28\pm0.02$ & $2.54\pm0.02$ & $46.93\pm0.26$ & $449.88\pm9.65$ \\ 
$10.0-14.0$ & $8.21\pm0.03$ & $5.33\pm0.03$ & $26.00\pm0.37$ & $474.03\pm5.53$ \\ 
$14.0-18.0$ & $5.68\pm0.03$ & $4.06\pm0.03$ & $42.36\pm0.30$ & $452.64\pm11.79$ \\ 
\hline
\end{tabular}
\caption{The best fit parameters for the profile fitting of the eclipse, including the persistent count rate, eclipse count rate, duration of ingress and egress, and the duration of the eclipse, were found for the stacked light curves of all the eclipses, folded by $T_{mid}$ across different energy ranges. All the light curves in this modeling are binned at 10 s across various energy bands due to the low count rates in individual energy bands.} 
\label{tab:folded_eclipse}
\end{table*}
\begin{figure}[t!]
    \centering
    \includegraphics[width=.7\linewidth]{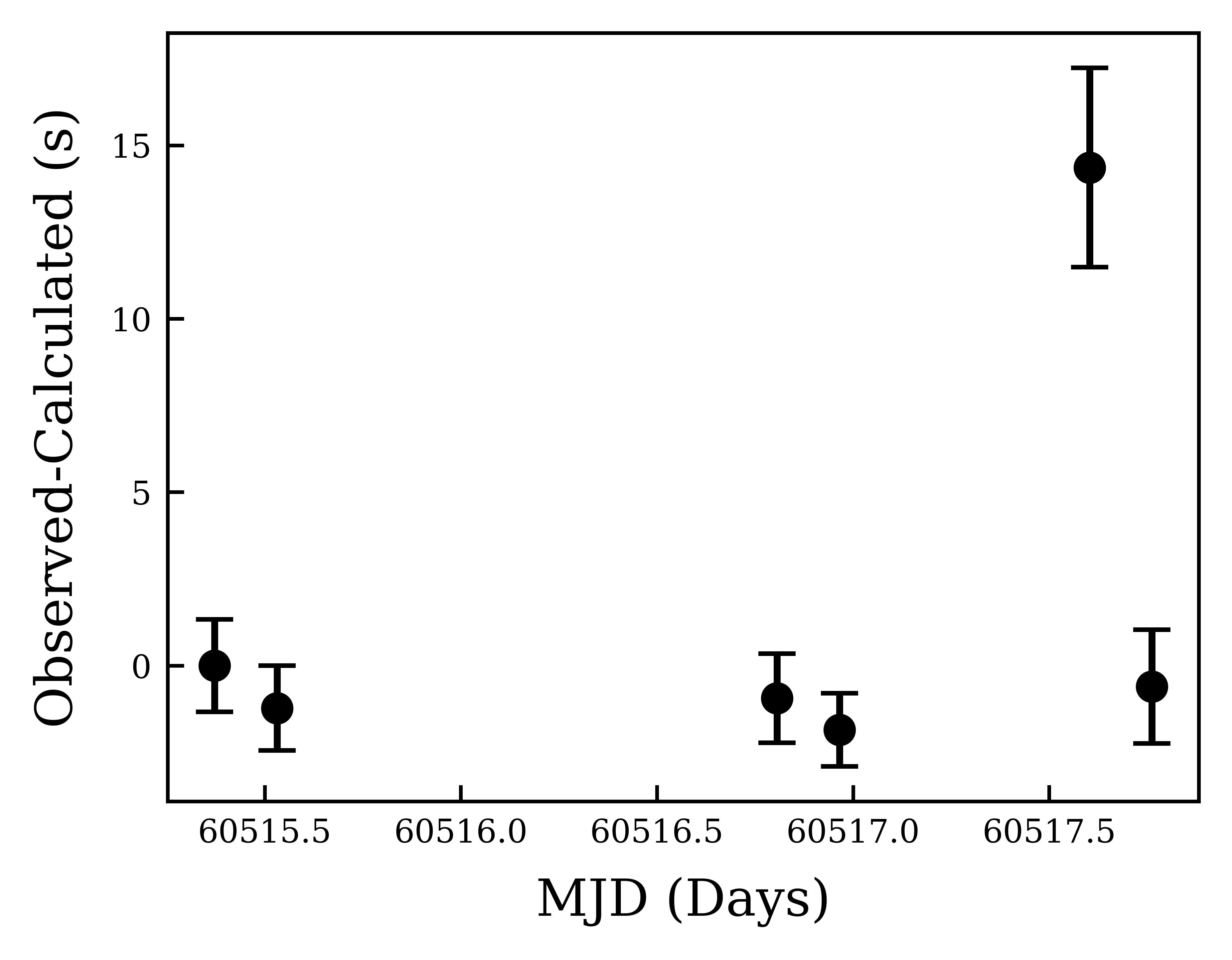}
    \caption{Mid-eclipse timing residuals for observed eclipses of EXO~0748--676 for the six eclipses observed for an orbital period of 0.15933782601~days mentioned in \cite{wolff_eclipse_2009}. Eclipse~5 shows a large residual, potentially caused by fitting artefacts due to a flash-like phenomenon that occurred right after the eclipse egress.}
    \label{fig:o-c}
\end{figure}

We aimed to identify changes in the orbital period of the binary system during the observation duration, using an orbital period of 0.15933782601 days, as noted in \cite{wolff_eclipse_2009}. This period was determined by \cite{wolff_eclipse_2009} based on observations between MJD 48107 and 51610, and we have taken this as our trial orbital period. Considering the mid-point of the first eclipse as our reference point, we computed the eclipse occurrence time in our sample. After that, we applied the orbital period to predict $t_{mid}$ of the subsequent eclipses. Next, we subtracted the observed mid-eclipse time from our predicted timing. Figure~\ref{fig:o-c} shows the difference between the observed and calculated mid-eclipse times, plotted against the mid-eclipse time in MJD. Eclipse 5 deviates from the expected orbital period. This discrepancy may be attributed to a fitting artifact caused by a flare-like phenomenon that occurred just after the eclipse, complicating the fitting process. The event is not because of a burst during the eclipse phase \citep{knight_type_2025, rikame_thermonuclear_2025}, as it does not follow a clear burst-like profile. There are incidents of flaring events reported after the eclipses \citep{buisson_dips_2021}. We have observed it after only one eclipse, so we lack sufficient samples for further analysis. Therefore, for further discussions, we will exclude Eclipse 5. The five eclipses do not vary much in the observations and predictions, indicating that the system follows the same orbital period throughout the observation period. This stability is expected in a binary orbit, which typically does not exhibit significant deviations over three days. To provide further insights into the evolution of the binary orbit, we performed the eclipse profile modeling for eclipses observed from \emph{NICER}, which will be discussed in more detail in the discussion section.

\section{Discussion}
\label{sec:discussion}
This paper reports the detection and analysis of three thermonuclear X-ray bursts and six eclipses detected from the NS LMXB EXO 0748-676 during its 2024 outburst using \emph{AstroSat}.

\subsection{Burst}

Joint time-resolved spectral analysis showcased the global changes during the burst evolution covering a wide energy range of $0.7-18$ keV. In this section, we will discuss the results obtained from the spectral and timing analysis of the bursts.

All three LAXPC detected bursts are relatively longer duration bursts with a relatively long rise time of $\sim5$~s and a decay time of a minute. To gain more understanding about the nature of the burst, we estimated the associated mass accretion rate based on the relation between the mass accretion rate ($\dot{M}$) and observed flux ($F$) described in \citep{johnston_multi-epoch_2020} 
\begin{equation}
    \dot{M}\:=\: \frac{(1+z)^34\pi d^2F\xi}{z\times c^2}
\end{equation}
where $\dot{M}$ is the mass accretion rate measured on the NS surface, $\xi$ is the anisotropy factor, $F$ is the persistent flux before the burst, and $d$ is the distance to the source. Here, $z$ is the gravitational redshift around the neutron star, defined as 
\begin{equation}
    1+z\:=\: \left(1-\frac{2GM_{NS}}{R_{NS}c^2}\right)^{-1/2}
\end{equation}
where $R_{NS}$ and $M_{NS}$ are the radius and mass of the neutron star, respectively. We considered the emission to be isotropic so that $\xi=1$. We used the persistent flux values in Table~\ref{tab:pre_burst} to estimate the mass accretion rate before each burst. We determined the distance to the source using the two PRE bursts and obtained the average value of $7.42\pm0.53\:\rm kpc$ (it will be discussed later). Assuming that the neutron star has a mass of $2.1\pm0.28\rm\:M_\odot$ and a radius of $13.8\pm1.8$ km \citep{ozel_soft_2006}, we estimated the pre-burst mass accretion rate as $1.05\pm0.08\times 10^{-10}\rm\: M_\odot\:yr^{-1}$, $1.15\pm0.09\times 10^{-10}\rm\: M_\odot\:yr^{-1}$, and $1.03\pm0.08\times 10^{-10}\rm\: M_\odot\:yr^{-1}$ respectively for B1, B2, and B3. Our estimates are consistent with the previous estimates of $\dot{M}<2\times 10^{-10}\rm \:M_\odot \:yr^{-1}$ for mixed H/He burning \citep{strohmayer_new_2003}
Since EXO~0748 proposed to have an M-dwarf companion of mass $\sim$ $0.4\rm \:M_\odot$ \citep{parmar_discovery_1986, knight_eclipse_2022}, which possesses a convective interior that fills the Roche Lobe, it can transfer mixed H/He into the NS. Presence of He also suggests that in the past, the companion star may have been massive enough to burn helium in its core. When it transferred matter to the neutron star, it increased its mass and accelerated its rotation to a high spin of 552 Hz \citep{galloway_discovery_2010}.

\begin{figure} 
    \centering
    \includegraphics[width=.5\linewidth]{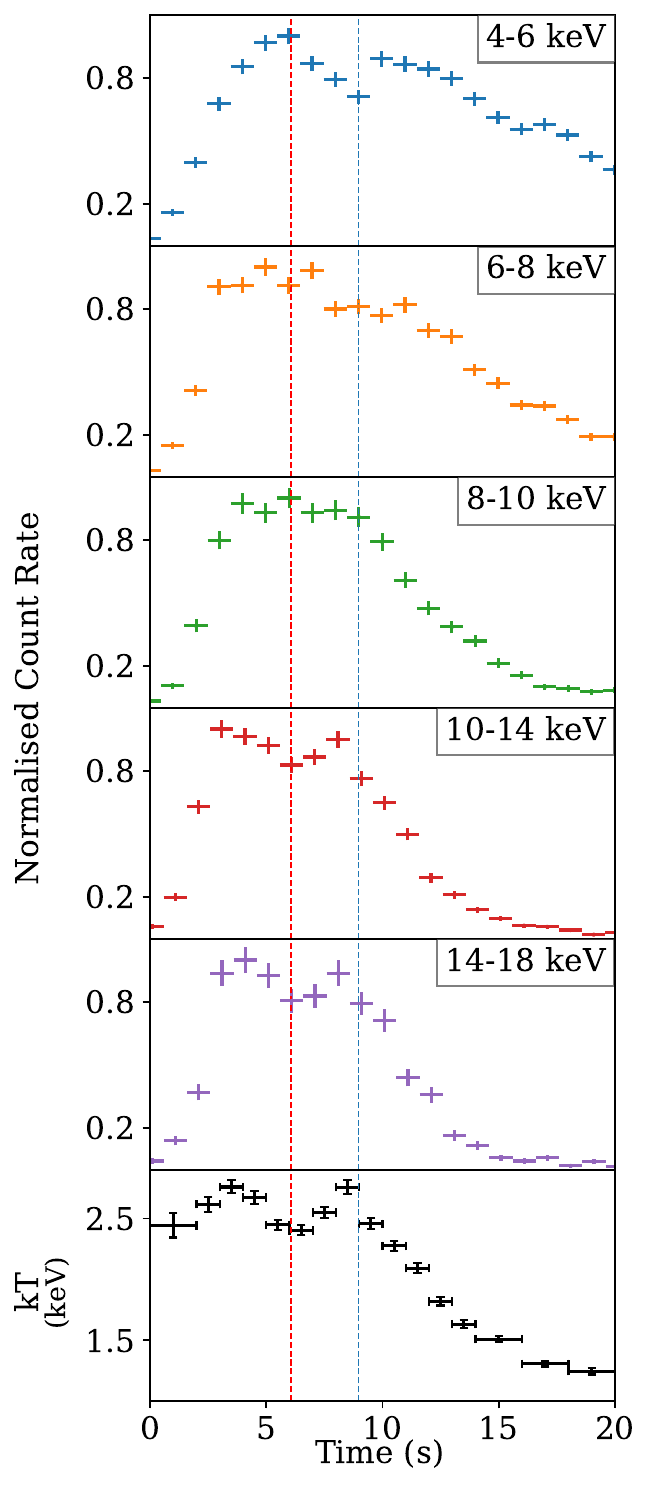}
    \caption{A comparison of the changes in light curves across different energy bands is presented in relation to the temperature evolution (bottom panel) obtained from time-resolved spectroscopy for B1 during the PRE phase. The red vertical line indicates the time at which the minimum count was observed during the peak for the higher energy bands (10-14 keV and 14-18 keV), while the blue vertical line represents the minimum count observed for the lower energy band (4-6 keV).}
    \label{fig:peak_b1}
\end{figure}

From the pre-burst spectral analysis presented in Table~\ref{tab:pre_burst}, we can observe a gradual change in the $\Gamma$ from 1.4 to 1.9 between B1 and B3, respectively. A change in $\Gamma$ value indicates that the pre-burst spectrum became softer as time progressed. There is a slight increase in the pre-burst flux during the B2, which does not overlap with the flux values of B1 and B3 within the error bars, indicating that the accretion rate increased slightly around that specific epoch, resulting in the faintest burst observed among all three bursts. Notably, the anti-correlation between persistent flux and burst peak flux has been discussed in several other studies as well \citep{chakraborty_x-ray_2011, den_hartog_burst-properties_2003, degenaar_accretion_2018}. If the accretion rate is higher, the conditions for the nuclear reaction will be attained in less time, resulting in less intense bursts.

\begin{figure*}
    \centering
    \includegraphics[width=0.32\textwidth]{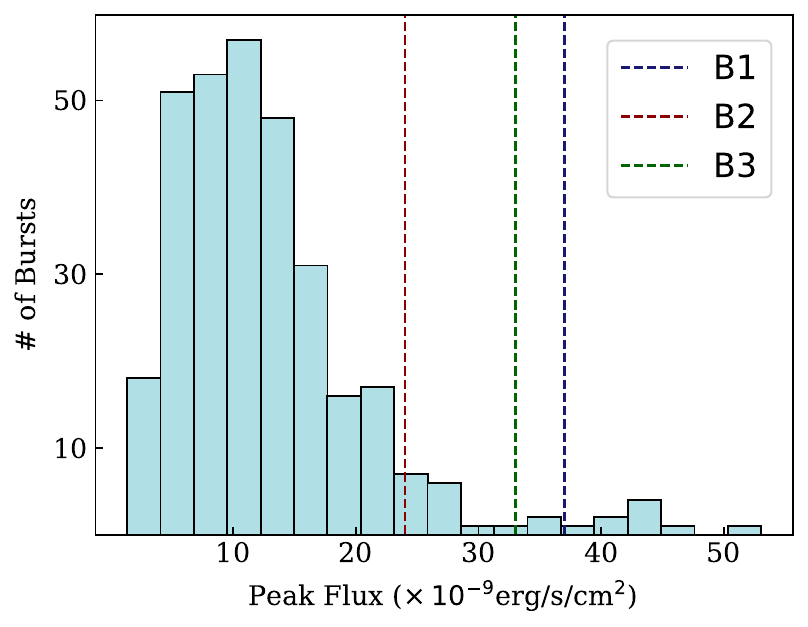}
    \includegraphics[width=0.32\textwidth]{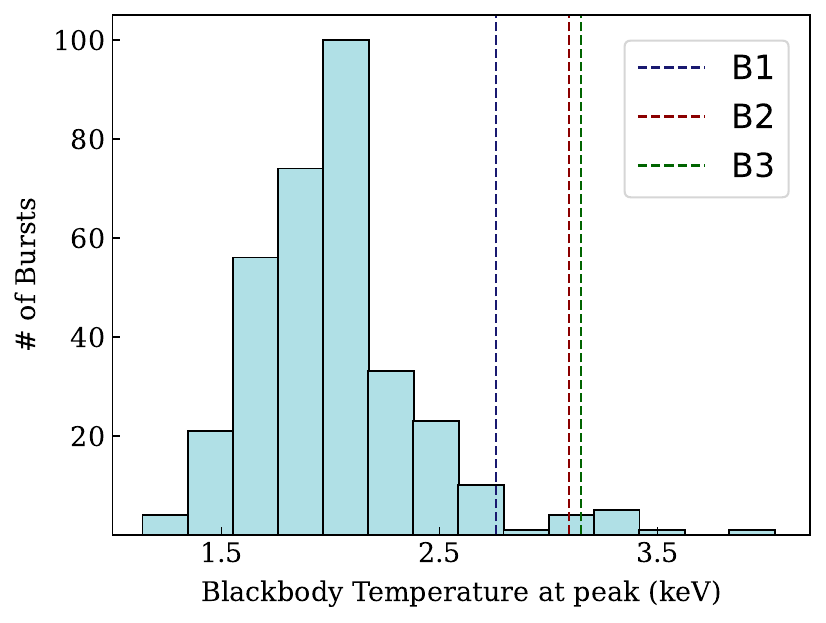}
    \includegraphics[width=0.32\textwidth]{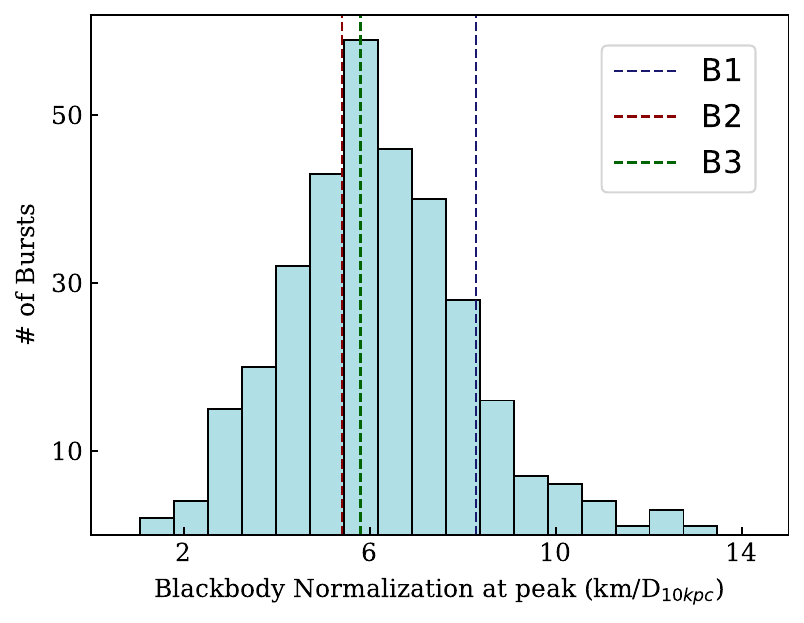}
    \caption{The histograms for the observed peak flux, blackbody temperature, and blackbody normalization values for EXO~0748--676, based on the MINBAR catalogue, which documented the source's previous outburst. Vertical dashed lines represent the values obtained from the time-resolved spectral analysis of individual bursts, labeled B1, B2, and B3, which are indicated in blue, red, and green colors, respectively. }
    \label{fig:minbar}
\end{figure*}

 Figure~\ref{fig:burst_sxt_laxpc} shows the reprocessed burst emission from B1 detected in the FUV band using the UVIT onboard \emph{AstroSat}. Unfortunately, UVIT caught only the decay part of the burst, and the low SNR ratio of the UV light curve obtained hindered detailed correlation studies between the burst and reprocessed emission in the X-ray and UV bands. It could be better to have high cadence simultaneous observations with higher SNR in future follow-up with \emph{AstroSat}, so that we can properly map and understand the neutron star environment through techniques like reverberation mapping. Reprocessed bursts were previously reported from EXO 0748 during its previous outburst \citep{hynes_multiwavelength_2006} and the 2024 outburst \citep{knight_simultaneous_2024}.

From the time-resolved spectroscopic analysis of the LAXPC only burst data (Figure~\ref{fig:trs_burst}), it is evident that both B1 and B3 show PRE nature. 
The temperature exhibits a decrease, and the radius ($Norm.$) reaches the maxima at $\sim 3-6$ for B1 and $\sim 3-5$ s for B3, signifying the phase where the radiation pressure exceeds the gravitational pull of the neutron star, resulting in the expansion of the burning nuclear ocean radially upward from the surface of the neutron star.
When the radiation pressure decreases just below the gravitational pull, the expanded envelope approaches the lowest temperature and maximum radius (around 6~s and 5~s, respectively, for B1 and B3), and thereafter, it starts to contract. During the contraction of the layer, an increase in temperature and a decrease in radius can be seen (6-8~s for B1 and 5-7~s for B3). When it touches the neutron star surface, it will reach a maximum temperature value (2.75 keV for B1 at 8~s and 3.14 keV for B3 at 7~s), and following that, it starts to decay, usually as we see in normal burst tail emission. The temperature profile will have a double-peaked structure as observed in B1 and B3. During a PRE burst, the expanding photosphere causes a temperature drop, which makes the blackbody spectrum shift toward lower energies, creating a deficit of higher-energy photons. During contraction, the peak of the blackbody will return to where it started (higher temperature); consequently, this phase is seen as a double-peaked structure in the X-ray light curve of the burst. Figure~\ref{fig:peak_b1} represents the zoomed-in portion of the PRE phase of B1, and it is very clear that the 10-18 keV flux varies exactly the way the temperature changes. We can not detect any such change in the 6-10 keV range; in this range, the count rate stays roughly constant throughout the PRE phase. In the soft energy range (4-6 keV), we can see that the first peak appears during the maximum radius (or lowest temperature), which is because as the blackbody shifts towards the softer temperature, more photons are generated in the softer energy range, and we will see a peak at these energies during this time. During the contraction phase, as the peak of the blackbody moves towards hard energies, it will reduce the photon counts in soft energy, so we can see a reduction in softer energies till the touchdown of the photosphere. We will get the least intensity in soft energy ranges during this time. Once the photosphere touches the NS surface, the temperature starts to cool down, so that the peak of the blackbody shifts to lower energies, and we can see a sudden increase in the count rate, which results in the second peak in the soft energy.

Using the PRE nature, we tried to estimate the distance to the source using the peak PRE flux. From \cite{galloway_thermonuclear_2008}, the distance to the source is given by
\begin{eqnarray}
    d &=& 8.6 \left( \frac{F_{\mathrm{peak,PRE}}}{3 \times 10^{-8}\ \mathrm{ergs\ cm}^{-2}\ \mathrm{s}^{-1}} \right)^{-1/2}
     \left( \frac{M_{\mathrm{NS}}}{1.4\, M_\odot} \right)^{1/2} \nonumber\\ &&\times \left[ \frac{1 + z(R)}{1.31} \right]^{-1/2}
     (1 + X)^{-1/2}\ \mathrm{kpc}
\end{eqnarray}
where $F_{\mathrm{peak,PRE}}$ is the peak flux observed during PRE, $z(R)$ is the gravitational redshift, $M_{NS}$ is the mass of the neutron star, and $X$ is the mass fraction of hydrogen in the NS atmosphere. We considered the hydrogen abundance to be 0.7 (which is the solar abundance) and obtained the distance separately for the two PRE bursts, B1 and B3. We fixed the neutron star mass and radius values as discussed earlier. Taking the average of the distance values obtained for the two PRE bursts, we found the distance to be $7.42\pm0.53$ kpc for the given hydrogen abundance. Based on the previous literature, it is found that the distance to EXO~0748 is estimated to be between 5 and 7.9 kpc, obtained through different methods such as PRE and spectral modeling \citep{wolff_strong_2005, zhang_distance_2011, cheng_cooling_2017, zhang_suzaku_2016}. Our distance estimate also matches the previously reported values, except that in \cite{ozel_soft_2006}.  The discrepancy with \citep{ozel_soft_2006} can be attributed to the fact that our distance calculation depends on the mass and radius taken from \cite{ozel_soft_2006} and \cite{knight_eclipse_2022}, whereas the distance measurement in \cite{ozel_soft_2006} is done independently from the spectral features of a large number of PRE bursts observed using \emph{RXTE}. Our model also does not account for the unknown color factor, which is influenced by the neutron star atmosphere and introduces another source of uncertainty. 
It should be noted that the distance estimation from PRE is highly sensitive to the hydrogen mass fraction, and minor variations in $X$ can result in substantial changes in distance estimations.
 
Figure~\ref{fig:burst_sxt_laxpc} shows a lag between the primary peaks of burst B3 in the simultaneous SXT and LAXPC light curves, with the peak delayed in the SXT. We attribute this lag to the rapid evolution of the blackbody temperature during the burst peak. Time-resolved spectroscopy of B3 (see Figure~\ref{fig:trs_burst}) indicates a peak temperature of 3.22 keV, which shifts the bulk of the emission flux into the mid X-ray ranges. Consequently, the LAXPC instrument (3-80 keV) detects the initial peak, while the SXT (0.3-8 keV), which is more sensitive to lower energies, does not. When the burst cools down, the spectral peak shifts into the SXT band, producing a delayed peak in the light curve. This effect is also present, but less prominent, in burst B1 with a lower peak temperature of 2.76 keV. At this temperature, a larger fraction of the peak emission falls within the SXT energy range, and cooling results in a slight flux increase in the SXT light curve just before the decay begins. 

In Figure~\ref{fig:trs_burst} blackbody normalization of B1 remains constant during its maximum phase (between 20-30~s). This interval corresponds to a distinct plateau in the decay portion of the SXT light curve and a minor hump in the LAXPC light curve (see Figure~\ref{fig:burst_sxt_laxpc}). We propose that this feature represents the period during which the thermonuclear flame front achieved full surface coverage and burned steadily for a prolonged time before starting to diminish.

Figures~\ref{fig:trs_burst} and \ref{fig:jtrs_burst} show that during the PRE phase, the radius of the blackbody does not reach a global maximum value, indicating that the flame spreading is confined to a small area till this phase. Confinement of the flame spreading might be because of several factors, like piling up of matter in front of the flame front, or it might be because of the stalling of the burst hotspot due to the interaction of the magnetic field of the neutron star with the dynamic magnetic field associated with the moving charged particles in the flame. There, it undergoes local PRE, and when it touches down again, the flame starts to propagate and engulf the NS surface or a fraction of it. Confinement of the burst hotspot causes the burst to spread slowly than the usual time scale of the burst \citep{chakraborty_thermonuclear_2012}. Among the three bursts observed, B1 is the brightest but corresponds to a comparatively lower hotspot temperature. However, the hotspot is spread over a larger area for this burst, resulting in a larger burst flux. B1 and B3 undergo PRE, and both appear not to engulf the entire NS surface during the PRE phase. However, it should be noted that this assumption is only valid when the observed normalization exactly mimics the evolution of the hotspot radius. In reality, it may be influenced by various factors, including the color factor, which depends on the composition of the photosphere at the time the burst occurs \citep{bhattacharyya_systematic_2010}. 

From the broadband spectral analysis of the bursts B1 and B3 detected simultaneously in LAXPC and SXT in Figure~\ref{fig:jtrs_burst}, we have observed an enhancement of the persistent emission well represented by the $f_a$ method near the burst peak. Only some bins near the peak and decay portion of B1 satisfied the F-test condition of more than $3\sigma$ confidence level. Broadband spectral analysis covers a wide energy range from 0.7 - 18 keV, where the soft excess, possibly coming from the persistent emission, increases during the burst, and can be detected more clearly. In Figure~\ref{fig:jtrs_burst}, we observed that only the $f_a$ method displayed the PRE phase, while the standard method did not. This is because the $f_a$ method considers the soft excess caused by increased persistent emission, where the standard method does not consider this factor. When averaged over time, the standard method attempts to fit the data without acknowledging the soft excess. In contrast, the $f_a$ method incorporates the soft excess, allowing us to observe the PRE phase clearly.

From the time-resolved spectral analysis and the light curves, it is evident that B1 produces more flux than B3 at the peak of the burst. As shown in Figure~\ref{fig:trs_burst}, the normalization during the local PRE indicates that the photosphere of B1 is extending farther than that of B3. This greater extension of the photosphere increases the likelihood of interaction between the accretion disk and the burst photons, thereby enhancing the accretion rate during the peak of burst B1. Alternatively, it is possible that B3 could result in a reduced interaction between the burst photons and the accretion disk, which is supported by the fact that we could not significantly detect the excess through the $f_a$ method in B3. The increase in accretion rate would not be sustained for a long time as the amount of photons ejected decreases after the burst peak, which will result in low disc-burst interaction and hence, no further increase in the accretion rate can be seen during the decay portion \citep{guver_burstdisk_2022}. 

Figure~\ref{fig:burst_sxt_laxpc}  shows the presence of a secondary burst peak detected during the decay phase of B3, which was reported for many sources previously \citep{guver_thermonuclear_2021, jaisawal_nicer_2019, bhattacharyya_non-pre_2006}. However, no detection of such a burst has been reported in the literature for EXO~0748 during its first outburst. Detection of burst triplets has been reported \citep{boirin_discovery_2007}, but those bursts are distinct bursts with very short separation between them, so it is not the same as the secondary peaks we are discussing here. We went through MINBAR data and identified several bursts detected from the source that contain signatures of a secondary peak (MINBAR ID: 3014, 3065, 3068, 3139, 3499, 3557, 3621, 3656, all observed using \emph{RXTE/PCA}  \citep{galloway_multi-instrument_2020}). Recent \emph{NuSTAR} observations during the 2024 outburst of EXO~0748 reported detections of bursts exhibiting a clear secondary peak that dominates in the soft X-rays \citep{subba_eclipse_2024}. Our observations indicate that a secondary peak occurs later in the B3 data, particularly in lower energy bands. The secondary peak is clearly evident in the SXT data (see Figure~\ref{fig:burst_sxt_laxpc}), where the secondary peak reaches a count rate of two-thirds of the primary peak intensity. From the SXT light curve and the time-resolved spectral analysis, we can conclude that the secondary peak is not associated with the PRE, as it occurred $\sim30$~s after the beginning of the burst and the PRE phase ends $\sim20$~s before the secondary peak. From the time-resolved spectroscopy, it is evident that there is no significant temperature increase during the secondary peak. However, the radius expands rapidly, contrary to the behaviour observed in the previous cases. 
Specifically, it can be noted that the secondary peaks reported by \cite{guver_thermonuclear_2021} and \cite{jaisawal_nicer_2019} have an apparent temperature rise associated with them. These studies discuss the physical scenarios that can generate the secondary peaks. One possible cause for the secondary peaks is a re-ignition of leftover material. However, we eliminated this possibility because such an event would have increased the temperature, which did not happen.
The alternate explanation for the secondary peak involves the stalling of the flame front. Stalling of the flame front can arise due to the interaction of the high magnetic field offered by the neutron star with the magnetic field developed by the moving ionic particles in the flame and the Coriolis force effects on the flame front in higher ignition latitude. When the flame front halts briefly, the restricting forces slowly dominate, eventually causing rapid expansion of the flame front without increasing the net temperature \citep{bhattacharyya_thermonuclear_2007}.

\begin{figure} [t!]
    \centering
    \includegraphics[width=0.8\linewidth]{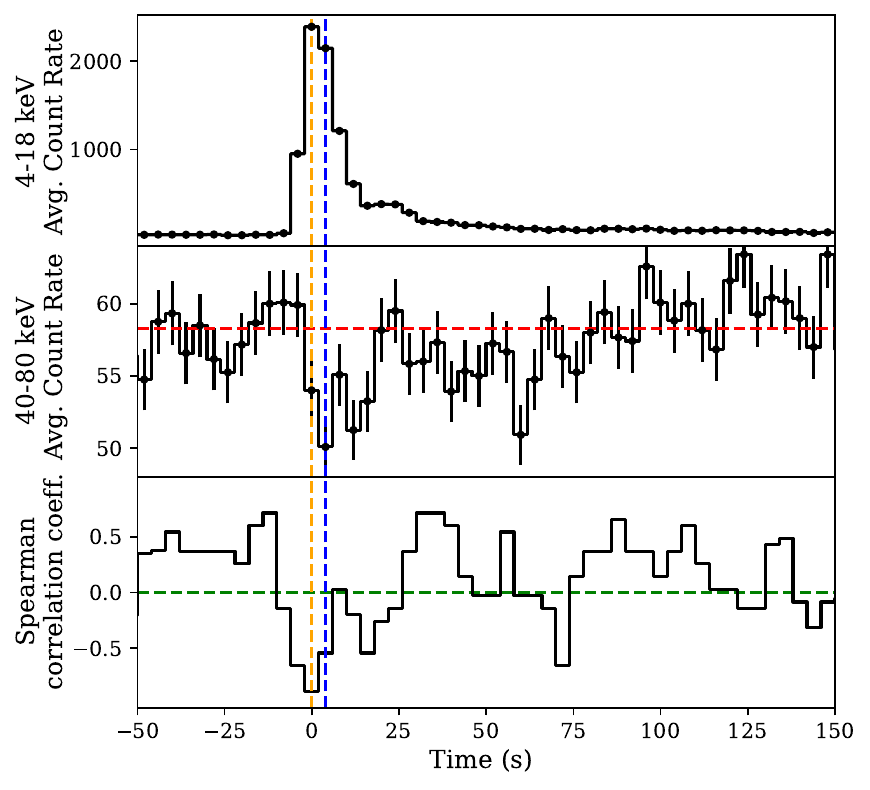}
    \caption{Panel 1 shows the 4~s binned light curve of the combined burst portions in the 4-18 keV range. Panel 2 shows the corresponding portion of the 40-80 keV light curve for the three bursts, binned over 4~s. Panel 3 presents the dynamic Spearman correlation coefficient for the high (40-80 keV) and low (4-18 keV) energy portions. The orange vertical line indicates the peak of the burst, and the blue vertical line shows the maximum deficit; both lines are separated by 4~s. The Red horizontal line in the second panel represents the persistent count rate prior to 100~s of the burst in the high-energy range of 40-80 keV. The green horizontal line in the third panel indicates a value of zero, meaning there is neither a correlation nor an anti-correlation present.}
    \label{fig:def_comb}
\end{figure}

We observed a rise in the temperature during the very end portion of the burst, which is usually not seen during a thermonuclear X-ray burst. It can be due to the statistical uncertainty while fitting the data or the low counts in the burst tail. We also got a persistent emission increase during the tail using the $f_a$ method, which the F-test statistically verifies. The temperature increase may be associated with the accretion disk returning to its pre-burst state following the burst, which had a significant impact on the persistent emission.


EXO~0748 was observed regularly with different instruments such as \emph{RXTE, XMM-Newton, INTEGRAL} during its previous outburst \citep{wolff_eclipse_2009,galloway_multi-instrument_2020}. 357 thermonuclear bursts from EXO~0748 were reported in the MINBAR catalog, making it the fourth frequent bursting source listed. We compared the three bursts observed during the 2024 outburst by \emph{AstroSat} with the 357 bursts observed in the previous outburst, and the comparison is given in Figure~\ref{fig:minbar}. The cyan-colored histograms show the bursts recorded during the previous outburst, categorized by different peak flux levels, blackbody temperatures, and blackbody normalizations at the peak. The dashed lines indicate the bursts observed by \emph{AstroSat}. The three bursts observed using \emph{AstroSat} during the 2024 outburst are stronger than most bursts recorded in the MINBAR catalog. Less than 25 bursts in the MINBAR catalog have peak flux greater than the least powerful burst among the three reported in this work. A similar trend can be seen in temperature. Our observations were taken in the initial phase of the outburst, and it was observed by \cite{guver_nicer_2021} that bursts in the initial outburst phase are powerful. Suppose we assume that the binary system does not undergo many changes between the two outbursts. In that case, the three bursts we observed here are more powerful than most of the bursts in MINBAR because they were observed in the initial phases of the outburst.

As reported in Figure~\ref{fig:defi_burst}, we observed a significantly high energy deficit during the burst. To understand the overall trend, we combined all three bursts in the low energy range (4-18 keV) and the corresponding high energy range (40-80 keV) into 4~s binned light curves. We combined the light curves so that the individual burst peaks would align. We calculated the Spearman correlation coefficient between the high-energy and low-energy counterparts of the light curve within a 28~s window. We then moved this window forward by 4~s from the start of the previous one, repeating this process for the entire length of the light curve. The result of this analysis is represented in the bottom panel of Figure~\ref{fig:def_comb}. We found a clear deficit in the high energy (40-80 keV) part of the light curve during the peak of the burst, and the corresponding Spearman coefficient signified the clear anti-correlation between the high energy and low energy intensity. The deficit in the hard energy is due to the cooling of the corona through Compton scattering of the burst photons with the high-energy electrons present in the corona \citep{maccarone_spectral_2003,ji_hard_2014}. It has been previously reported for various sources \citep{kashyap_probing_2022, guver_burstdisk_2022}. We observed a delay of about 4~s between the burst peak in the 4-18 keV and the maximum deficit in 40-80 keV; the coarser time binning of the light curve limits the resolution. This hard lag matches the reported 1-3~s \citep{chen_insighthxmt_2024}.

We observed a secondary dip after the burst as well at $\sim 60$~s after the burst start, and it is unclear whether it is due to some statistical fluctuations or of physical origin. Nevertheless, the time scale of the secondary dip matches the time when we observed the temperature increase seen in the burst tail from the time-resolved spectroscopy. Hence, this renders the scenario of the secondary dip arising from statistical fluctuations less favorable. In the alternate scenario, as the temperature increases at the burst's end, more seed photons are available for interaction with the corona. Because of this enhanced interaction, Compton cooling of the corona occurs, decreasing the high-energy photons. 

\subsection{Eclipse}
Eclipses in EXO 0748 have been observed since its discovery back in 1985, and that makes it a great candidate for probing the changes in the binary orbital parameters. We have performed the timing analysis associated with the eclipse. Due to poor statistics, background dominance, and low intensity during the eclipse, we could not conduct the spectral analysis for the eclipse regions. 

In Figure~\ref{fig:eclipse_stacked}, we tried to understand the variation in the ingress and the egress time of the eclipses detected during the LAXPC observation of EXO~0748. In \cite{knight_eclipse_2022}, it was shown that ingress and egress time show energy dependence in the energy range 0.2-10 keV. We could not detect any significant energy dependency of the ingress or egress time in the energy range of 4.0-18.0 keV. This might be because of LAXPC's comparatively poor spectral resolution. As we obtained six eclipses over a short duration, as in Figure~\ref{fig:o-c}, obtaining the binary period derivative associated with the system in the current phase is hard. The binary period derivative can be estimated once we have more eclipse observations from the source using different instruments over a longer span during the ongoing outburst.

\begin{figure}[ht!] 
    \centering
    \includegraphics[width=.7\linewidth]{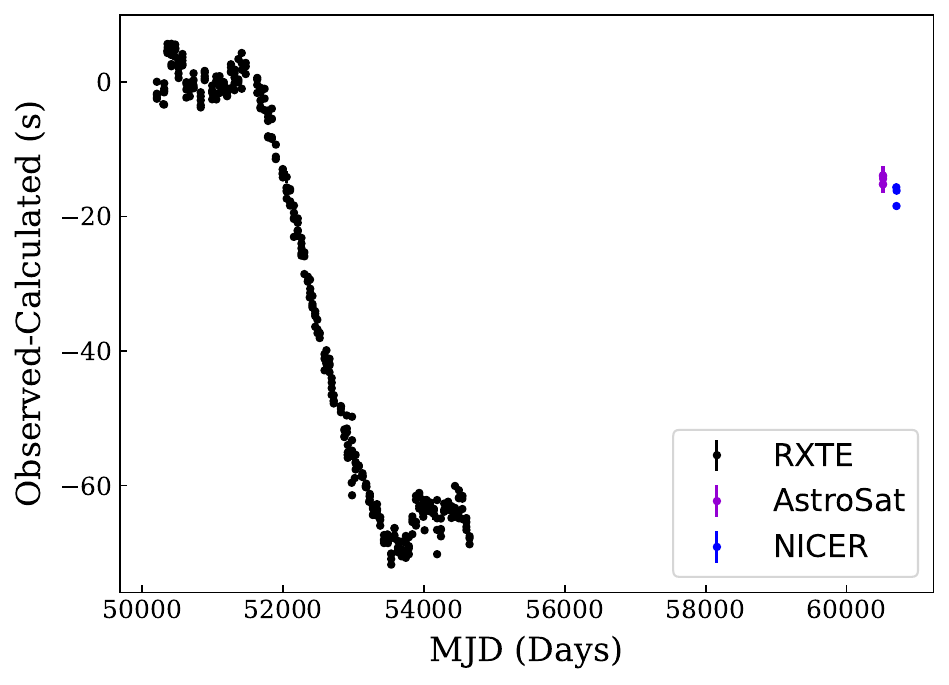}
    \caption{Mid-eclipse timing residuals for observed eclipses of EXO~0748--676 for an orbital period of 0.15933782601 days for different observatories, including \emph{RXTE} data (black) from the previous outburst \citep{wolff_eclipse_2009}, \emph{AstroSat} data (purple) from this study and \emph{NICER} data (blue) in the current outburst, spanning from 1996-2025. \emph{AstroSat} and \emph{NICER} data are separated by $\sim 200$ days, and the orbital period variation is clearly visible in the plot.}
    \label{fig:o-cRXTE}
\end{figure}

We processed the \emph{NICER} data using standard criteria with \texttt{NICERDAS} and the calibration file \texttt{xti20240206} \citep{den_herder_neutron_2016}. After applying a barycentric correction with the \texttt{heasoft} tool \texttt{barycorr}, we identified three eclipses in the data.

Eclipses in EXO 0748 have been studied in detail during the previous outburst using \emph{RXTE} and several other instruments by \cite{wolff_eclipse_2009}. They mentioned that EXO~0748 exhibits different epochs in which the orbital period changes in the order of milliseconds. We combined our detected eclipses and also the eclipses obtained in the \emph{NICER} data with the eclipse data reported in \cite{wolff_eclipse_2009}, and computed the residual in the observed and the calculated mid eclipse time, which is shown in Figure~\ref{fig:o-cRXTE}. \emph{AstroSat} and \emph{NICER} data is separated by $\sim 200$~days. It can be seen that, at the beginning of the current outburst (shown by the purple and blue symbols), the binary system was not in the same epoch as it was before the quiescence and \emph{NICER} data verifies that result. Hence, we can assume that the source is in a different epoch, currently in a phase where the orbital period is decreasing. \cite{wolff_eclipse_2009} concluded that the residuals in the calculated and observed mid-time of the eclipses are due to the magnetic field cycling of the secondary star. Our results also support their findings despite having fewer data points. It follows the same old pattern of decreasing orbital period as shown between 51610 MJD and 53363 MJD. We assume that the secondary star reversed its magnetic field in quiescence, leading to several changes in orbital period evolution. Currently, the system has returned to a phase similar to that seen in the above-mentioned time interval.


\section*{Acknowledgements}

This work uses data from the \emph{AstroSat} mission of the Indian Space Research Organization (ISRO), archived at the Indian Space Science Data Centre (ISSDC). We thank the LAXPC POC team, the SXT POC team and the UVIT POC team for their support, timely release of data, and provision of the necessary software tools. This work has made use of software provided by HEASARC. SB acknowledges financial support by the Fulbright-Nehru Academic \& Professional Excellence Award (Research), sponsored by the U.S. Department of State and the United States-India Educational Foundation (grant number: 3062/F-N APE/2024; program number: G-1-00005). 


\bibliographystyle{elsarticle-harv} 
\bibliography{reference}

\begin{thebibliography}{67}
\expandafter\ifx\csname natexlab\endcsname\relax\def\natexlab#1{#1}\fi
\providecommand{\url}[1]{\texttt{#1}}
\providecommand{\href}[2]{#2}
\providecommand{\path}[1]{#1}
\providecommand{\DOIprefix}{doi:}
\providecommand{\ArXivprefix}{arXiv:}
\providecommand{\URLprefix}{URL: }
\providecommand{\Pubmedprefix}{pmid:}
\providecommand{\doi}[1]{\href{http://dx.doi.org/#1}{\path{#1}}}
\providecommand{\Pubmed}[1]{\href{pmid:#1}{\path{#1}}}
\providecommand{\bibinfo}[2]{#2}
\ifx\xfnm\relax \def\xfnm[#1]{\unskip,\space#1}\fi
\bibitem[{Antia et~al.(2017)Antia, Yadav, Agrawal, Chauhan, Manchanda, Chitnis, Paul, Dedhia, Shah, Gujar, Katoch, Kurhade, Madhwani, Manojkumar, Nikam, Pandya, Parmar, Pawar, Pahari, Misra, Navalgund, Pandiyan, Sharma and Subbarao}]{antia_calibration_2017}
\bibinfo{author}{Antia, H.M.}, \bibinfo{author}{Yadav, J.S.}, \bibinfo{author}{Agrawal, P.C.}, \bibinfo{author}{Chauhan, J.V.}, \bibinfo{author}{Manchanda, R.K.}, \bibinfo{author}{Chitnis, V.}, \bibinfo{author}{Paul, B.}, \bibinfo{author}{Dedhia, D.}, \bibinfo{author}{Shah, P.}, \bibinfo{author}{Gujar, V.M.}, \bibinfo{author}{Katoch, T.}, \bibinfo{author}{Kurhade, V.N.}, \bibinfo{author}{Madhwani, P.}, \bibinfo{author}{Manojkumar, T.K.}, \bibinfo{author}{Nikam, V.A.}, \bibinfo{author}{Pandya, A.S.}, \bibinfo{author}{Parmar, J.V.}, \bibinfo{author}{Pawar, D.M.}, \bibinfo{author}{Pahari, M.}, \bibinfo{author}{Misra, R.}, \bibinfo{author}{Navalgund, K.H.}, \bibinfo{author}{Pandiyan, R.}, \bibinfo{author}{Sharma, K.S.}, \bibinfo{author}{Subbarao, K.}, \bibinfo{year}{2017}.
\newblock \bibinfo{title}{Calibration of the {Large} {Area} {X}-{Ray} {Proportional} {Counter} ({LAXPC}) {Instrument} on board {AstroSat}}.
\newblock \bibinfo{journal}{The Astrophysical Journal Supplement Series} \bibinfo{volume}{231}, \bibinfo{pages}{10}.
\newblock \URLprefix \url{https://iopscience.iop.org/article/10.3847/1538-4365/aa7a0e}, \DOIprefix\doi{10.3847/1538-4365/aa7a0e}.
\bibitem[{Bahramian and Degenaar(2023)}]{bambi_low-mass_2023}
\bibinfo{author}{Bahramian, A.}, \bibinfo{author}{Degenaar, N.}, \bibinfo{year}{2023}.
\newblock \bibinfo{title}{Low-{Mass} {X}-ray {Binaries}}, in: \bibinfo{editor}{Bambi, C.}, \bibinfo{editor}{Santangelo, A.} (Eds.), \bibinfo{booktitle}{Handbook of {X}-ray and {Gamma}-ray {Astrophysics}}. \bibinfo{publisher}{Springer Nature Singapore}, \bibinfo{address}{Singapore}, pp. \bibinfo{pages}{1--62}.
\newblock \URLprefix \url{https://link.springer.com/10.1007/978-981-16-4544-0_94-1}, \DOIprefix\doi{10.1007/978-981-16-4544-0_94-1}.
\bibitem[{Bhattacharya et~al.(2024)Bhattacharya, Bhattacharyya and Shaw}]{bhattacharya_xmm-newton_2024}
\bibinfo{author}{Bhattacharya, S.}, \bibinfo{author}{Bhattacharyya, S.}, \bibinfo{author}{Shaw, G.}, \bibinfo{year}{2024}.
\newblock \bibinfo{title}{{XMM}-{Newton} high-resolution spectroscopy of {EXO} 0748-676 after its re-emergence from a long quiescence}.
\newblock \URLprefix \url{http://arxiv.org/abs/2408.02715}, \DOIprefix\doi{10.48550/arXiv.2408.02715}. \bibinfo{note}{arXiv:2408.02715 [astro-ph]}.
\bibitem[{Bhattacharyya(2010)}]{bhattacharyya_measurement_2010}
\bibinfo{author}{Bhattacharyya, S.}, \bibinfo{year}{2010}.
\newblock \bibinfo{title}{Measurement of neutron star parameters: {A} review of methods for low-mass {X}-ray binaries}.
\newblock \bibinfo{journal}{Advances in Space Research} \bibinfo{volume}{45}, \bibinfo{pages}{949--978}.
\newblock \URLprefix \url{https://linkinghub.elsevier.com/retrieve/pii/S0273117710000451}, \DOIprefix\doi{10.1016/j.asr.2010.01.010}.
\bibitem[{Bhattacharyya et~al.(2010)Bhattacharyya, Miller and Galloway}]{bhattacharyya_systematic_2010}
\bibinfo{author}{Bhattacharyya, S.}, \bibinfo{author}{Miller, M.C.}, \bibinfo{author}{Galloway, D.K.}, \bibinfo{year}{2010}.
\newblock \bibinfo{title}{Systematic variation in the apparent burning area of thermonuclear bursts and its implication for neutron star radius measurement}.
\newblock \bibinfo{journal}{Monthly Notices of the Royal Astronomical Society} \bibinfo{volume}{401}, \bibinfo{pages}{2--6}.
\newblock \URLprefix \url{https://academic.oup.com/mnras/article-lookup/doi/10.1111/j.1365-2966.2009.15632.x}, \DOIprefix\doi{10.1111/j.1365-2966.2009.15632.x}.
\bibitem[{Bhattacharyya and Strohmayer(2006)}]{bhattacharyya_non-pre_2006}
\bibinfo{author}{Bhattacharyya, S.}, \bibinfo{author}{Strohmayer, T.E.}, \bibinfo{year}{2006}.
\newblock \bibinfo{title}{A {Non}-{PRE} {Double}-peaked {Burst} from {4U} 1636-536: {Evidence} of {Burning} {Front} {Propagation}}.
\newblock \bibinfo{journal}{The Astrophysical Journal} \bibinfo{volume}{636}, \bibinfo{pages}{L121--L124}.
\newblock \URLprefix \url{https://iopscience.iop.org/article/10.1086/500199}, \DOIprefix\doi{10.1086/500199}.
\bibitem[{Bhattacharyya and Strohmayer(2007)}]{bhattacharyya_thermonuclear_2007}
\bibinfo{author}{Bhattacharyya, S.}, \bibinfo{author}{Strohmayer, T.E.}, \bibinfo{year}{2007}.
\newblock \bibinfo{title}{Thermonuclear {Flame} {Spreading} on {Rapidly} {Spinning} {Neutron} {Stars}: {Indications} of the {Coriolis} {Force}?}
\newblock \bibinfo{journal}{The Astrophysical Journal} \bibinfo{volume}{666}, \bibinfo{pages}{L85--L88}.
\newblock \URLprefix \url{https://iopscience.iop.org/article/10.1086/521790}, \DOIprefix\doi{10.1086/521790}.
\bibitem[{Boirin et~al.(2007)Boirin, Keek, Méndez, Cumming, Zand, Cottam, Paerels and Lewin}]{boirin_discovery_2007}
\bibinfo{author}{Boirin, L.}, \bibinfo{author}{Keek, L.}, \bibinfo{author}{Méndez, M.}, \bibinfo{author}{Cumming, A.}, \bibinfo{author}{Zand, J.J.M.I.T.}, \bibinfo{author}{Cottam, J.}, \bibinfo{author}{Paerels, F.}, \bibinfo{author}{Lewin, W.H.G.}, \bibinfo{year}{2007}.
\newblock \bibinfo{title}{Discovery of {X}-ray burst triplets in {EXO} 0748-676}.
\newblock \bibinfo{journal}{Astronomy \& Astrophysics} \bibinfo{volume}{465}, \bibinfo{pages}{559--573}.
\newblock \URLprefix \url{http://www.aanda.org/10.1051/0004-6361:20066204}, \DOIprefix\doi{10.1051/0004-6361:20066204}.
\bibitem[{Buisson et~al.(2020)Buisson, Altamirano, Bult, Mancuso, Güver, Jaisawal, Hare, Albayati, Arzoumanian, Castro Segura, Chakrabarty, Gandhi, Guillot, Homan, Gendreau, Jiang, Malacaria, Miller, Özbey Arabacı, Remillard, Strohmayer, Tombesi, Tomsick, Vincentelli and Walton}]{buisson_discovery_2020}
\bibinfo{author}{Buisson, D.J.K.}, \bibinfo{author}{Altamirano, D.}, \bibinfo{author}{Bult, P.}, \bibinfo{author}{Mancuso, G.C.}, \bibinfo{author}{Güver, T.}, \bibinfo{author}{Jaisawal, G.K.}, \bibinfo{author}{Hare, J.}, \bibinfo{author}{Albayati, A.C.}, \bibinfo{author}{Arzoumanian, Z.}, \bibinfo{author}{Castro Segura, N.}, \bibinfo{author}{Chakrabarty, D.}, \bibinfo{author}{Gandhi, P.}, \bibinfo{author}{Guillot, S.}, \bibinfo{author}{Homan, J.}, \bibinfo{author}{Gendreau, K.C.}, \bibinfo{author}{Jiang, J.}, \bibinfo{author}{Malacaria, C.}, \bibinfo{author}{Miller, J.M.}, \bibinfo{author}{Özbey Arabacı, M.}, \bibinfo{author}{Remillard, R.}, \bibinfo{author}{Strohmayer, T.E.}, \bibinfo{author}{Tombesi, F.}, \bibinfo{author}{Tomsick, J.A.}, \bibinfo{author}{Vincentelli, F.M.}, \bibinfo{author}{Walton, D.J.}, \bibinfo{year}{2020}.
\newblock \bibinfo{title}{Discovery of thermonuclear ({Type} {I}) {X}-ray bursts in the {X}-ray binary {Swift} {J1858}.6–0814 observed with \textit{{NICER}} and \textit{{NuSTAR}}}.
\newblock \bibinfo{journal}{Monthly Notices of the Royal Astronomical Society} \bibinfo{volume}{499}, \bibinfo{pages}{793--803}.
\newblock \URLprefix \url{https://academic.oup.com/mnras/article/499/1/793/5903708}, \DOIprefix\doi{10.1093/mnras/staa2749}.
\bibitem[{Buisson et~al.(2021)Buisson, Altamirano, Padilla, Arzoumanian, Bult, Segura, Charles, Degenaar, Trigo, Eijnden, Fogantini, Gandhi, Gendreau, Hare, Homan, Knigge, Malacaria, Mendez, Darias, Ng, Arabaci, Remillard, Strohmayer, Tombesi, Tomsick, Vincentelli and Walton}]{buisson_dips_2021}
\bibinfo{author}{Buisson, D.J.K.}, \bibinfo{author}{Altamirano, D.}, \bibinfo{author}{Padilla, M.A.}, \bibinfo{author}{Arzoumanian, Z.}, \bibinfo{author}{Bult, P.}, \bibinfo{author}{Segura, N.C.}, \bibinfo{author}{Charles, P.A.}, \bibinfo{author}{Degenaar, N.}, \bibinfo{author}{Trigo, M.D.}, \bibinfo{author}{Eijnden, J.v.d.}, \bibinfo{author}{Fogantini, F.}, \bibinfo{author}{Gandhi, P.}, \bibinfo{author}{Gendreau, K.}, \bibinfo{author}{Hare, J.}, \bibinfo{author}{Homan, J.}, \bibinfo{author}{Knigge, C.}, \bibinfo{author}{Malacaria, C.}, \bibinfo{author}{Mendez, M.}, \bibinfo{author}{Darias, T.M.}, \bibinfo{author}{Ng, M.}, \bibinfo{author}{Arabaci, M.O.}, \bibinfo{author}{Remillard, R.}, \bibinfo{author}{Strohmayer, T.E.}, \bibinfo{author}{Tombesi, F.}, \bibinfo{author}{Tomsick, J.A.}, \bibinfo{author}{Vincentelli, F.}, \bibinfo{author}{Walton, D.J.}, \bibinfo{year}{2021}.
\newblock \bibinfo{title}{Dips and eclipses in the {X}-ray binary {Swift} {J1858}.6-0814 observed with {NICER}}.
\newblock \bibinfo{journal}{Monthly Notices of the Royal Astronomical Society} \bibinfo{volume}{503}, \bibinfo{pages}{5600--5610}.
\newblock \URLprefix \url{http://arxiv.org/abs/2103.12787}, \DOIprefix\doi{10.1093/mnras/stab863}. \bibinfo{note}{arXiv:2103.12787 [astro-ph]}.
\bibitem[{Chakraborty and Bhattacharyya(2011)}]{chakraborty_x-ray_2011}
\bibinfo{author}{Chakraborty, M.}, \bibinfo{author}{Bhattacharyya, S.}, \bibinfo{year}{2011}.
\newblock \bibinfo{title}{X-{RAY} {BURSTS} {FROM} {THE} {TERZAN} 5 {TRANSIENT} {IGR} {J17480}-2446: {NUCLEAR} {RATHER} {THAN} {GRAVITATIONAL}}.
\newblock \bibinfo{journal}{The Astrophysical Journal} \bibinfo{volume}{730}, \bibinfo{pages}{L23}.
\newblock \URLprefix \url{https://iopscience.iop.org/article/10.1088/2041-8205/730/2/L23}, \DOIprefix\doi{10.1088/2041-8205/730/2/L23}.
\bibitem[{Chakraborty and Bhattacharyya(2012)}]{chakraborty_thermonuclear_2012}
\bibinfo{author}{Chakraborty, M.}, \bibinfo{author}{Bhattacharyya, S.}, \bibinfo{year}{2012}.
\newblock \bibinfo{title}{Thermonuclear {X}-ray bursts from the 401-{Hz} accreting pulsar {IGR} {J17498}-2921: indication of burning in confined regions}.
\newblock \bibinfo{journal}{Monthly Notices of the Royal Astronomical Society} \bibinfo{volume}{422}, \bibinfo{pages}{2351--2356}.
\newblock \URLprefix \url{https://academic.oup.com/mnras/article-lookup/doi/10.1111/j.1365-2966.2012.20786.x}, \DOIprefix\doi{10.1111/j.1365-2966.2012.20786.x}.
\bibitem[{Chen et~al.(2024)Chen, Zhang, Ji, Zhang, Peng, Kong, Chang, Shui, Tao, Qu, Ge and Li}]{chen_insighthxmt_2024}
\bibinfo{author}{Chen, Y.P.}, \bibinfo{author}{Zhang, S.}, \bibinfo{author}{Ji, L.}, \bibinfo{author}{Zhang, S.N.}, \bibinfo{author}{Peng, J.Q.}, \bibinfo{author}{Kong, L.D.}, \bibinfo{author}{Chang, Z.}, \bibinfo{author}{Shui, Q.C.}, \bibinfo{author}{Tao, L.}, \bibinfo{author}{Qu, J.L.}, \bibinfo{author}{Ge, M.Y.}, \bibinfo{author}{Li, J.}, \bibinfo{year}{2024}.
\newblock \bibinfo{title}{\textit{{Insight}–{HXMT}} observations of thermonuclear {X}-ray bursts from {4U} 1608–52 in the low/hard state: the energy-dependent hard {X}-ray deficit and cooling saturation of the corona}.
\newblock \bibinfo{journal}{Monthly Notices of the Royal Astronomical Society} \bibinfo{volume}{531}, \bibinfo{pages}{1756--1764}.
\newblock \URLprefix \url{https://academic.oup.com/mnras/article/531/1/1756/7676186}, \DOIprefix\doi{10.1093/mnras/stae1257}.
\bibitem[{Cheng et~al.(2017)Cheng, Méndez, Díaz-Trigo and Costantini}]{cheng_cooling_2017}
\bibinfo{author}{Cheng, Z.}, \bibinfo{author}{Méndez, M.}, \bibinfo{author}{Díaz-Trigo, M.}, \bibinfo{author}{Costantini, E.}, \bibinfo{year}{2017}.
\newblock \bibinfo{title}{The cooling, mass and radius of the neutron star in {EXO} 0748-676 in quiescence with {XMM}-{Newton}}.
\newblock \bibinfo{journal}{Monthly Notices of the Royal Astronomical Society} \bibinfo{volume}{471}, \bibinfo{pages}{2605--2615}.
\newblock \URLprefix \url{http://academic.oup.com/mnras/article/471/3/2605/3866924/The-cooling-mass-and-radius-of-the-neutron-star-in}, \DOIprefix\doi{10.1093/mnras/stx1452}.
\bibitem[{Churazov et~al.(1996)Churazov, Gilfanov, Forman and Jones}]{churazov_mapping_1996}
\bibinfo{author}{Churazov, E.}, \bibinfo{author}{Gilfanov, M.}, \bibinfo{author}{Forman, W.}, \bibinfo{author}{Jones, C.}, \bibinfo{year}{1996}.
\newblock \bibinfo{title}{Mapping the {Gas} {Temperature} {Distribution} in {Extended} {X}-{Ray} {Sources} and {Spectral} {Analysis} in the {Case} of {Low} {Statistics}: {Application} to {ASCA} {Observations} of {Clusters} of {Galaxies}}.
\newblock \bibinfo{journal}{The Astrophysical Journal} \bibinfo{volume}{471}, \bibinfo{pages}{673}.
\newblock \URLprefix \url{https://ui.adsabs.harvard.edu/abs/1996ApJ...471..673C}, \DOIprefix\doi{10.1086/177997}. \bibinfo{note}{publisher: IOP ADS Bibcode: 1996ApJ...471..673C}.
\bibitem[{Degenaar et~al.(2018)Degenaar, Ballantyne, Belloni, Chakraborty, Chen, Ji, Kretschmar, Kuulkers, Li, Maccarone, Malzac, Zhang and Zhang}]{degenaar_accretion_2018}
\bibinfo{author}{Degenaar, N.}, \bibinfo{author}{Ballantyne, D.R.}, \bibinfo{author}{Belloni, T.}, \bibinfo{author}{Chakraborty, M.}, \bibinfo{author}{Chen, Y.P.}, \bibinfo{author}{Ji, L.}, \bibinfo{author}{Kretschmar, P.}, \bibinfo{author}{Kuulkers, E.}, \bibinfo{author}{Li, J.}, \bibinfo{author}{Maccarone, T.J.}, \bibinfo{author}{Malzac, J.}, \bibinfo{author}{Zhang, S.}, \bibinfo{author}{Zhang, S.N.}, \bibinfo{year}{2018}.
\newblock \bibinfo{title}{Accretion {Disks} and {Coronae} in the {X}-{Ray} {Flashlight}}.
\newblock \bibinfo{journal}{Space Science Reviews} \bibinfo{volume}{214}, \bibinfo{pages}{15}.
\newblock \URLprefix \url{http://link.springer.com/10.1007/s11214-017-0448-3}, \DOIprefix\doi{10.1007/s11214-017-0448-3}.
\bibitem[{Degenaar et~al.(2009)Degenaar, Wijnands, Wolff, Ray, Wood, Homan, Lewin, Jonker, Cackett, Miller and Brown}]{degenaar_chandra_2009}
\bibinfo{author}{Degenaar, N.}, \bibinfo{author}{Wijnands, R.}, \bibinfo{author}{Wolff, M.T.}, \bibinfo{author}{Ray, P.S.}, \bibinfo{author}{Wood, K.S.}, \bibinfo{author}{Homan, J.}, \bibinfo{author}{Lewin, W.H.G.}, \bibinfo{author}{Jonker, P.G.}, \bibinfo{author}{Cackett, E.M.}, \bibinfo{author}{Miller, J.M.}, \bibinfo{author}{Brown, E.F.}, \bibinfo{year}{2009}.
\newblock \bibinfo{title}{\textit{{Chandra}} and \textit{{Swift}} observations of the quasi-persistent neutron star transient {EXO} 0748—676 back to quiescence}.
\newblock \bibinfo{journal}{Monthly Notices of the Royal Astronomical Society: Letters} \bibinfo{volume}{396}, \bibinfo{pages}{L26--L30}.
\newblock \URLprefix \url{https://academic.oup.com/mnrasl/article/396/1/L26/979943}, \DOIprefix\doi{10.1111/j.1745-3933.2009.00655.x}.
\bibitem[{Den~Hartog et~al.(2003)Den~Hartog, {J. J. M In 'T Zand}, Kuulkers, Cornelisse, Heise, Bazzano, Cocchi, Natalucci and Ubertini}]{den_hartog_burst-properties_2003}
\bibinfo{author}{Den~Hartog, P.R.}, \bibinfo{author}{{J. J. M In 'T Zand}}, \bibinfo{author}{Kuulkers, E.}, \bibinfo{author}{Cornelisse, R.}, \bibinfo{author}{Heise, J.}, \bibinfo{author}{Bazzano, A.}, \bibinfo{author}{Cocchi, M.}, \bibinfo{author}{Natalucci, L.}, \bibinfo{author}{Ubertini, P.}, \bibinfo{year}{2003}.
\newblock \bibinfo{title}{Burst-properties as a function of mass accretion rate in {GX} 3+1}.
\newblock \bibinfo{journal}{Astronomy \& Astrophysics} \bibinfo{volume}{400}, \bibinfo{pages}{633--641}.
\newblock \URLprefix \url{http://www.aanda.org/10.1051/0004-6361:20030038}, \DOIprefix\doi{10.1051/0004-6361:20030038}.
\bibitem[{Fragile et~al.(2020)Fragile, Ballantyne and Blankenship}]{fragile_interactions_2020}
\bibinfo{author}{Fragile, P.C.}, \bibinfo{author}{Ballantyne, D.R.}, \bibinfo{author}{Blankenship, A.}, \bibinfo{year}{2020}.
\newblock \bibinfo{title}{Interactions of type {I} {X}-ray bursts with thin accretion disks}.
\newblock \bibinfo{journal}{Nature Astronomy} \bibinfo{volume}{4}, \bibinfo{pages}{541--546}.
\newblock \URLprefix \url{https://www.nature.com/articles/s41550-019-0987-5}, \DOIprefix\doi{10.1038/s41550-019-0987-5}.
\bibitem[{Fragile et~al.(2018)Fragile, Ballantyne, Maccarone and Witry}]{fragile_simulating_2018}
\bibinfo{author}{Fragile, P.C.}, \bibinfo{author}{Ballantyne, D.R.}, \bibinfo{author}{Maccarone, T.J.}, \bibinfo{author}{Witry, J.W.L.}, \bibinfo{year}{2018}.
\newblock \bibinfo{title}{Simulating the {Collapse} of a {Thick} {Accretion} {Disk} due to a {Type} {I} {X}-{Ray} {Burst} from a {Neutron} {Star}}.
\newblock \bibinfo{journal}{The Astrophysical Journal Letters} \bibinfo{volume}{867}, \bibinfo{pages}{L28}.
\newblock \URLprefix \url{https://iopscience.iop.org/article/10.3847/2041-8213/aaeb99}, \DOIprefix\doi{10.3847/2041-8213/aaeb99}.
\bibitem[{Galloway et~al.(2020)Galloway, In~’T~Zand, Chenevez, Wörpel, Keek, Ootes, Watts, Gisler, Sanchez-Fernandez and Kuulkers}]{galloway_multi-instrument_2020}
\bibinfo{author}{Galloway, D.K.}, \bibinfo{author}{In~’T~Zand, J.}, \bibinfo{author}{Chenevez, J.}, \bibinfo{author}{Wörpel, H.}, \bibinfo{author}{Keek, L.}, \bibinfo{author}{Ootes, L.}, \bibinfo{author}{Watts, A.L.}, \bibinfo{author}{Gisler, L.}, \bibinfo{author}{Sanchez-Fernandez, C.}, \bibinfo{author}{Kuulkers, E.}, \bibinfo{year}{2020}.
\newblock \bibinfo{title}{The {Multi}-{INstrument} {Burst} {ARchive} ({MINBAR})}.
\newblock \bibinfo{journal}{The Astrophysical Journal Supplement Series} \bibinfo{volume}{249}, \bibinfo{pages}{32}.
\newblock \URLprefix \url{https://iopscience.iop.org/article/10.3847/1538-4365/ab9f2e}, \DOIprefix\doi{10.3847/1538-4365/ab9f2e}.
\bibitem[{Galloway et~al.(2010)Galloway, Lin, Chakrabarty and Hartman}]{galloway_discovery_2010}
\bibinfo{author}{Galloway, D.K.}, \bibinfo{author}{Lin, J.}, \bibinfo{author}{Chakrabarty, D.}, \bibinfo{author}{Hartman, J.M.}, \bibinfo{year}{2010}.
\newblock \bibinfo{title}{{DISCOVERY} {OF} {A} 552 {Hz} {BURST} {OSCILLATION} {IN} {THE} {LOW}-{MASS} {X}-{RAY} {BINARY} {EXO} 0748–676}.
\newblock \bibinfo{journal}{The Astrophysical Journal} \bibinfo{volume}{711}, \bibinfo{pages}{L148--L151}.
\newblock \URLprefix \url{https://iopscience.iop.org/article/10.1088/2041-8205/711/2/L148}, \DOIprefix\doi{10.1088/2041-8205/711/2/L148}.
\bibitem[{Galloway et~al.(2008)Galloway, Muno, Hartman, Psaltis and Chakrabarty}]{galloway_thermonuclear_2008}
\bibinfo{author}{Galloway, D.K.}, \bibinfo{author}{Muno, M.P.}, \bibinfo{author}{Hartman, J.M.}, \bibinfo{author}{Psaltis, D.}, \bibinfo{author}{Chakrabarty, D.}, \bibinfo{year}{2008}.
\newblock \bibinfo{title}{Thermonuclear ({Type} {I}) {X}‐{Ray} {Bursts} {Observed} by the \textit{{Rossi} {X}‐{Ray} {Timing} {Explorer}}}.
\newblock \bibinfo{journal}{The Astrophysical Journal Supplement Series} \bibinfo{volume}{179}, \bibinfo{pages}{360--422}.
\newblock \URLprefix \url{https://iopscience.iop.org/article/10.1086/592044}, \DOIprefix\doi{10.1086/592044}.
\bibitem[{Gendreau et~al.(2016)Gendreau, Arzoumanian, Adkins, Albert, Anders, Aylward, Baker, Balsamo, Bamford, Benegalrao, Berry, Bhalwani, Black, Blaurock, Bronke, Brown, Budinoff, Cantwell, Cazeau, Chen, Clement, Colangelo, Coleman, Coopersmith, Dehaven, Doty, Egan, Enoto, Fan, Ferro, Foster, Galassi, Gallo, Green, Grosh, Ha, Hasouneh, Heefner, Hestnes, Hoge, Jacobs, Jørgensen, Kaiser, Kellogg, Kenyon, Koenecke, Kozon, LaMarr, Lambertson, Larson, Lentine, Lewis, Lilly, Liu, Malonis, Manthripragada, Markwardt, Matonak, Mcginnis, Miller, Mitchell, Mitchell, Mohammed, Monroe, Montt De~Garcia, Mulé, Nagao, Ngo, Norris, Norwood, Novotka, Okajima, Olsen, Onyeachu, Orosco, Peterson, Pevear, Pham, Pollard, Pope, Powers, Powers, Price, Prigozhin, Ramirez, Reid, Remillard, Rogstad, Rosecrans, Rowe, Sager, Sanders, Savadkin, Saylor, Schaeffer, Schweiss, Semper, Serlemitsos, Shackelford, Soong, Struebel, Vezie, Villasenor, Winternitz, Wofford, Wright, Yang and Yu}]{den_herder_neutron_2016}
\bibinfo{author}{Gendreau, K.C.}, \bibinfo{author}{Arzoumanian, Z.}, \bibinfo{author}{Adkins, P.W.}, \bibinfo{author}{Albert, C.L.}, \bibinfo{author}{Anders, J.F.}, \bibinfo{author}{Aylward, A.T.}, \bibinfo{author}{Baker, C.L.}, \bibinfo{author}{Balsamo, E.R.}, \bibinfo{author}{Bamford, W.A.}, \bibinfo{author}{Benegalrao, S.S.}, \bibinfo{author}{Berry, D.L.}, \bibinfo{author}{Bhalwani, S.}, \bibinfo{author}{Black, J.K.}, \bibinfo{author}{Blaurock, C.}, \bibinfo{author}{Bronke, G.M.}, \bibinfo{author}{Brown, G.L.}, \bibinfo{author}{Budinoff, J.G.}, \bibinfo{author}{Cantwell, J.D.}, \bibinfo{author}{Cazeau, T.}, \bibinfo{author}{Chen, P.T.}, \bibinfo{author}{Clement, T.G.}, \bibinfo{author}{Colangelo, A.T.}, \bibinfo{author}{Coleman, J.S.}, \bibinfo{author}{Coopersmith, J.D.}, \bibinfo{author}{Dehaven, W.E.}, \bibinfo{author}{Doty, J.P.}, \bibinfo{author}{Egan, M.D.}, \bibinfo{author}{Enoto, T.}, \bibinfo{author}{Fan, T.W.}, \bibinfo{author}{Ferro, D.M.}, \bibinfo{author}{Foster, R.}, \bibinfo{author}{Galassi,
  N.M.}, \bibinfo{author}{Gallo, L.D.}, \bibinfo{author}{Green, C.M.}, \bibinfo{author}{Grosh, D.}, \bibinfo{author}{Ha, K.Q.}, \bibinfo{author}{Hasouneh, M.A.}, \bibinfo{author}{Heefner, K.B.}, \bibinfo{author}{Hestnes, P.}, \bibinfo{author}{Hoge, L.J.}, \bibinfo{author}{Jacobs, T.M.}, \bibinfo{author}{Jørgensen, J.L.}, \bibinfo{author}{Kaiser, M.A.}, \bibinfo{author}{Kellogg, J.W.}, \bibinfo{author}{Kenyon, S.J.}, \bibinfo{author}{Koenecke, R.G.}, \bibinfo{author}{Kozon, R.P.}, \bibinfo{author}{LaMarr, B.}, \bibinfo{author}{Lambertson, M.D.}, \bibinfo{author}{Larson, A.M.}, \bibinfo{author}{Lentine, S.}, \bibinfo{author}{Lewis, J.H.}, \bibinfo{author}{Lilly, M.G.}, \bibinfo{author}{Liu, K.A.}, \bibinfo{author}{Malonis, A.}, \bibinfo{author}{Manthripragada, S.S.}, \bibinfo{author}{Markwardt, C.B.}, \bibinfo{author}{Matonak, B.D.}, \bibinfo{author}{Mcginnis, I.E.}, \bibinfo{author}{Miller, R.L.}, \bibinfo{author}{Mitchell, A.L.}, \bibinfo{author}{Mitchell, J.W.}, \bibinfo{author}{Mohammed, J.S.},
  \bibinfo{author}{Monroe, C.A.}, \bibinfo{author}{Montt De~Garcia, K.M.}, \bibinfo{author}{Mulé, P.D.}, \bibinfo{author}{Nagao, L.T.}, \bibinfo{author}{Ngo, S.N.}, \bibinfo{author}{Norris, E.D.}, \bibinfo{author}{Norwood, D.A.}, \bibinfo{author}{Novotka, J.}, \bibinfo{author}{Okajima, T.}, \bibinfo{author}{Olsen, L.G.}, \bibinfo{author}{Onyeachu, C.O.}, \bibinfo{author}{Orosco, H.Y.}, \bibinfo{author}{Peterson, J.R.}, \bibinfo{author}{Pevear, K.N.}, \bibinfo{author}{Pham, K.K.}, \bibinfo{author}{Pollard, S.E.}, \bibinfo{author}{Pope, J.S.}, \bibinfo{author}{Powers, D.F.}, \bibinfo{author}{Powers, C.E.}, \bibinfo{author}{Price, S.R.}, \bibinfo{author}{Prigozhin, G.Y.}, \bibinfo{author}{Ramirez, J.B.}, \bibinfo{author}{Reid, W.J.}, \bibinfo{author}{Remillard, R.A.}, \bibinfo{author}{Rogstad, E.M.}, \bibinfo{author}{Rosecrans, G.P.}, \bibinfo{author}{Rowe, J.N.}, \bibinfo{author}{Sager, J.A.}, \bibinfo{author}{Sanders, C.A.}, \bibinfo{author}{Savadkin, B.}, \bibinfo{author}{Saylor, M.R.},
  \bibinfo{author}{Schaeffer, A.F.}, \bibinfo{author}{Schweiss, N.S.}, \bibinfo{author}{Semper, S.R.}, \bibinfo{author}{Serlemitsos, P.J.}, \bibinfo{author}{Shackelford, L.V.}, \bibinfo{author}{Soong, Y.}, \bibinfo{author}{Struebel, J.}, \bibinfo{author}{Vezie, M.L.}, \bibinfo{author}{Villasenor, J.S.}, \bibinfo{author}{Winternitz, L.B.}, \bibinfo{author}{Wofford, G.I.}, \bibinfo{author}{Wright, M.R.}, \bibinfo{author}{Yang, M.Y.}, \bibinfo{author}{Yu, W.H.}, \bibinfo{year}{2016}.
\newblock \bibinfo{title}{The {Neutron} star {Interior} {Composition} {Explorer} ({NICER}): design and development}, \bibinfo{address}{Edinburgh, United Kingdom}. p. \bibinfo{pages}{99051H}.
\newblock \URLprefix \url{http://proceedings.spiedigitallibrary.org/proceeding.aspx?doi=10.1117/12.2231304}, \DOIprefix\doi{10.1117/12.2231304}.
\bibitem[{Guver et~al.(2022)Guver, Bostanci, Boztepe, Gogus, Bult, Kashyap, Chakraborty, Ballantyne, Ludlam, Malacaria, Jaisawal, Strohmayer, Guillot and Ng}]{guver_burstdisk_2022}
\bibinfo{author}{Guver, T.}, \bibinfo{author}{Bostanci, Z.F.}, \bibinfo{author}{Boztepe, T.}, \bibinfo{author}{Gogus, E.}, \bibinfo{author}{Bult, P.}, \bibinfo{author}{Kashyap, U.}, \bibinfo{author}{Chakraborty, M.}, \bibinfo{author}{Ballantyne, D.R.}, \bibinfo{author}{Ludlam, R.M.}, \bibinfo{author}{Malacaria, C.}, \bibinfo{author}{Jaisawal, G.K.}, \bibinfo{author}{Strohmayer, T.E.}, \bibinfo{author}{Guillot, S.}, \bibinfo{author}{Ng, M.}, \bibinfo{year}{2022}.
\newblock \bibinfo{title}{Burst–{Disk} {Interaction} in {4U} 1636–536 as {Observed} by {NICER}}.
\newblock \bibinfo{journal}{The Astrophysical Journal} \bibinfo{volume}{935}, \bibinfo{pages}{154}.
\newblock \URLprefix \url{https://iopscience.iop.org/article/10.3847/1538-4357/ac8106}, \DOIprefix\doi{10.3847/1538-4357/ac8106}.
\bibitem[{Güver et~al.(2021a)Güver, Boztepe, Ballantyne, Bostancı, Bult, Jaisawal, Göğüş, Strohmayer, Altamirano, Guillot and Chakrabarty}]{guver_nicer_2021}
\bibinfo{author}{Güver, T.}, \bibinfo{author}{Boztepe, T.}, \bibinfo{author}{Ballantyne, D.R.}, \bibinfo{author}{Bostancı, Z.F.}, \bibinfo{author}{Bult, P.}, \bibinfo{author}{Jaisawal, G.K.}, \bibinfo{author}{Göğüş, E.}, \bibinfo{author}{Strohmayer, T.E.}, \bibinfo{author}{Altamirano, D.}, \bibinfo{author}{Guillot, S.}, \bibinfo{author}{Chakrabarty, D.}, \bibinfo{year}{2021}a.
\newblock \bibinfo{title}{A {NICER} look at thermonuclear {X}-ray bursts from {Aql} {X}-1}.
\newblock \bibinfo{journal}{Monthly Notices of the Royal Astronomical Society} \bibinfo{volume}{510}, \bibinfo{pages}{1577--1596}.
\newblock \URLprefix \url{https://academic.oup.com/mnras/article/510/2/1577/6442240}, \DOIprefix\doi{10.1093/mnras/stab3422}.
\bibitem[{Güver et~al.(2021b)Güver, Boztepe, Göğüş, Chakraborty, Strohmayer, Bult, Altamirano, Jaisawal, Kocabıyık, Malacaria, Kashyap, Gendreau, Arzoumanian and Chakrabarty}]{guver_thermonuclear_2021}
\bibinfo{author}{Güver, T.}, \bibinfo{author}{Boztepe, T.}, \bibinfo{author}{Göğüş, E.}, \bibinfo{author}{Chakraborty, M.}, \bibinfo{author}{Strohmayer, T.E.}, \bibinfo{author}{Bult, P.}, \bibinfo{author}{Altamirano, D.}, \bibinfo{author}{Jaisawal, G.K.}, \bibinfo{author}{Kocabıyık, T.}, \bibinfo{author}{Malacaria, C.}, \bibinfo{author}{Kashyap, U.}, \bibinfo{author}{Gendreau, K.C.}, \bibinfo{author}{Arzoumanian, Z.}, \bibinfo{author}{Chakrabarty, D.}, \bibinfo{year}{2021}b.
\newblock \bibinfo{title}{Thermonuclear {X}-{Ray} {Bursts} with {Late} {Secondary} {Peaks} {Observed} from {4U} 1608–52}.
\newblock \bibinfo{journal}{The Astrophysical Journal} \bibinfo{volume}{910}, \bibinfo{pages}{37}.
\newblock \URLprefix \url{https://iopscience.iop.org/article/10.3847/1538-4357/abe1ae}, \DOIprefix\doi{10.3847/1538-4357/abe1ae}.
\bibitem[{Homan and Van Der~Klis(2000)}]{homan_695_2000}
\bibinfo{author}{Homan, J.}, \bibinfo{author}{Van Der~Klis, M.}, \bibinfo{year}{2000}.
\newblock \bibinfo{title}{A 695 {Hz} {Quasi}‐periodic {Oscillation} in the {Low}‐{Mass} {X}‐{Ray} {Binary} {EXO} 0748-676}.
\newblock \bibinfo{journal}{The Astrophysical Journal} \bibinfo{volume}{539}, \bibinfo{pages}{847--850}.
\newblock \URLprefix \url{https://iopscience.iop.org/article/10.1086/309280}, \DOIprefix\doi{10.1086/309280}.
\bibitem[{Homan et~al.(2003)Homan, Wijnands and Van Den~Berg}]{homan_xmm-newton_2003}
\bibinfo{author}{Homan, J.}, \bibinfo{author}{Wijnands, R.}, \bibinfo{author}{Van Den~Berg, M.}, \bibinfo{year}{2003}.
\newblock \bibinfo{title}{{XMM}-{Newton} light curves of the low-mass {X}-ray binary {EXO} 0748–676: {Dips}, eclipses, and bursts}.
\newblock \bibinfo{journal}{Astronomy \& Astrophysics} \bibinfo{volume}{412}, \bibinfo{pages}{799--812}.
\newblock \URLprefix \url{http://www.aanda.org/10.1051/0004-6361:20031484}, \DOIprefix\doi{10.1051/0004-6361:20031484}.
\bibitem[{Hynes et~al.(2006)Hynes, Horne, O’Brien, Haswell, Robinson, King, Charles and Pearson}]{hynes_multiwavelength_2006}
\bibinfo{author}{Hynes, R.I.}, \bibinfo{author}{Horne, K.}, \bibinfo{author}{O’Brien, K.}, \bibinfo{author}{Haswell, C.A.}, \bibinfo{author}{Robinson, E.L.}, \bibinfo{author}{King, A.R.}, \bibinfo{author}{Charles, P.A.}, \bibinfo{author}{Pearson, K.J.}, \bibinfo{year}{2006}.
\newblock \bibinfo{title}{Multiwavelength {Observations} of {EXO} 0748–676. {I}. {Reprocessing} of {X}‐{Ray} {Bursts}}.
\newblock \bibinfo{journal}{The Astrophysical Journal} \bibinfo{volume}{648}, \bibinfo{pages}{1156--1168}.
\newblock \URLprefix \url{https://iopscience.iop.org/article/10.1086/505592}, \DOIprefix\doi{10.1086/505592}.
\bibitem[{In~’T~Zand et~al.(2013)In~’T~Zand, Galloway, Marshall, Ballantyne, Jonker, Paerels, Palmer, Patruno and Weinberg}]{in_t_zand_bright_2013}
\bibinfo{author}{In~’T~Zand, J.J.M.}, \bibinfo{author}{Galloway, D.K.}, \bibinfo{author}{Marshall, H.L.}, \bibinfo{author}{Ballantyne, D.R.}, \bibinfo{author}{Jonker, P.G.}, \bibinfo{author}{Paerels, F.B.S.}, \bibinfo{author}{Palmer, D.M.}, \bibinfo{author}{Patruno, A.}, \bibinfo{author}{Weinberg, N.N.}, \bibinfo{year}{2013}.
\newblock \bibinfo{title}{A bright thermonuclear {X}-ray burst simultaneously observed with \textit{{Chandra}} and {RXTE}}.
\newblock \bibinfo{journal}{A\&A} \bibinfo{volume}{553}, \bibinfo{pages}{A83}.
\newblock \URLprefix \url{http://www.aanda.org/10.1051/0004-6361/201321056}, \DOIprefix\doi{10.1051/0004-6361/201321056}.
\bibitem[{Jaisawal et~al.(2019)Jaisawal, Chenevez, Bult, In’T~Zand, Galloway, Strohmayer, Güver, Adkins, Altamirano, Arzoumanian, Chakrabarty, Coopersmith, Gendreau, Guillot, Keek, Ludlam and Malacaria}]{jaisawal_nicer_2019}
\bibinfo{author}{Jaisawal, G.K.}, \bibinfo{author}{Chenevez, J.}, \bibinfo{author}{Bult, P.}, \bibinfo{author}{In’T~Zand, J.J.M.}, \bibinfo{author}{Galloway, D.K.}, \bibinfo{author}{Strohmayer, T.E.}, \bibinfo{author}{Güver, T.}, \bibinfo{author}{Adkins, P.}, \bibinfo{author}{Altamirano, D.}, \bibinfo{author}{Arzoumanian, Z.}, \bibinfo{author}{Chakrabarty, D.}, \bibinfo{author}{Coopersmith, J.}, \bibinfo{author}{Gendreau, K.C.}, \bibinfo{author}{Guillot, S.}, \bibinfo{author}{Keek, L.}, \bibinfo{author}{Ludlam, R.M.}, \bibinfo{author}{Malacaria, C.}, \bibinfo{year}{2019}.
\newblock \bibinfo{title}{{NICER} {Observes} a {Secondary} {Peak} in the {Decay} of a {Thermonuclear} {Burst} from {4U} 1608–52}.
\newblock \bibinfo{journal}{The Astrophysical Journal} \bibinfo{volume}{883}, \bibinfo{pages}{61}.
\newblock \URLprefix \url{https://iopscience.iop.org/article/10.3847/1538-4357/ab3a37}, \DOIprefix\doi{10.3847/1538-4357/ab3a37}.
\bibitem[{Ji et~al.(2014)Ji, Zhang, Chen, Zhang, Torres, Kretschmar and Li}]{ji_hard_2014}
\bibinfo{author}{Ji, L.}, \bibinfo{author}{Zhang, S.}, \bibinfo{author}{Chen, Y.}, \bibinfo{author}{Zhang, S.N.}, \bibinfo{author}{Torres, D.F.}, \bibinfo{author}{Kretschmar, P.}, \bibinfo{author}{Li, J.}, \bibinfo{year}{2014}.
\newblock \bibinfo{title}{{THE} {HARD} {X}-{RAY} {SHORTAGES} {PROMPTED} {BY} {THE} {CLOCK} {BURSTS} {IN} {GS} 1826-238}.
\newblock \bibinfo{journal}{The Astrophysical Journal} \bibinfo{volume}{782}, \bibinfo{pages}{40}.
\newblock \URLprefix \url{https://iopscience.iop.org/article/10.1088/0004-637X/782/1/40}, \DOIprefix\doi{10.1088/0004-637x/782/1/40}. \bibinfo{note}{publisher: American Astronomical Society}.
\bibitem[{Johnston et~al.(2020)Johnston, Heger and Galloway}]{johnston_multi-epoch_2020}
\bibinfo{author}{Johnston, Z.}, \bibinfo{author}{Heger, A.}, \bibinfo{author}{Galloway, D.K.}, \bibinfo{year}{2020}.
\newblock \bibinfo{title}{Multi-epoch {X}-ray burst modelling: {MCMC} with large grids of {1D} simulations}.
\newblock \bibinfo{journal}{Monthly Notices of the Royal Astronomical Society} \bibinfo{volume}{494}, \bibinfo{pages}{4576--4589}.
\newblock \URLprefix \url{https://academic.oup.com/mnras/article/494/3/4576/5822790}, \DOIprefix\doi{10.1093/mnras/staa1054}.
\bibitem[{Joseph et~al.(2021)Joseph, Stalin, Tandon and Ghosh}]{joseph_curvit_2021}
\bibinfo{author}{Joseph, P.}, \bibinfo{author}{Stalin, C.S.}, \bibinfo{author}{Tandon, S.N.}, \bibinfo{author}{Ghosh, S.K.}, \bibinfo{year}{2021}.
\newblock \bibinfo{title}{Curvit: {An} open-source {Python} package to generate light curves from {UVIT} data}.
\newblock \bibinfo{journal}{J Astrophys Astron} \bibinfo{volume}{42}, \bibinfo{pages}{25}.
\newblock \URLprefix \url{https://link.springer.com/10.1007/s12036-020-09680-5}, \DOIprefix\doi{10.1007/s12036-020-09680-5}.
\bibitem[{{Kashyap} et~al.(2024){Kashyap}, {Aromal}, {Chakraborty}, {Bhattacharyya} and {Maccarone}}]{Kashyap2024}
\bibinfo{author}{{Kashyap}, U.}, \bibinfo{author}{{Aromal}, P.}, \bibinfo{author}{{Chakraborty}, M.}, \bibinfo{author}{{Bhattacharyya}, S.}, \bibinfo{author}{{Maccarone}, T.J.}, \bibinfo{year}{2024}.
\newblock \bibinfo{title}{{AstroSat detects 3 Type-I bursts from EXO 0748-676}}.
\newblock \bibinfo{journal}{The Astronomer's Telegram} \bibinfo{volume}{16753}, \bibinfo{pages}{1}.
\bibitem[{Kashyap et~al.(2022)Kashyap, Chakraborty and Bhattacharyya}]{kashyap_probing_2022}
\bibinfo{author}{Kashyap, U.}, \bibinfo{author}{Chakraborty, M.}, \bibinfo{author}{Bhattacharyya, S.}, \bibinfo{year}{2022}.
\newblock \bibinfo{title}{Probing spectral and temporal evolution of the neutron star low-mass {X}-ray binary {4U} 1724–30 with \textit{{AstroSat}}}.
\newblock \bibinfo{journal}{Monthly Notices of the Royal Astronomical Society} \bibinfo{volume}{512}, \bibinfo{pages}{6180--6191}.
\newblock \URLprefix \url{https://academic.oup.com/mnras/article/512/4/6180/6563897}, \DOIprefix\doi{10.1093/mnras/stac908}.
\bibitem[{Kashyap et~al.(2021)Kashyap, Ram, Guver and Chakraborty}]{kashyap_broadband_2021}
\bibinfo{author}{Kashyap, U.}, \bibinfo{author}{Ram, B.}, \bibinfo{author}{Guver, T.}, \bibinfo{author}{Chakraborty, M.}, \bibinfo{year}{2021}.
\newblock \bibinfo{title}{Broadband time-resolved spectroscopy of thermonuclear {X}-ray bursts from {4U} 1636-536 using {AstroSat}}.
\newblock \bibinfo{journal}{Monthly Notices of the Royal Astronomical Society} \bibinfo{volume}{509}, \bibinfo{pages}{3989--4007}.
\newblock \URLprefix \url{http://arxiv.org/abs/2109.14631}, \DOIprefix\doi{10.1093/mnras/stab2838}. \bibinfo{note}{arXiv:2109.14631 [astro-ph]}.
\bibitem[{Keek et~al.(2018)Keek, Arzoumanian, Chakrabarty, Chenevez, Gendreau, Guillot, Güver, Homan, Jaisawal, LaMarr, Lamb, Mahmoodifar, Markwardt, Okajima, Strohmayer and In~’T~Zand}]{keek_nicer_2018}
\bibinfo{author}{Keek, L.}, \bibinfo{author}{Arzoumanian, Z.}, \bibinfo{author}{Chakrabarty, D.}, \bibinfo{author}{Chenevez, J.}, \bibinfo{author}{Gendreau, K.C.}, \bibinfo{author}{Guillot, S.}, \bibinfo{author}{Güver, T.}, \bibinfo{author}{Homan, J.}, \bibinfo{author}{Jaisawal, G.K.}, \bibinfo{author}{LaMarr, B.}, \bibinfo{author}{Lamb, F.K.}, \bibinfo{author}{Mahmoodifar, S.}, \bibinfo{author}{Markwardt, C.B.}, \bibinfo{author}{Okajima, T.}, \bibinfo{author}{Strohmayer, T.E.}, \bibinfo{author}{In~’T~Zand, J.J.M.}, \bibinfo{year}{2018}.
\newblock \bibinfo{title}{{NICER} {Detection} of {Strong} {Photospheric} {Expansion} during a {Thermonuclear} {X}-{Ray} {Burst} from {4U} 1820–30}.
\newblock \bibinfo{journal}{The Astrophysical Journal Letters} \bibinfo{volume}{856}, \bibinfo{pages}{L37}.
\newblock \URLprefix \url{https://iopscience.iop.org/article/10.3847/2041-8213/aab904}, \DOIprefix\doi{10.3847/2041-8213/aab904}.
\bibitem[{Knight et~al.(2022)Knight, Ingram, Middleton and Drake}]{knight_eclipse_2022}
\bibinfo{author}{Knight, A.H.}, \bibinfo{author}{Ingram, A.}, \bibinfo{author}{Middleton, M.}, \bibinfo{author}{Drake, J.}, \bibinfo{year}{2022}.
\newblock \bibinfo{title}{Eclipse mapping of {EXO} 0748–676: evidence for a massive neutron star}.
\newblock \bibinfo{journal}{Monthly Notices of the Royal Astronomical Society} \bibinfo{volume}{510}, \bibinfo{pages}{4736--4756}.
\newblock \URLprefix \url{https://academic.oup.com/mnras/article/510/4/4736/6481638}, \DOIprefix\doi{10.1093/mnras/stab3722}.
\bibitem[{Knight et~al.(2024)Knight, Rhodes, Buisson, Matthews, Segura, Ingram, Middleton and Roberts}]{knight_simultaneous_2024}
\bibinfo{author}{Knight, A.H.}, \bibinfo{author}{Rhodes, L.}, \bibinfo{author}{Buisson, D.J.K.}, \bibinfo{author}{Matthews, J.H.}, \bibinfo{author}{Segura, N.C.}, \bibinfo{author}{Ingram, A.}, \bibinfo{author}{Middleton, M.}, \bibinfo{author}{Roberts, T.P.}, \bibinfo{year}{2024}.
\newblock \bibinfo{title}{Simultaneous {Optical} and {X}-ray {Detection} of a {Thermonuclear} {Burst} in the 2024 {Outburst} of {EXO} 0748-676}.
\newblock \URLprefix \url{http://arxiv.org/abs/2411.03269}, \DOIprefix\doi{10.48550/arXiv.2411.03269}. \bibinfo{note}{arXiv:2411.03269 [astro-ph]}.
\bibitem[{Knight et~al.(2025)Knight, van den Eijnden, Ingram, Matthews, Motta, Middleton, Mancuso, Buisson, Altamirano, Fender and Roberts}]{knight_type_2025}
\bibinfo{author}{Knight, A.H.}, \bibinfo{author}{van den Eijnden, J.}, \bibinfo{author}{Ingram, A.}, \bibinfo{author}{Matthews, J.H.}, \bibinfo{author}{Motta, S.E.}, \bibinfo{author}{Middleton, M.}, \bibinfo{author}{Mancuso, G.C.}, \bibinfo{author}{Buisson, D.J.K.}, \bibinfo{author}{Altamirano, D.}, \bibinfo{author}{Fender, R.}, \bibinfo{author}{Roberts, T.P.}, \bibinfo{year}{2025}.
\newblock \bibinfo{title}{Type {I} {X}-ray burst emission reflected into the eclipses of {EXO} 0748-676}.
\newblock \bibinfo{journal}{Monthly Notices of the Royal Astronomical Society} \bibinfo{volume}{538}, \bibinfo{pages}{2058--2074}.
\newblock \URLprefix \url{https://academic.oup.com/mnras/article/538/3/2058/8063945}, \DOIprefix\doi{10.1093/mnras/staf395}.
\bibitem[{Kuulkers et~al.(2003)Kuulkers, Den~Hartog, In~'T~Zand, Verbunt, Harris and Cocchi}]{kuulkers_photospheric_2003}
\bibinfo{author}{Kuulkers, E.}, \bibinfo{author}{Den~Hartog, P.R.}, \bibinfo{author}{In~'T~Zand, J.J.M.}, \bibinfo{author}{Verbunt, F.W.M.}, \bibinfo{author}{Harris, W.E.}, \bibinfo{author}{Cocchi, M.}, \bibinfo{year}{2003}.
\newblock \bibinfo{title}{Photospheric radius expansion {X}-ray bursts as standard candles}.
\newblock \bibinfo{journal}{Astronomy \& Astrophysics} \bibinfo{volume}{399}, \bibinfo{pages}{663--680}.
\newblock \URLprefix \url{http://www.aanda.org/10.1051/0004-6361:20021781}, \DOIprefix\doi{10.1051/0004-6361:20021781}.
\bibitem[{Lewin et~al.(1993)Lewin, Van~Paradijs and Taam}]{lewin_x-ray_1993}
\bibinfo{author}{Lewin, W.H.G.}, \bibinfo{author}{Van~Paradijs, J.}, \bibinfo{author}{Taam, R.E.}, \bibinfo{year}{1993}.
\newblock \bibinfo{title}{X-ray bursts}.
\newblock \bibinfo{journal}{Space Science Reviews} \bibinfo{volume}{62}, \bibinfo{pages}{223--389}.
\newblock \URLprefix \url{http://link.springer.com/10.1007/BF00196124}, \DOIprefix\doi{10.1007/BF00196124}.
\bibitem[{Maccarone and Coppi(2003)}]{maccarone_spectral_2003}
\bibinfo{author}{Maccarone, T.J.}, \bibinfo{author}{Coppi, P.S.}, \bibinfo{year}{2003}.
\newblock \bibinfo{title}{Spectral {Fits} to the 1999 {Aql} {X}-1 {Outburst} {Data}}.
\newblock \bibinfo{journal}{A\&A} \bibinfo{volume}{399}, \bibinfo{pages}{1151--1157}.
\newblock \URLprefix \url{http://arxiv.org/abs/astro-ph/0301091}, \DOIprefix\doi{10.1051/0004-6361:20021881}. \bibinfo{note}{arXiv:astro-ph/0301091}.
\bibitem[{Mancuso et~al.(2019)Mancuso, Altamirano, García, Lyu, Méndez, Combi, Díaz-Trigo and in’t Zand}]{mancuso_discovery_2019}
\bibinfo{author}{Mancuso, G.C.}, \bibinfo{author}{Altamirano, D.}, \bibinfo{author}{García, F.}, \bibinfo{author}{Lyu, M.}, \bibinfo{author}{Méndez, M.}, \bibinfo{author}{Combi, J.A.}, \bibinfo{author}{Díaz-Trigo, M.}, \bibinfo{author}{in’t Zand, J.J.M.}, \bibinfo{year}{2019}.
\newblock \bibinfo{title}{Discovery of millihertz quasi-periodic oscillations in the {X}-ray binary {EXO} 0748-676}.
\newblock \bibinfo{journal}{Monthly Notices of the Royal Astronomical Society: Letters} \bibinfo{volume}{486}, \bibinfo{pages}{L74--L79}.
\newblock \URLprefix \url{https://academic.oup.com/mnrasl/article/486/1/L74/5487900}, \DOIprefix\doi{10.1093/mnrasl/slz057}.
\bibitem[{Parmar et~al.(1985)Parmar, Gottwald, Haberl, Giommi and White}]{parmar_exo0748-676_1985}
\bibinfo{author}{Parmar, A.N.}, \bibinfo{author}{Gottwald, M.}, \bibinfo{author}{Haberl, F.}, \bibinfo{author}{Giommi, P.}, \bibinfo{author}{White, N.E.}, \bibinfo{year}{1985}.
\newblock \bibinfo{title}{{EXO}:0748-676 - an {Exciting} {New} {X}-{Ray} {Transient}}, p. \bibinfo{pages}{119}.
\newblock \URLprefix \url{https://ui.adsabs.harvard.edu/abs/1985ESASP.236..119P}. \bibinfo{note}{aDS Bibcode: 1985ESASP.236..119P}.
\bibitem[{Parmar et~al.(1986)Parmar, White, Giommi and Gottwald}]{parmar_discovery_1986}
\bibinfo{author}{Parmar, A.N.}, \bibinfo{author}{White, N.E.}, \bibinfo{author}{Giommi, P.}, \bibinfo{author}{Gottwald, M.}, \bibinfo{year}{1986}.
\newblock \bibinfo{title}{The discovery of 3.8 hour periodic intensity dips and eclipses from the transient low-mass {X}-ray binary {EXO} 0748-676}.
\newblock \bibinfo{journal}{The Astrophysical Journal} \bibinfo{volume}{308}, \bibinfo{pages}{199}.
\newblock \URLprefix \url{http://adsabs.harvard.edu/doi/10.1086/164490}, \DOIprefix\doi{10.1086/164490}.
\bibitem[{Payne and Melatos(2006)}]{payne_frequency_2006}
\bibinfo{author}{Payne, D.J.B.}, \bibinfo{author}{Melatos, A.}, \bibinfo{year}{2006}.
\newblock \bibinfo{title}{Frequency {Spectrum} of {Gravitational} {Radiation} from {Global} {Hydromagnetic} {Oscillations} of a {Magnetically} {Confined} {Mountain} on an {Accreting} {Neutron} {Star}}.
\newblock \bibinfo{journal}{The Astrophysical Journal} \bibinfo{volume}{641}, \bibinfo{pages}{471--478}.
\newblock \URLprefix \url{https://iopscience.iop.org/article/10.1086/498855}, \DOIprefix\doi{10.1086/498855}.
\bibitem[{Rikame et~al.(2025)Rikame, Paul, Sharma and Jithesh}]{rikame_thermonuclear_2025}
\bibinfo{author}{Rikame, K.}, \bibinfo{author}{Paul, B.}, \bibinfo{author}{Sharma, R.}, \bibinfo{author}{Jithesh, V.}, \bibinfo{year}{2025}.
\newblock \bibinfo{title}{Thermonuclear {X}-ray bursts across the eclipse transitions in the {LMXBs} {EXO} 0748-676 and {XTE} {J1710}-281}.
\newblock \URLprefix \url{http://arxiv.org/abs/2509.13299}, \DOIprefix\doi{10.48550/arXiv.2509.13299}. \bibinfo{note}{arXiv:2509.13299 [astro-ph]}.
\bibitem[{Schaefer(2025)}]{schaefer_evolutionary_2025}
\bibinfo{author}{Schaefer, B.E.}, \bibinfo{year}{2025}.
\newblock \bibinfo{title}{Evolutionary {Period} {Changes} for 25 {X}-ray {Binaries} and the {Measurement} of an {Empirical} {Universal} {Law} for {Angular} {Momentum} {Loss} in {Accreting} {Binaries}}.
\newblock \URLprefix \url{http://arxiv.org/abs/2507.13515}, \DOIprefix\doi{10.48550/arXiv.2507.13515}. \bibinfo{note}{arXiv:2507.13515 [astro-ph]}.
\bibitem[{Singh et~al.(2017)Singh, Stewart, Westergaard, Bhattacharayya, Chandra, Chitnis, Dewangan, Kothare, Mirza, Mukerjee, Navalkar, Shah, Abbey, Beardmore, Kotak, Kamble, Vishwakarama, Pathare, Risbud, Koyande, Stevenson, Bicknell, Crawford, Hansford, Peters, Sykes, Agarwal, Sebastian, Rajarajan, Nagesh, Narendra, Ramesh, Rai, Navalgund, Sarma, Pandiyan, Subbarao, Gupta, Thakkar, Singh and Bajpai}]{singh_soft_2017}
\bibinfo{author}{Singh, K.P.}, \bibinfo{author}{Stewart, G.C.}, \bibinfo{author}{Westergaard, N.J.}, \bibinfo{author}{Bhattacharayya, S.}, \bibinfo{author}{Chandra, S.}, \bibinfo{author}{Chitnis, V.R.}, \bibinfo{author}{Dewangan, G.C.}, \bibinfo{author}{Kothare, A.T.}, \bibinfo{author}{Mirza, I.M.}, \bibinfo{author}{Mukerjee, K.}, \bibinfo{author}{Navalkar, V.}, \bibinfo{author}{Shah, H.}, \bibinfo{author}{Abbey, A.F.}, \bibinfo{author}{Beardmore, A.P.}, \bibinfo{author}{Kotak, S.}, \bibinfo{author}{Kamble, N.}, \bibinfo{author}{Vishwakarama, S.}, \bibinfo{author}{Pathare, D.P.}, \bibinfo{author}{Risbud, V.M.}, \bibinfo{author}{Koyande, J.P.}, \bibinfo{author}{Stevenson, T.}, \bibinfo{author}{Bicknell, C.}, \bibinfo{author}{Crawford, T.}, \bibinfo{author}{Hansford, G.}, \bibinfo{author}{Peters, G.}, \bibinfo{author}{Sykes, J.}, \bibinfo{author}{Agarwal, P.}, \bibinfo{author}{Sebastian, M.}, \bibinfo{author}{Rajarajan, A.}, \bibinfo{author}{Nagesh, G.}, \bibinfo{author}{Narendra, S.}, \bibinfo{author}{Ramesh,
  M.}, \bibinfo{author}{Rai, R.}, \bibinfo{author}{Navalgund, K.H.}, \bibinfo{author}{Sarma, K.S.}, \bibinfo{author}{Pandiyan, R.}, \bibinfo{author}{Subbarao, K.}, \bibinfo{author}{Gupta, T.}, \bibinfo{author}{Thakkar, N.}, \bibinfo{author}{Singh, A.K.}, \bibinfo{author}{Bajpai, A.}, \bibinfo{year}{2017}.
\newblock \bibinfo{title}{Soft {X}-ray {Focusing} {Telescope} {Aboard} {AstroSat}: {Design}, {Characteristics} and {Performance}}.
\newblock \bibinfo{journal}{Journal of Astrophysics and Astronomy} \bibinfo{volume}{38}, \bibinfo{pages}{29}.
\newblock \URLprefix \url{http://link.springer.com/10.1007/s12036-017-9448-7}, \DOIprefix\doi{10.1007/s12036-017-9448-7}.
\bibitem[{Strohmayer and Bildsten(2003)}]{strohmayer_new_2003}
\bibinfo{author}{Strohmayer, T.}, \bibinfo{author}{Bildsten, L.}, \bibinfo{year}{2003}.
\newblock \bibinfo{title}{New {Views} of {Thermonuclear} {Bursts}}.
\newblock \URLprefix \url{http://arxiv.org/abs/astro-ph/0301544}, \DOIprefix\doi{10.48550/arXiv.astro-ph/0301544}. \bibinfo{note}{arXiv:astro-ph/0301544}.
\bibitem[{Subba et~al.(2024)Subba, Subba, Paul, Sharma and Ghimiray}]{subba_eclipse_2024}
\bibinfo{author}{Subba, N.}, \bibinfo{author}{Subba, N.}, \bibinfo{author}{Paul, J.}, \bibinfo{author}{Sharma, P.}, \bibinfo{author}{Ghimiray, M.}, \bibinfo{year}{2024}.
\newblock \bibinfo{title}{Eclipse {Dynamics} and {X}-ray {Burst} {Characteristics} in the {Low}-{Mass} {X}-ray {Binary} {EXO} 0748-676}.
\newblock \URLprefix \url{http://arxiv.org/abs/2410.06201}, \DOIprefix\doi{10.48550/arXiv.2410.06201}. \bibinfo{note}{arXiv:2410.06201 [astro-ph]}.
\bibitem[{Tandon et~al.(2017)Tandon, Subramaniam, Girish, Postma, Sankarasubramanian, Sriram, Stalin, Mondal, Sahu, Joseph, Hutchings, Ghosh, Barve, George, Kamath, Kathiravan, Kumar, Lancelot, Leahy, Mahesh, Mohan, Nagabhushana, Pati, Kameswara~Rao, Sreedhar and Sreekumar}]{tandon_-orbit_2017}
\bibinfo{author}{Tandon, S.N.}, \bibinfo{author}{Subramaniam, A.}, \bibinfo{author}{Girish, V.}, \bibinfo{author}{Postma, J.}, \bibinfo{author}{Sankarasubramanian, K.}, \bibinfo{author}{Sriram, S.}, \bibinfo{author}{Stalin, C.S.}, \bibinfo{author}{Mondal, C.}, \bibinfo{author}{Sahu, S.}, \bibinfo{author}{Joseph, P.}, \bibinfo{author}{Hutchings, J.}, \bibinfo{author}{Ghosh, S.K.}, \bibinfo{author}{Barve, I.V.}, \bibinfo{author}{George, K.}, \bibinfo{author}{Kamath, P.U.}, \bibinfo{author}{Kathiravan, S.}, \bibinfo{author}{Kumar, A.}, \bibinfo{author}{Lancelot, J.P.}, \bibinfo{author}{Leahy, D.}, \bibinfo{author}{Mahesh, P.K.}, \bibinfo{author}{Mohan, R.}, \bibinfo{author}{Nagabhushana, S.}, \bibinfo{author}{Pati, A.K.}, \bibinfo{author}{Kameswara~Rao, N.}, \bibinfo{author}{Sreedhar, Y.H.}, \bibinfo{author}{Sreekumar, P.}, \bibinfo{year}{2017}.
\newblock \bibinfo{title}{In-orbit {Calibrations} of the {Ultraviolet} {Imaging} {Telescope}}.
\newblock \bibinfo{journal}{The Astronomical Journal} \bibinfo{volume}{154}, \bibinfo{pages}{128}.
\newblock \URLprefix \url{https://iopscience.iop.org/article/10.3847/1538-3881/aa8451}, \DOIprefix\doi{10.3847/1538-3881/aa8451}.
\bibitem[{Villarreal and Strohmayer(2004)}]{villarreal_discovery_2004}
\bibinfo{author}{Villarreal, A.R.}, \bibinfo{author}{Strohmayer, T.E.}, \bibinfo{year}{2004}.
\newblock \bibinfo{title}{Discovery of the {Neutron} {Star} {Spin} {Frequency} in {EXO} 0748-676}.
\newblock \bibinfo{journal}{The Astrophysical Journal} \bibinfo{volume}{614}, \bibinfo{pages}{L121--L124}.
\newblock \URLprefix \url{https://iopscience.iop.org/article/10.1086/425737}, \DOIprefix\doi{10.1086/425737}.
\bibitem[{Virtanen et~al.(2020)Virtanen, Gommers, Oliphant, Haberland, Reddy, Cournapeau, Burovski, Peterson, Weckesser, Bright, Van Der~Walt, Brett, Wilson, Millman, Mayorov, Nelson, Jones, Kern, Larson, Carey, Polat, Feng, Moore, VanderPlas, Laxalde, Perktold, Cimrman, Henriksen, Quintero, Harris, Archibald, Ribeiro, Pedregosa, Van~Mulbregt, {SciPy 1.0 Contributors}, Vijaykumar, Bardelli, Rothberg, Hilboll, Kloeckner, Scopatz, Lee, Rokem, Woods, Fulton, Masson, Häggström, Fitzgerald, Nicholson, Hagen, Pasechnik, Olivetti, Martin, Wieser, Silva, Lenders, Wilhelm, Young, Price, Ingold, Allen, Lee, Audren, Probst, Dietrich, Silterra, Webber, Slavič, Nothman, Buchner, Kulick, Schönberger, De~Miranda~Cardoso, Reimer, Harrington, Rodríguez, Nunez-Iglesias, Kuczynski, Tritz, Thoma, Newville, Kümmerer, Bolingbroke, Tartre, Pak, Smith, Nowaczyk, Shebanov, Pavlyk, Brodtkorb, Lee, McGibbon, Feldbauer, Lewis, Tygier, Sievert, Vigna, Peterson, More, Pudlik, Oshima, Pingel, Robitaille, Spura, Jones, Cera, Leslie,
  Zito, Krauss, Upadhyay, Halchenko and Vázquez-Baeza}]{virtanen_scipy_2020}
\bibinfo{author}{Virtanen, P.}, \bibinfo{author}{Gommers, R.}, \bibinfo{author}{Oliphant, T.E.}, \bibinfo{author}{Haberland, M.}, \bibinfo{author}{Reddy, T.}, \bibinfo{author}{Cournapeau, D.}, \bibinfo{author}{Burovski, E.}, \bibinfo{author}{Peterson, P.}, \bibinfo{author}{Weckesser, W.}, \bibinfo{author}{Bright, J.}, \bibinfo{author}{Van Der~Walt, S.J.}, \bibinfo{author}{Brett, M.}, \bibinfo{author}{Wilson, J.}, \bibinfo{author}{Millman, K.J.}, \bibinfo{author}{Mayorov, N.}, \bibinfo{author}{Nelson, A.R.J.}, \bibinfo{author}{Jones, E.}, \bibinfo{author}{Kern, R.}, \bibinfo{author}{Larson, E.}, \bibinfo{author}{Carey, C.J.}, \bibinfo{author}{Polat, I.}, \bibinfo{author}{Feng, Y.}, \bibinfo{author}{Moore, E.W.}, \bibinfo{author}{VanderPlas, J.}, \bibinfo{author}{Laxalde, D.}, \bibinfo{author}{Perktold, J.}, \bibinfo{author}{Cimrman, R.}, \bibinfo{author}{Henriksen, I.}, \bibinfo{author}{Quintero, E.A.}, \bibinfo{author}{Harris, C.R.}, \bibinfo{author}{Archibald, A.M.}, \bibinfo{author}{Ribeiro, A.H.},
  \bibinfo{author}{Pedregosa, F.}, \bibinfo{author}{Van~Mulbregt, P.}, \bibinfo{author}{{SciPy 1.0 Contributors}}, \bibinfo{author}{Vijaykumar, A.}, \bibinfo{author}{Bardelli, A.P.}, \bibinfo{author}{Rothberg, A.}, \bibinfo{author}{Hilboll, A.}, \bibinfo{author}{Kloeckner, A.}, \bibinfo{author}{Scopatz, A.}, \bibinfo{author}{Lee, A.}, \bibinfo{author}{Rokem, A.}, \bibinfo{author}{Woods, C.N.}, \bibinfo{author}{Fulton, C.}, \bibinfo{author}{Masson, C.}, \bibinfo{author}{Häggström, C.}, \bibinfo{author}{Fitzgerald, C.}, \bibinfo{author}{Nicholson, D.A.}, \bibinfo{author}{Hagen, D.R.}, \bibinfo{author}{Pasechnik, D.V.}, \bibinfo{author}{Olivetti, E.}, \bibinfo{author}{Martin, E.}, \bibinfo{author}{Wieser, E.}, \bibinfo{author}{Silva, F.}, \bibinfo{author}{Lenders, F.}, \bibinfo{author}{Wilhelm, F.}, \bibinfo{author}{Young, G.}, \bibinfo{author}{Price, G.A.}, \bibinfo{author}{Ingold, G.L.}, \bibinfo{author}{Allen, G.E.}, \bibinfo{author}{Lee, G.R.}, \bibinfo{author}{Audren, H.}, \bibinfo{author}{Probst, I.},
  \bibinfo{author}{Dietrich, J.P.}, \bibinfo{author}{Silterra, J.}, \bibinfo{author}{Webber, J.T.}, \bibinfo{author}{Slavič, J.}, \bibinfo{author}{Nothman, J.}, \bibinfo{author}{Buchner, J.}, \bibinfo{author}{Kulick, J.}, \bibinfo{author}{Schönberger, J.L.}, \bibinfo{author}{De~Miranda~Cardoso, J.V.}, \bibinfo{author}{Reimer, J.}, \bibinfo{author}{Harrington, J.}, \bibinfo{author}{Rodríguez, J.L.C.}, \bibinfo{author}{Nunez-Iglesias, J.}, \bibinfo{author}{Kuczynski, J.}, \bibinfo{author}{Tritz, K.}, \bibinfo{author}{Thoma, M.}, \bibinfo{author}{Newville, M.}, \bibinfo{author}{Kümmerer, M.}, \bibinfo{author}{Bolingbroke, M.}, \bibinfo{author}{Tartre, M.}, \bibinfo{author}{Pak, M.}, \bibinfo{author}{Smith, N.J.}, \bibinfo{author}{Nowaczyk, N.}, \bibinfo{author}{Shebanov, N.}, \bibinfo{author}{Pavlyk, O.}, \bibinfo{author}{Brodtkorb, P.A.}, \bibinfo{author}{Lee, P.}, \bibinfo{author}{McGibbon, R.T.}, \bibinfo{author}{Feldbauer, R.}, \bibinfo{author}{Lewis, S.}, \bibinfo{author}{Tygier, S.},
  \bibinfo{author}{Sievert, S.}, \bibinfo{author}{Vigna, S.}, \bibinfo{author}{Peterson, S.}, \bibinfo{author}{More, S.}, \bibinfo{author}{Pudlik, T.}, \bibinfo{author}{Oshima, T.}, \bibinfo{author}{Pingel, T.J.}, \bibinfo{author}{Robitaille, T.P.}, \bibinfo{author}{Spura, T.}, \bibinfo{author}{Jones, T.R.}, \bibinfo{author}{Cera, T.}, \bibinfo{author}{Leslie, T.}, \bibinfo{author}{Zito, T.}, \bibinfo{author}{Krauss, T.}, \bibinfo{author}{Upadhyay, U.}, \bibinfo{author}{Halchenko, Y.O.}, \bibinfo{author}{Vázquez-Baeza, Y.}, \bibinfo{year}{2020}.
\newblock \bibinfo{title}{{SciPy} 1.0: fundamental algorithms for scientific computing in {Python}}.
\newblock \bibinfo{journal}{Nat Methods} \bibinfo{volume}{17}, \bibinfo{pages}{261--272}.
\newblock \URLprefix \url{https://www.nature.com/articles/s41592-019-0686-2}, \DOIprefix\doi{10.1038/s41592-019-0686-2}.
\bibitem[{Wilms et~al.(2000)Wilms, Allen and McCray}]{wilms_absorption_2000}
\bibinfo{author}{Wilms, J.}, \bibinfo{author}{Allen, A.}, \bibinfo{author}{McCray, R.}, \bibinfo{year}{2000}.
\newblock \bibinfo{title}{On the {Absorption} of {X}-rays in the {Interstellar} {Medium}}.
\newblock \bibinfo{journal}{ApJ} \bibinfo{volume}{542}, \bibinfo{pages}{914--924}.
\newblock \URLprefix \url{http://arxiv.org/abs/astro-ph/0008425}, \DOIprefix\doi{10.1086/317016}. \bibinfo{note}{arXiv:astro-ph/0008425}.
\bibitem[{Wolff et~al.(2005)Wolff, Becker, Ray and Wood}]{wolff_strong_2005}
\bibinfo{author}{Wolff, M.T.}, \bibinfo{author}{Becker, P.A.}, \bibinfo{author}{Ray, P.S.}, \bibinfo{author}{Wood, K.S.}, \bibinfo{year}{2005}.
\newblock \bibinfo{title}{A {Strong} {X}‐{Ray} {Burst} from the {Low}‐{Mass} {X}‐{Ray} {Binary} {EXO} 0748-676}.
\newblock \bibinfo{journal}{The Astrophysical Journal} \bibinfo{volume}{632}, \bibinfo{pages}{1099--1103}.
\newblock \URLprefix \url{https://iopscience.iop.org/article/10.1086/444348}, \DOIprefix\doi{10.1086/444348}.
\bibitem[{Wolff et~al.(2009)Wolff, Ray, Wood and Hertz}]{wolff_eclipse_2009}
\bibinfo{author}{Wolff, M.T.}, \bibinfo{author}{Ray, P.S.}, \bibinfo{author}{Wood, K.S.}, \bibinfo{author}{Hertz, P.L.}, \bibinfo{year}{2009}.
\newblock \bibinfo{title}{{ECLIPSE} {TIMINGS} {OF} {THE} {TRANSIENT} {LOW}-{MASS} {X}-{RAY} {BINARY} {EXO} 0748–676. {IV}. {THE} \textit{{ROSSI} {X}-{RAY} {TIMING} {EXPLORER}} {ECLIPSES}}.
\newblock \bibinfo{journal}{The Astrophysical Journal Supplement Series} \bibinfo{volume}{183}, \bibinfo{pages}{156--170}.
\newblock \URLprefix \url{https://iopscience.iop.org/article/10.1088/0067-0049/183/1/156}, \DOIprefix\doi{10.1088/0067-0049/183/1/156}.
\bibitem[{Worpel et~al.(2015)Worpel, Galloway and Price}]{worpel_evidence_2015}
\bibinfo{author}{Worpel, H.}, \bibinfo{author}{Galloway, D.K.}, \bibinfo{author}{Price, D.J.}, \bibinfo{year}{2015}.
\newblock \bibinfo{title}{{EVIDENCE} {FOR} {ENHANCED} {PERSISTENT} {EMISSION} {DURING} {SUB}-{EDDINGTON} {THERMONUCLEAR} {BURSTS}}.
\newblock \bibinfo{journal}{The Astrophysical Journal} \bibinfo{volume}{801}, \bibinfo{pages}{60}.
\newblock \URLprefix \url{https://iopscience.iop.org/article/10.1088/0004-637X/801/1/60}, \DOIprefix\doi{10.1088/0004-637X/801/1/60}.
\bibitem[{Yadav et~al.(2017)Yadav, Agrawal, Antia, Manchanda, Paul and Misra}]{yadav_large_2017}
\bibinfo{author}{Yadav, J.S.}, \bibinfo{author}{Agrawal, P.C.}, \bibinfo{author}{Antia, H.M.}, \bibinfo{author}{Manchanda, R.K.}, \bibinfo{author}{Paul, B.}, \bibinfo{author}{Misra, R.}, \bibinfo{year}{2017}.
\newblock \bibinfo{title}{Large {Area} {X}-ray {Proportional} {Counter} instrument on {AstroSat}}.
\newblock \bibinfo{journal}{Current Science} \bibinfo{volume}{113}, \bibinfo{pages}{591}.
\newblock \URLprefix \url{https://ui.adsabs.harvard.edu/abs/2017CSci..113..591Y}, \DOIprefix\doi{10.18520/cs/v113/i04/591-594}. \bibinfo{note}{aDS Bibcode: 2017CSci..113..591Y}.
\bibitem[{Yan and Xie(2018)}]{yan_decades-long_2018}
\bibinfo{author}{Yan, Z.}, \bibinfo{author}{Xie, F.G.}, \bibinfo{year}{2018}.
\newblock \bibinfo{title}{A decades-long fast-rise-exponential-decay flare in low-luminosity {AGN} {NGC} 7213}.
\newblock \bibinfo{journal}{Monthly Notices of the Royal Astronomical Society} \bibinfo{volume}{475}, \bibinfo{pages}{1190--1197}.
\newblock \URLprefix \url{https://academic.oup.com/mnras/article/475/1/1190/4768276}, \DOIprefix\doi{10.1093/mnras/stx3259}.
\bibitem[{Zanon et~al.(2025)Zanon, Ambrosino, Illiano, Papitto, Israel, Zelati, Stella, Salvo, Campana, Benevento, Vago, Baglio, Casella, D'Avanzo, Martino, Imbrogno, Placa and Motta}]{zanon_ultraviolet_2025}
\bibinfo{author}{Zanon, A.M.}, \bibinfo{author}{Ambrosino, F.}, \bibinfo{author}{Illiano, G.}, \bibinfo{author}{Papitto, A.}, \bibinfo{author}{Israel, G.L.}, \bibinfo{author}{Zelati, F.C.}, \bibinfo{author}{Stella, L.}, \bibinfo{author}{Salvo, T.D.}, \bibinfo{author}{Campana, S.}, \bibinfo{author}{Benevento, G.}, \bibinfo{author}{Vago, N.O.P.}, \bibinfo{author}{Baglio, M.C.}, \bibinfo{author}{Casella, P.}, \bibinfo{author}{D'Avanzo, P.}, \bibinfo{author}{Martino, D.d.}, \bibinfo{author}{Imbrogno, M.}, \bibinfo{author}{Placa, R.L.}, \bibinfo{author}{Motta, S.E.}, \bibinfo{year}{2025}.
\newblock \bibinfo{title}{An ultraviolet burst oscillation candidate from the low-mass {X}-ray binary {EXO} 0748-676}.
\newblock \URLprefix \url{http://arxiv.org/abs/2509.10115}, \DOIprefix\doi{10.48550/arXiv.2509.10115}. \bibinfo{note}{arXiv:2509.10115 [astro-ph]}.
\bibitem[{Zhang et~al.(2011)Zhang, Méndez, Jonker and Hiemstra}]{zhang_distance_2011}
\bibinfo{author}{Zhang, G.}, \bibinfo{author}{Méndez, M.}, \bibinfo{author}{Jonker, P.}, \bibinfo{author}{Hiemstra, B.}, \bibinfo{year}{2011}.
\newblock \bibinfo{title}{The distance and internal composition of the neutron star in {EXO} 0748-676 with {XMM}-{Newton}: {EXO} 0748-676}.
\newblock \bibinfo{journal}{Monthly Notices of the Royal Astronomical Society} \bibinfo{volume}{414}, \bibinfo{pages}{1077--1081}.
\newblock \URLprefix \url{https://academic.oup.com/mnras/article-lookup/doi/10.1111/j.1365-2966.2011.18443.x}, \DOIprefix\doi{10.1111/j.1365-2966.2011.18443.x}.
\bibitem[{Zhang et~al.(2016)Zhang, Sakurai, Makishima, Nakazawa, Ono, Yamada and Xu}]{zhang_suzaku_2016}
\bibinfo{author}{Zhang, Z.}, \bibinfo{author}{Sakurai, S.}, \bibinfo{author}{Makishima, K.}, \bibinfo{author}{Nakazawa, K.}, \bibinfo{author}{Ono, K.}, \bibinfo{author}{Yamada, S.}, \bibinfo{author}{Xu, H.}, \bibinfo{year}{2016}.
\newblock \bibinfo{title}{\textit{{SUZAKU}} {OBSERVATION} {OF} {THE} {HIGH}-{INCLINATION} {BINARY} {EXO} 0748–676 {IN} {THE} {HARD} {STATE}}.
\newblock \bibinfo{journal}{The Astrophysical Journal} \bibinfo{volume}{823}, \bibinfo{pages}{131}.
\newblock \URLprefix \url{https://iopscience.iop.org/article/10.3847/0004-637X/823/2/131}, \DOIprefix\doi{10.3847/0004-637X/823/2/131}.
\bibitem[{Özel(2006)}]{ozel_soft_2006}
\bibinfo{author}{Özel, F.}, \bibinfo{year}{2006}.
\newblock \bibinfo{title}{Soft equations of state for neutron-star matter ruled out by {EXO} 0748 - 676}.
\newblock \bibinfo{journal}{Nature} \bibinfo{volume}{441}, \bibinfo{pages}{1115--1117}.
\newblock \URLprefix \url{https://www.nature.com/articles/nature04858}, \DOIprefix\doi{10.1038/nature04858}.

\end{thebibliography}

\end{document}